\documentclass{jfm}
\usepackage{graphicx}
\usepackage{epstopdf, epsfig}
\usepackage{bm}
\usepackage{amsmath}
\usepackage{caption}
\usepackage{subcaption}
\usepackage[LGRgreek]{mathastext}

\shorttitle{Shape dependence on the rise and sink of a wiggling intruder}
\shortauthor{Bitang K. Tripura, Sonu Kumar, Vamsi Krishna Reddy Anyam and K. Anki Reddy}

\title{The rise and sink dependence on the shape of a horizontally wiggling intruder}
\author{Bitang Kwrung Tripura\aff{1}
	Sonu Kumar\aff{2},
	 Vamsi Krishna Reddy Anyam\aff{3}
	\and K. Anki Reddy\aff{4}\corresp{\email{anki.reddy@iittp.ac.in}}}
\affiliation{\aff{1} Department of Chemical Engineering, Indian Institute of Technology, Guwahati, Assam, 781039, India
\aff{2} Department of Chemical and Biomolecular Engineering, University of Notre Dame, Notre Dame, Indiana 46556, USA
\aff{3} Department of Chemical Engineering, Indian Institute of Science, Bangalore 560 012, India
\aff{4}  Department of Chemical Engineering, Indian Institute of Technology, Tirupati, AndraPradesh, 517619, India}

\begin{document}

\maketitle

\begin{abstract}
	
We investigate the effect of shape and orientation for a horizontally oscillating intruder on its vertical dynamics in a granular medium via Discrete Element Method (DEM) simulations. Five distinct intruder shapes were considered in this study: a disk ($OS1$), a square ($OS2$), two rectangles with aspect ratios $>1$ ($OS3$) and $<1$ ($OS4$) respectively, and an equilateral triangle ($OS5$). The vertical velocity of the oscillating object was observed to be a function of amplitude and frequency $(f)$/time period $(T)$ of oscillation and the shape of the object. The dynamics of the motion are modelled with the help of a cavity-based model, which can empirically produce the regimes observed by incorporating the free-fall of the particles filling the cavity. The cavity-based model assumes that there is a point on the surface of each oscillating object unique to its shape that any grain crossing it will irreversibly reach the bottom of the cavity. This leads to the creation of a bed of particles at the bottom of the cavity, which increases the intruder's elevation as it oscillates, leading to its vertical rise in the granular medium. In addition, we observed a minimum amplitude ($A_{min}$) below which the vertical position of the intruder remains unaltered for various $T$. Among all the shapes, disk (OS1) displays the highest rising rate while triangle (OS5) shows the highest sinking rate. The cavity model can integrate the effect of shape and guess the efficiency of one shape over another in the rising dynamics. We have also presented the time-averaged pressure $P$, velocity $V$, and area fraction $\Phi$ fields around the intruder shapes.

\end{abstract}

\section{\label{sec:intro} Introduction}

Animals native to deserts have an extreme shortage of food and water, making any wastage of energy detrimental to them. As such, their bodies have undergone evolutionary changes to have the optimal shape required to walk on the surface or even exhibit subsurface motion with minimum loss of energy. For example, sand lizards \citep{maladen2009undulatory, goldman2010wiggling, sharpe2013environmental} and sidewinder snakes \citep{marvi2014sidewinding, ZHU2021104629} employ a wiggling motion to move in the sand. A comprehensive understanding of this kind of motion has various applications in sub-terrain robotics  \citep{raibert2008bigdog, zhang2013ground}, in rescue operations through sand or snow \citep{URQUHART201023}, and in mining related operations.

The subsurface motion of animals has garnered interest in the fields of earth-sciences  \citep{PhysRevLett.109.238302,PhysRevE.89.012201}, biology \citep{maladen2009undulatory, mazouchova2010utilization, RevModPhys.86.943,  sharpe2015controlled}, and engineering \citep{Hu10081, meirion2010empirical, li2013terradynamics} owing to its interesting dynamics. To understand the mechanism behind their movement through the granular medium, researchers \citep{maladen2009undulatory, sharpe2013environmental} conducted laboratory experiments on the creatures. \citet{maladen2009undulatory} studied how a sand lizard moves within the granular media with the help of high-speed x-ray imaging. Surprisingly, they noticed that the lizard does not use its limbs for propulsion, but it generates thrust by propagating a wavy motion of its body once it is under the surface. \citet{sharpe2013environmental} investigated the neuromechanical movement of sandfish lizard during its walking above the surface or sand swimming under the surface of a granular medium with the help of visible light imaging using electromyogram (EMG). The authors reported that the frictional forces play a prominent role in resisting the intrusion of the creatures. Moreover, they observed that the resistance to the undulatory motion of the creatures increases with the depth. Despite the resistance, the sand-dwelling creatures have adopted certain mechanisms to move within the bed of granular particles. \citet{maladen2011undulatory, maladen2011granular} have tried to comprehend this phenomenon by developing sand swimming robots. The ratio of forwarding speed to the wave speed of a robot was noticed \citep{maladen2011undulatory} to increases with an increase in the inter-particle friction but decreases with an increase in the particle-robot friction. In addition, \cite{maladen2011granular} investigated various head shapes of the robot and their influence on the drag-induced lift. The authors noticed that for certain wedge shapes of the robot, it reaches the free surface whereas, for other ones, it sinks to the bottom, indicating the importance of the robot's shape on its rise/sink dynamics. Apart from these studies, to gain a better insight into the creatures movement within granular media, they have been modelled as self-energized objects.

Few recent studies focused on understanding the dynamics of a self-energized object in granular media either translating (\citep{ding2011drag,guillard2014lift, kumar2017scaling,Bitang2021drag}) or oscillating horizontally in both two (\cite{PhysRevE.94.062906}) and three-dimensional systems (\cite{ping2020sink},\cite{liu2021locomotion}). \cite{PhysRevE.94.062906} explained the mechanism behind the rise and sink of a horizontally oscillating circular intruder in a two-dimensional system with the help of a mathematical model. They stated that the amplitude and frequency at which the intruder object oscillates mainly contribute to its rise or sink. They have also reported a minimum oscillation amplitude ($A_{min}$) below which the intruder does not sink irrespective of its frequency. \cite{ping2020sink} noticed that the intruder rises at low amplitude rates and sinks at high amplitude rates. They also found that an intruder might either rise or sink or remain at the same initial vertical position depending on its oscillating frequency and amplitude in a three-dimensional system. Further, the authors explained the driving mechanisms such as fluidization or cavity formation behind the intruder's vertical dynamics. 

The previous studies were concerned mostly with the sole effect of the intruder's horizontal wiggling motion on its vertical dynamics. However, a coupled effect of wiggling motion along with other properties of the intruder can enhance the rise/sink dynamics. To this extent, \cite{liu2021locomotion} studied the combined effect of rotation and oscillation of a spherical intruder on its rise/sink dependence within a three-dimensional granular media. The effect of intruder's shape, coupled with its wiggling motion has not been studied to the best of our knowledge. In this regard, we numerically investigated how the shape of the intruder affects its vertical motion induced by horizontal wiggling. We studied how the oscillation parameters affect their vertical dynamics for each of these shapes. Further, we probed the influence of the properties of the granular medium, such as the coefficient of friction on the intruder's vertical motion. Finally, we proposed a mathematical model that predicts the intruder's rise/sink dynamics depending on its shape and the oscillation parameters.
\begin{figure}
	\centering
	\begin{minipage}{0.65\textwidth}\hspace*{-1cm}
		\includegraphics[width=1.25\textwidth]{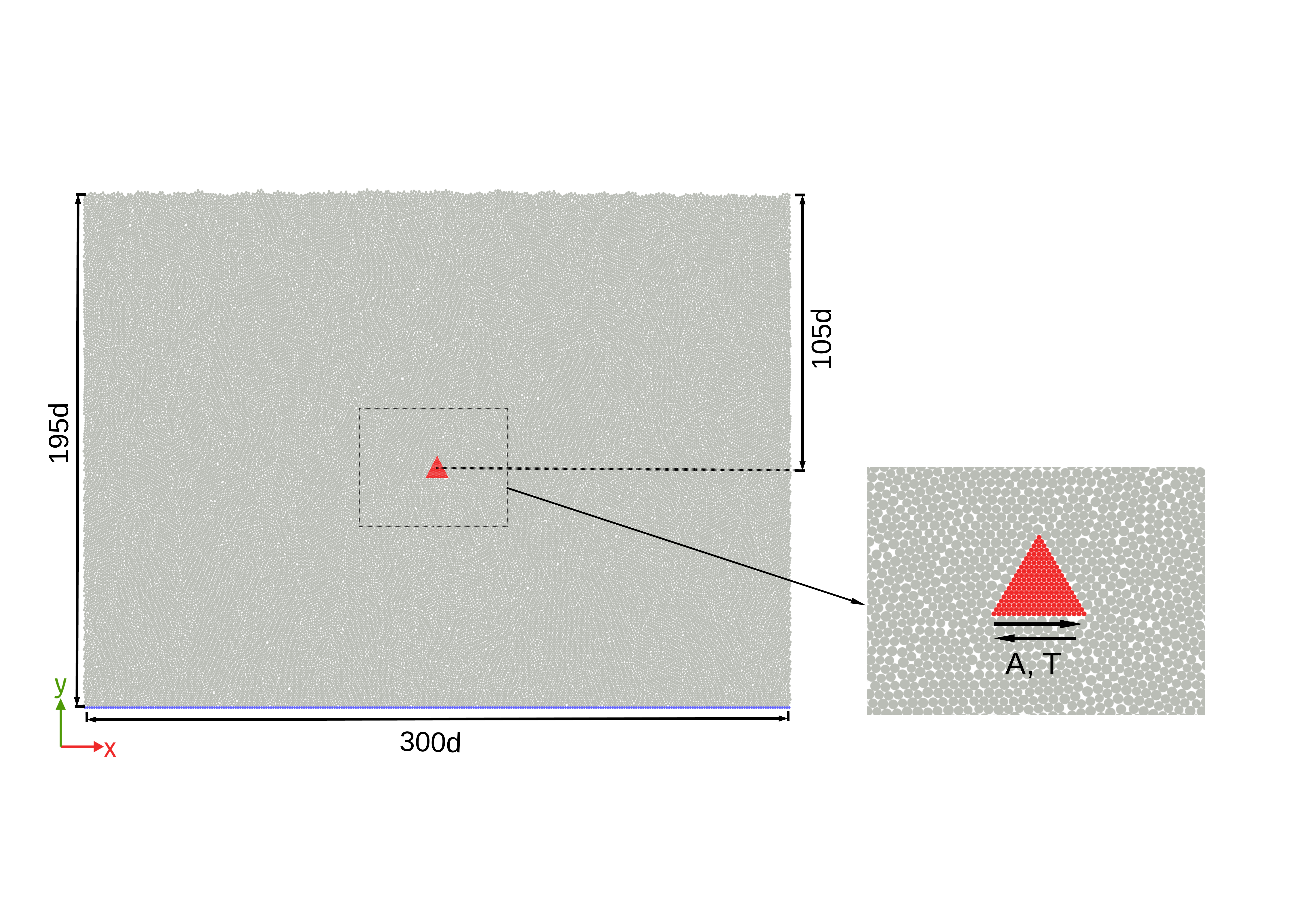} 
		
	\end{minipage}\vfill\vspace{-1cm}
	\begin{minipage}{0.55\textwidth}\hspace*{-1cm}
		\includegraphics[width=1.25\textwidth]{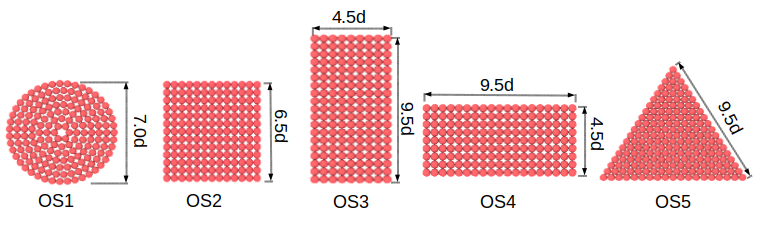} 
		
	\end{minipage}
	\caption{Initial configuration with one of the intruders (triangle) placed in the center (taken as origin) of the system and at a depth $h=105d$ from the free surface. The base is made of particles of size $1d$ (blue particles) at $y=0$ while the top surface is free to move and periodic boundary conditions are applied in $x$ direction. The intruder shapes that are considered in our study are disk $OS1$, square $OS2$, rectangle with its longer dimension along the $y$-axis $OS3$, rectangle with its longer dimension along the $x$-axis $OS4$, and equilateral triangle $OS5$. The intruder is oscillated horizontally with a velocity $V_{x}=A$ $sin( 2\pi t/T )$ where $A$, $T$ represents the amplitude and the time period of oscillation and $t$ is the time elapsed.}
\end{figure}
\section{\label{sec:simu_method} Simulation methodology}
To study the dynamics of a horizontally oscillating intruder within a two-dimensional granular medium, we used the discrete element method (\cite{cundall1979discrete}). Figure 1 shows the initial configuration of our system having dimensions of 300$d$ ${\times}$ 195$d$ along the $x$ and the $y$-axis respectively, where $d$ is the mean particle diameter. The intruder is placed at a depth of $105d$ from the free surface and at the centre of the simulation system. A set of particles are introduced uniformly at the top of the system which gets settled at the bottom due to gravity, while keeping the intruder stationary. Then, another set of particles are introduced and so on until the height of the granular bed reaches $195d$. The total number of particles comprising the system at the end is $N_t=63000$. The diameter of the particles ranges from $0.9d$ to $1.1d$ to avoid crystallization and while pouring and settling, we ensured that there are no overlaps among the particles. We further continued the simulation until the total kinetic energy of all the particles in the system have reached almost zero. At time $t=0$, the intruder is oscillated horizontally with a velocity $V_{x}=A$ $sin( 2\pi t/T )$ where $A$, $T$ represents the amplitude, and the time period of oscillation and $t$ is the time elapsed.
In our work, we have considered five different shaped intruders, which are pictorially represented in figure 1. Each of the individual intruder shapes is created by gluing a set of spherical particles having a diameter of $0.5d$ in such a way that their entire mass is the same for all the shapes. 

In the discrete element method (DEM), the positions and the velocities of particles are updated at regular intervals of time by integrating the equations of motion. The forces considered are the contact force and the body force. The normal $F_{ij}^n$ and the tangential $F_{ij}^t$ components of contact force between two particles $i$ and $j$ at a given time $t$, is calculated using the contact force model (\cite{Brilliantov,silbert2001granular}), as shown in equations 2.1 and 2.2. According to the model, two particles $i$ and $j$ are said to be in contact if the overlap is $\delta \geq0$, where $\delta={R_i+R_j}-{r_{ij}}$. Here, $R_{i}$, $R_j$ are the radii of the two contacting particles, and $r_{ij}$ is the instantaneous distance between the centres of the two particles.  

\begin{equation}
{F_{ij}^n=\sqrt{(R_{\textrm{eff}}\delta_{ij})} (K_n\delta_{ij}\hat{n}_{ij}-m_{\textrm{eff}}\gamma_n v_{ij}^n)}
\end{equation}
\begin{equation}
{F_{ij}^t=-min(\mu F_n,\sqrt{(R_{\textrm{eff}}\delta_{ij})} (K_t\Delta s_{t}+m_{\textrm{eff}}\gamma_t v_{ij}^t)}
\end{equation}

Here, $K_n = 2 \times 10^6 \rho dg$ and $K_t = 2.46 \times 10^6 \rho dg$ are the nonlinear spring constants, where `n' and `t' in the subscripts or superscripts correspond to normal and tangential components of respective parameters. Here, $v_{ij}$ is relative velocity, and $\hat{n}_{ij}$ is a unit vector in the direction of the line joining the centres of particles $i$ and $j$. The damping coefficients $\gamma_n$ and $\gamma_t$ are set to 2500$\sqrt{g/d^3}$ and $\Delta s_t$ is the tangential displacement vector between the two particles in contact. The effective mass $m_{\textrm{eff}}$ and the effective radius $R_{\textrm{eff}}$ for a given pair of contacts are expressed as $\frac{m_im_j}{m_i+m_j}$ and $\frac{R_iR_j}{R_i+R_j}$ respectively. The maximum value $F_{ij}^t$ can reach is $\mu F_{ij}^n$ because, at this threshold, the particles slip due to the dynamic friction. The time-step is $\Delta t = 1\times10^{-4}\sqrt{d/g}$ for all cases. The coefficient of friction $\mu$ is 0.1 for all the simulations unless specified. Each of the simulations was performed for a run time of $10^7 \Delta t$ to maintain parity among all the calculated results. The simulations are performed using dimensionless units, by considering acceleration due to gravity g, mean particle diameter $d$, and density $\rho$ as the basic units. The values of $d$, $\rho$, and $g$ are set to 1.0 without loss of generality. All the simulations are carried out using the Large-scale Atomic/Molecular Massively Parallel Simulator (LAMMPS) (https://www.lammps.org, \cite{PLIMPTON19951}), and for the visualization, OVITO package (\cite{ovito}) has been utilized.

\section{\label{sec:resdis} Results and Discussion}
\begin{figure}
	\centering
	\includegraphics[width=0.75\linewidth]{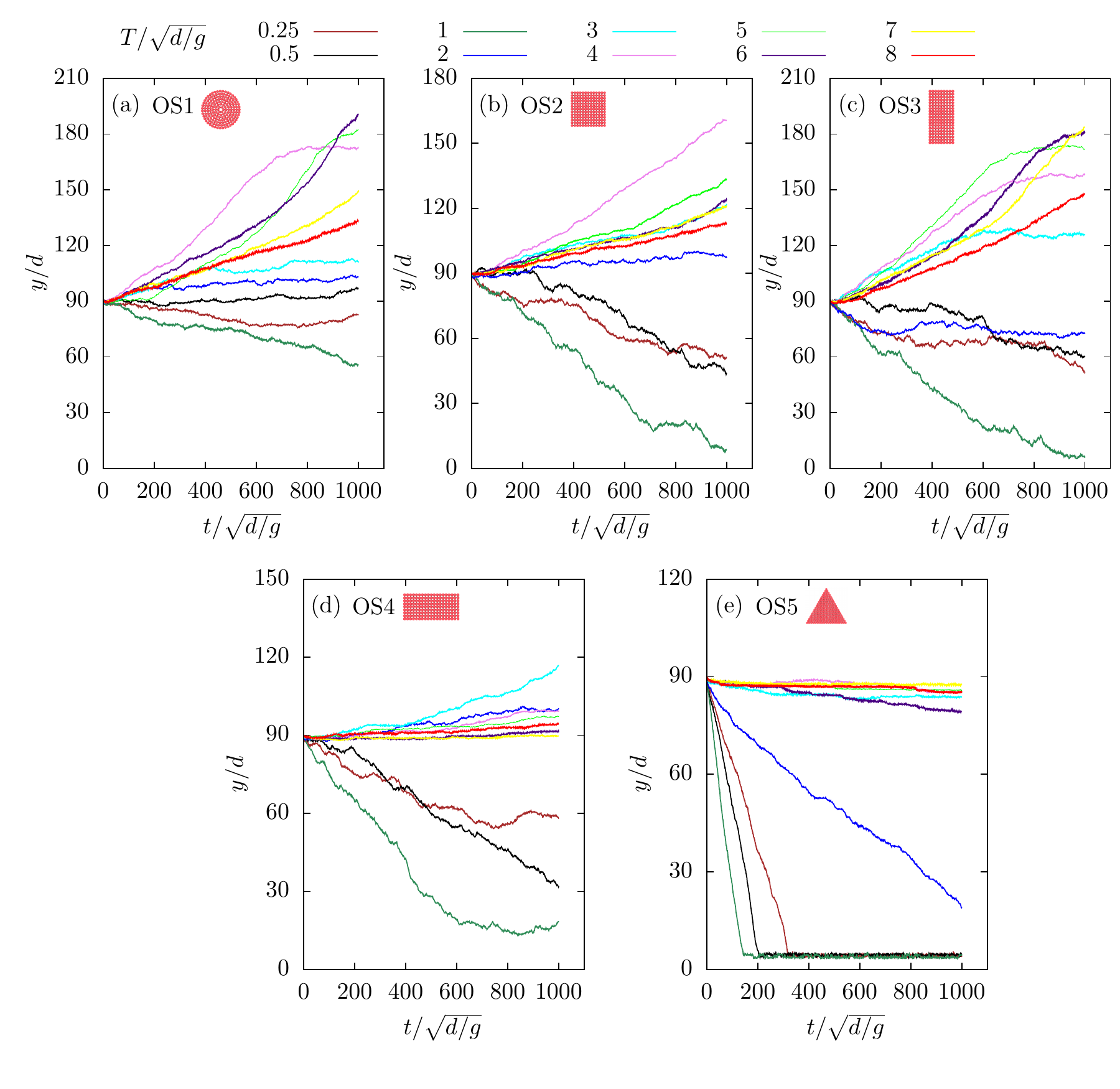}
	\caption{The vertical displacement of the intruder (a) disc, (b) square, (c) narrow-rectangle, (d) wide-rectangle, and (e) triangle shapes as a function of simulation time($t$) for an oscillation amplitude $A/d$= 3 and at different time period ($T/\sqrt{d/g}$).\label{fig:disp}}
\end{figure}
In this section, we presented the numerical results obtained for five different intruder shapes subjected to different oscillation parameters. We horizontally oscillated each intruder at different constant amplitudes $A/d=$ 0.5, 1, 3, 5 and at various constant time periods $T$ ranging from 0.25 to 8. This section is organized as follows: Subsection 3.1 demonstrates the vertical displacement of each intruder as a function of time at various time periods $T$ and for an amplitude $A/d=3$. The average rate of vertical displacement $V_y$ dependence on the intruder's shape at different $A/d$, $T$ and $\mu$ is discussed in subsection 3.2. In subsection 3.3, we showed the average rate of vertical displacement of the intruder as a function of the oscillating phase-angle $\phi$. In subsection 3.4, we elucidated the mean flow fields around the intruder. Lastly, in subsection 3.5, we have compared our results with the proposed mathematical model and explained the physical phenomenon leading to the rise and sink of the intruder for different parameters considered in the study.

\subsection{Vertical displacement of oscillating intruder}
\begin{figure}
	\centering
	\includegraphics[width=0.75\linewidth]{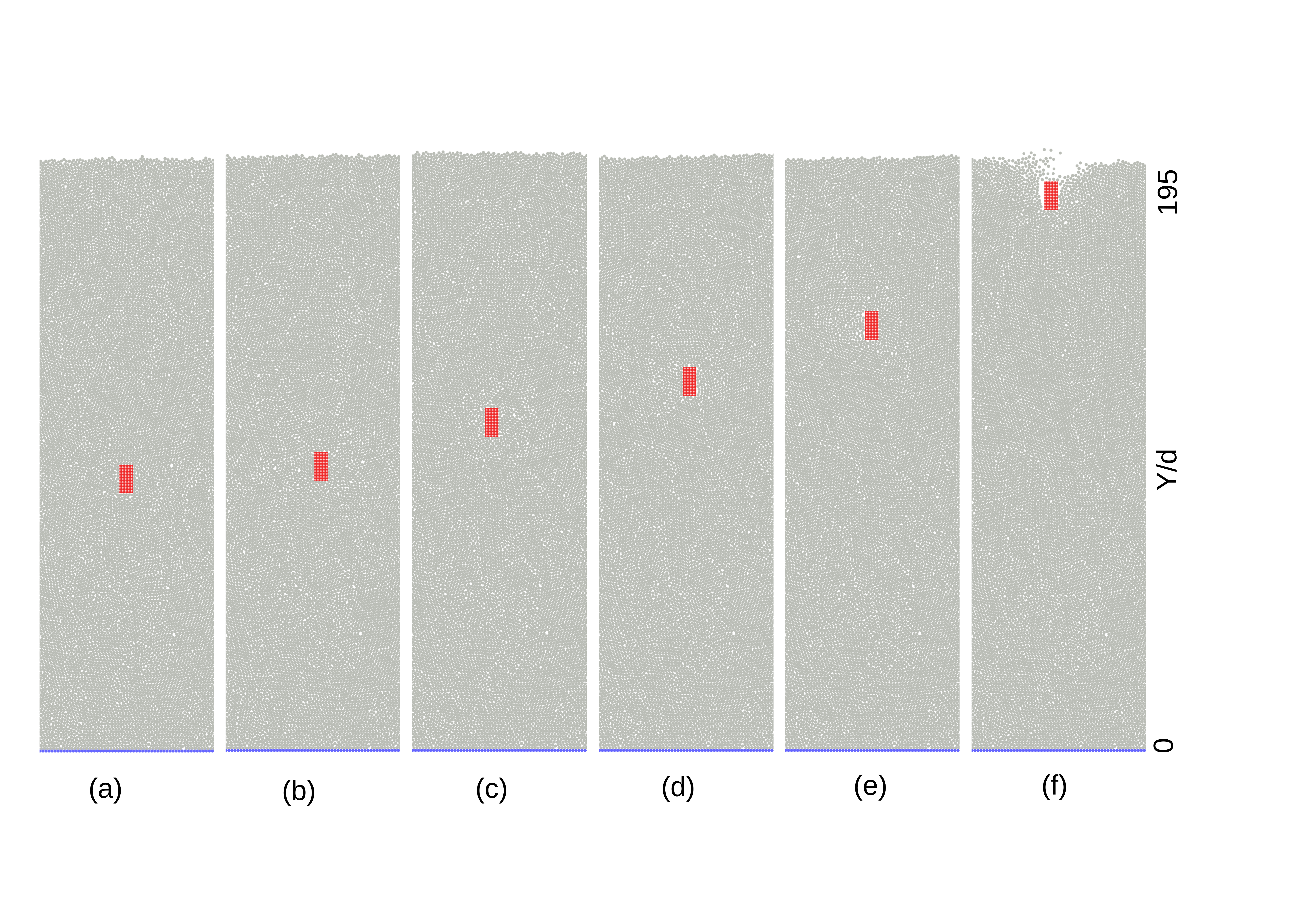}
	\label{fig:init}
	\vspace{-1cm}
	\caption{ A pictorial representation depicting the position of the narrow-rectangle $OS3$  for $A/d$=3 and $T/\sqrt{d/g}$=7 at different simulation time $t/\sqrt{d/g}$: (a) $0$ (b) $100$ (c) $300$ (d) $500$ (e) $700$ and (f) $1000$.}
\end{figure}
\begin{figure}
	\centering
	\includegraphics[width=0.75\linewidth]{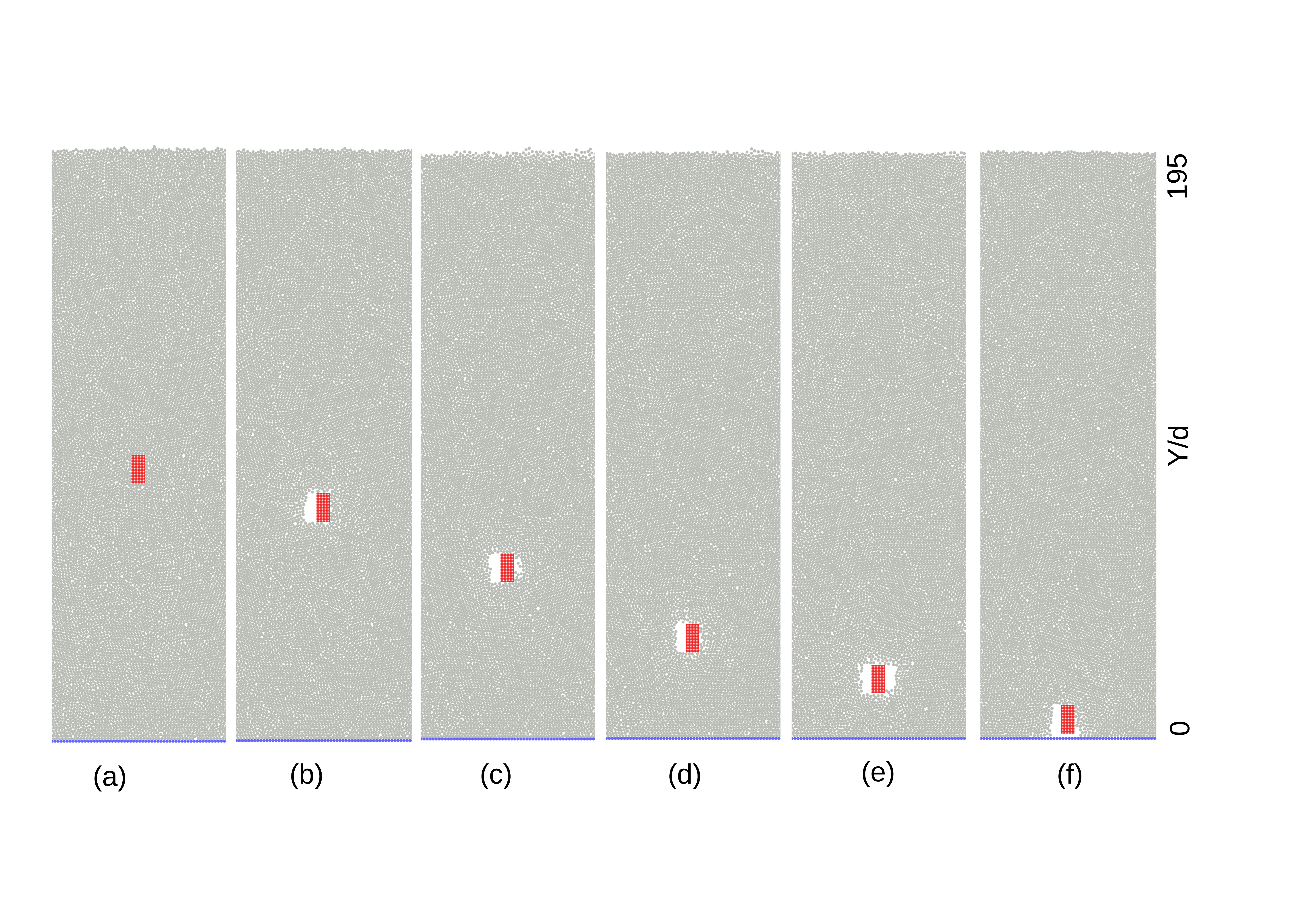}
	\label{fig:init}
	\vspace{-1cm}
	\caption{A pictorial representation depicting the position of the narrow-rectangle $OS3$ for $A/d$=3 and $T/\sqrt{d/g}$=1 at different simulation time $t/\sqrt{d/g}$: (a) $0$ (b) $100$ (c) $300$ (d) $500$ (e) $700$ and (f) $1000$.}
\end{figure}
\begin{figure}
	\centering	
	\includegraphics[width=0.75\linewidth]{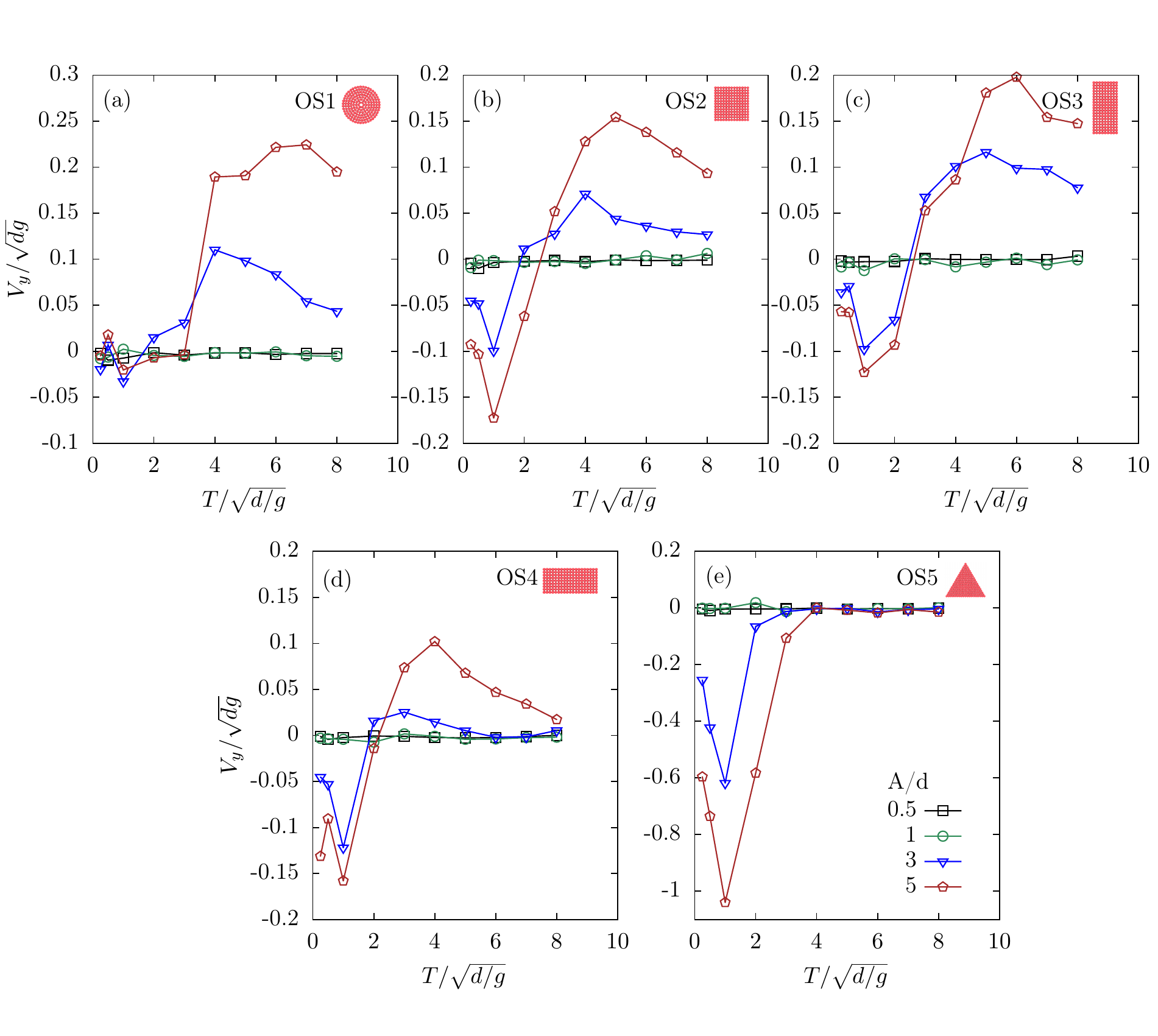}	
	\caption{The rise rate($V_{y}$ ) of the horizontally oscillating intruder as a function of its time-period($T$) for different amplitude $A$. Here, each plot corresponds to a different intruder shape: (a) disk, (b) square, (c) narrow-rectangle, (d) wide-rectangle, and (e) triangle.}
\end{figure}

In this subsection, we showed the vertical displacement of the intruders as a function of the simulation time $t$ for an amplitude of $A/d=3$ at different time-periods $T/\sqrt{d/g}$. Figure 2 a shows the vertical displacement of a disk-shaped intruder OS1 oscillated at different $T$. At time periods of oscillation $T \geq 2$, a wake is formed either to the left side of the intruder or to the right side, depending on its direction of motion. In this sense, when the intruder moves from left to right, the wake is developed to the left side of the intruder. When the intruder changes its direction of oscillation and starts to move from right to left, it interacts with the wake that is on its left side. When the intruder reaches the middle point of oscillation, the wake to its left vanishes, and simultaneously another wake evolves to the intruder's right side. When the intruder is traversing the granular medium, the particles above the wake region tries to fill the wake due to the action of gravity. In the process, a bed of particles is created on the edges of the wake region. When the intruder changes its direction of motion, it must climb the bed of particles to move ahead, and thus a vertical displacement of the intruder takes place. This is the reason behind the rise of the disk-shaped intruder at longer time periods. Interestingly, the intruder neither sinks nor rises at certain intermediate time-periods $T = 2$. At time periods $T$ ranging from 0.25 to 1, the wake is formed not only beside the intruder but also above and below it. This is due to its high oscillation frequency and, consequently, a greater momentum transfer to the surrounding particles. The presence of the wake below the intruder propels it to sink due to the gravity acting on it. Thus, the disk-shaped intruder sinks at shorter time periods or larger frequencies of oscillation. Similar behavior has been observed in both 2D \citep{PhysRevE.94.062906} and 3D \citep{ping2020sink} granular systems, where the disk (2D) and the spherical (3D) shaped intruders rise at lower frequencies and sink at higher frequencies of oscillation.

The square-shaped intruder shows similar behavior (figure 2 b) to that of the disk one (in supplementary information we have provided the links for the videos showing the rise and sink of the square shaped intruder when oscillated at time period of 5 and 1 within the granular medium). However, the square-shaped intruder is noticed to move to lower vertical distances than the disk one for longer time periods. This is due to the larger stress experienced by the square one ($OS2$) as compared to that of the disk one ($OS1$) from the particles above. The reason for it is due to the presence of a flat surface at the top of the square intruder and a curved one for the disk one. Figure 2 c shows the vertical displacement of a rectangle with a small base ($OS3$) whose behavior is like that of the disk and the square shapes. Though rectangle and square-shaped intruders have flat top surfaces, the former moves to a higher vertical distance than the latter one. The reason for this behavior is that the rectangle intruder ($OS3$) has a smaller top surface than the square one and hence a smaller stress from the particles above it. A pictorial representation of the rectangle intruder ($OS3$) is shown to depict this rise and sink phenomena at different simulation times $t/\sqrt{d/g}$ when oscillated at a higher time period ($T/\sqrt{d/g}=7$)  (figure 3) and at a lower one ($T/\sqrt{d/g}=1$) (figure 4) for a given amplitude $A/d=3$. Further, the rectangle with a larger base/top surface ($OS4$) moves upwards at a slower rate than (figure 2 d) that of the square one and the rectangle one with a smaller base ($OS3$). This is due to a larger top surface and consequently a higher stress from the particles above it. Interestingly, the sinking behavior is observed to be independent of the base length of the quadrilaterals ($OS2, OS3$ and $OS4$) considered in our study. The
triangular intruder behaved differently as compared to all the other shapes. In this sense, the triangular intruder is noticed to sink always irrespective of the time periods (figure 2 e). In this case, a bed of particles is not created near the edges of the wake owing to the intruder's inclined sides and sharp edges.  An intriguing observation is that the maximum vertical distance traversed by the intruder varies non-monotonically with the time period. Apart from the maximum vertical distance travelled by an intruder, how quickly an intruder can rise, or sink is equally important. This aspect of the intruder is discussed in the next subsection.

\subsection{Rate of rise of intruder with time-period}
\begin{figure}
	\centering	
	\includegraphics[width=0.75\linewidth]{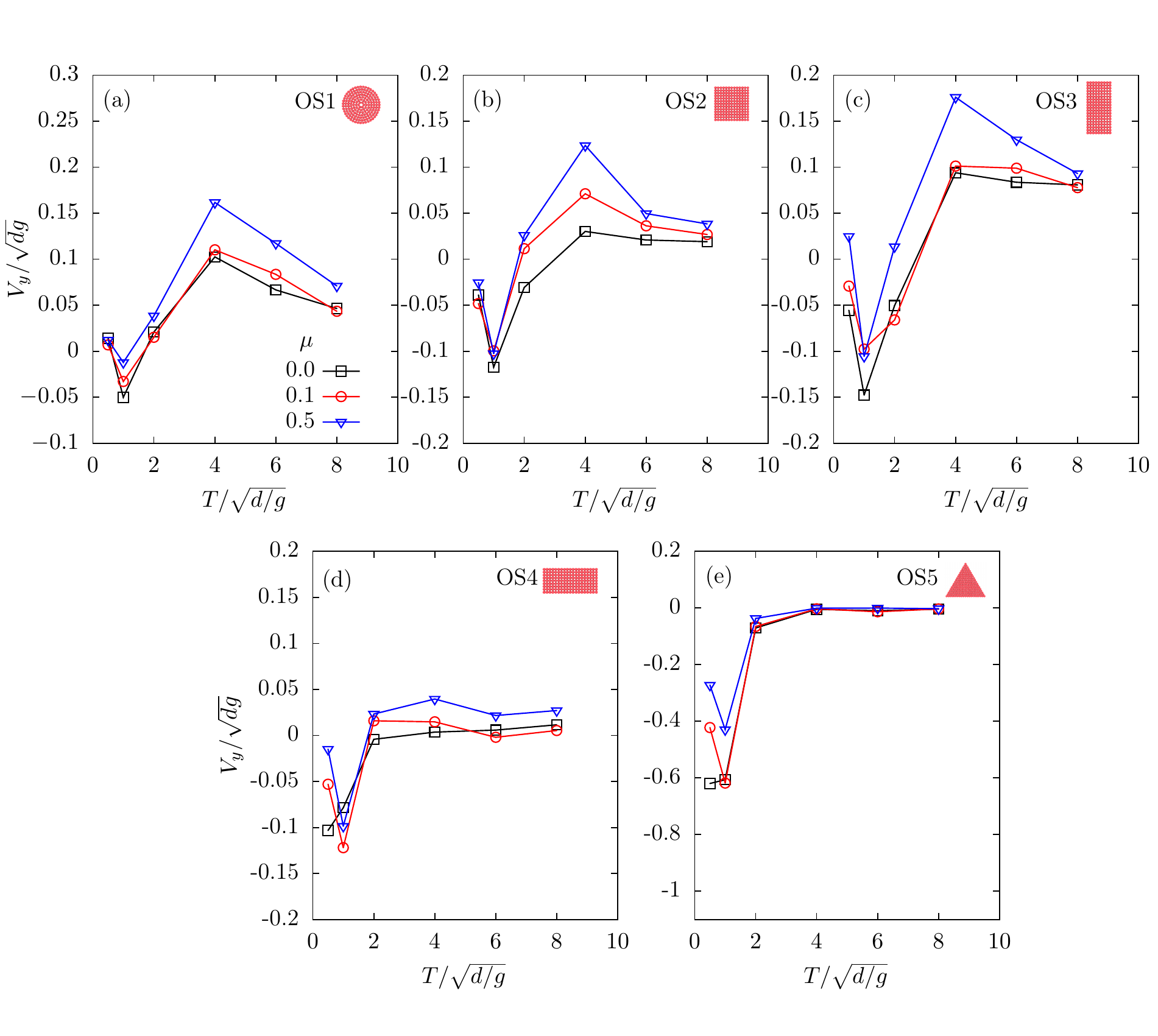}	
	\caption{The rise rate ($V_{y}$) of the horizontally oscillating intruder as a function of its time-period ($T$) for different friction coefficient $\mu$ and amplitude $A/d=3$. Here, each plot corresponds to a different intruder shape: (a) disk, (b) square, (c) narrow-rectangle, (d) wide-rectangle, and (e) triangle.}
\end{figure}

In this subsection, we explain how the vertical displacement rate of an intruder varies with its shape and oscillation parameters. The rate of change of vertical displacement is defined as $V_{y} = (y_{f}-y_{i})/(t_{f} - t_{i})$. Here, $y_i$ and $y_f$ are the initial and final vertical positions of the intruder at times $t_i$ and $t_f$. Note that $t_i$ is considered at $1/10^{th}$ of the total simulation time ( 1000$\sqrt{d/g}$). For some cases, we noticed the intruder reaches its final vertical position before the end of the simulation (1000 $\sqrt{d/g})$. An example of such behaviour can be seen in figure \ref{fig:disp} a for a circular intruder at time period $T=4$ for which $t_f=650$. For these type of cases, we have taken $t_f$ as the time at which the intruder is about to reach its final vertical position. Henceforth, the positive value of $V_y$ implies the intruder is rising and the negative value of $V_y$ corresponds to sinking. Figure 5 shows the intruder's rate of rise/sink as a function of time-period ($T$) and for amplitudes, $A/d =$ 0.5, 1, 3, and 5. The rate of rise/sink is negligible at all time periods for the amplitudes ($A$) of 0.5 and 1 for all the intruder shapes. Similar behavior has been reported for a oscillating disk in a previous work by \cite{PhysRevE.94.062906}, where the authors reported that to facilitate the vertical motion of the intruder, a minimum amplitude of oscillation is necessary. 

Figure 5 a shows a slightly negative $V_y$ for $A/d = 3$ and at $T\leq1$ for a disk-shaped intruder. This result is consistent with the negative vertical displacement observed in figure 2 a for $T\leq1$. At $T\geq2$, we observed a positive $V_y$ for the range of time periods considered in our study. Surprisingly, as the time-period increases, the rise rate first increases, reaches a maxima and then decreases. When the intruder changes its direction of motion, it is impelled to climb over a bed of particles leading to its rise at large time periods. Suppose, if the intruder is moving from left to right, then the above-mentioned bed of particles corresponds to the ones that are trying to fill the wake which is formed just to the right of the intruder. These particles could be the ones that fall from above the intruder or the ones that are adjacent to the wake. As the time period increases, more particles get piled up, and consequently, the intruder has to climb over a larger set of particles. This leads to an increase in its vertical position and its rise rate. As time period increases, we noticed a maxima and then a decrease in the rise rate. The reason for this is at larger time periods, greater number of particles gets piled up which dampens the upward motion of the intruder. In summary, the group of particles falling behind the intruder, or in other words, the size of the bed of particles behind it, is responsible for the non-monotonic behavior of the intruder's rise rate. Even at amplitude $A/d = 5$, the non-monotonic behavior in the rise rate is noticed. Moreover, the rise rate increases with an increase in the amplitude. 

The rate of rise/sink of a square intruder ($OS2$) as a function of the time period is shown in figure 5 b. At time periods $T\leq2$, amplitude $A/d = 3$, the square intruder has a negative rate of vertical displacement or sink rate. Interestingly, as the $T$ increases, the sink rate gradually increases reaching a maxima and then decreases. At $T < 2$, a wake is formed not only on either sides of the intruder but also above and below it as a result of a high frequency of oscillation. The action of gravity on the intruder coupled with a low area fraction of particles below it results in its sinking. As the time period increases, the size of the wake above the intruder decreases, and consequently, the number of particles interacting, or in other words, the stress exerted on the top surface of the intruder increases. This results in an increase in the sink rate. With a further increase in the time period, the wake size below the intruder decreases, resulting in the resistance to the downward motion of the intruder. In summary, the wake dynamics above and below the intruder are responsible for the sink rate's non-monotonic behavior. At time periods $T \geq 2$, the square intruder experiences mostly a positive rate of vertical displacement. The square-shaped intruder shows a non-monotonic rise rate as a function of time period at $T \geq 2$, similar to that of the disk-shaped intruder. However, the square-shaped intruder has lower rise rates than that of the disk-shaped intruder. This is because the curved base of the disk-shaped intruder enables it to ascend the slant bed of particles more quickly than that of square one.

Figure 5 c displays the rise/sink rate of a  rectangle-shaped ($OS3$) intruder as a function of time period. The rectangle-shaped intruder has a smaller base and a larger height. The rectangle-shaped intruder shows qualitatively similar behavior to that of the square-shaped one. However, at $T\leq2$, the sink rate is smaller for the rectangle-shaped intruder ($OS3$) as compared to that of the square-shaped intruder($OS2$). This is due to a smaller top surface of the rectangle-shaped intruder ($4.5d$) than that of the square-shaped one ($6.5d$), which yields in smaller stress experienced by the top surface of $OS3$ than that of $OS2$. Due to a small base, $OS3$ can climb the slant bed of particles more easily than that of $OS2$. This results in slightly higher rise rates for the $OS3$ shape than the $OS2$ one. For the rectangle-shaped intruder with a larger base and a smaller height ($OS4$), an
opposite behavior is witnessed (figure 5 d). In this sense, the rise rate is smaller for $OS4$ as compared to that of $OS3$ and $OS2$. Moreover, the magnitude of the sink rate is greater for $OS4$ than that of $OS2$ and $OS3$ owing to the larger top surface of $OS4$. The rate of
vertical displacement of the equilateral-triangle-shaped intruder ($OS5$) is shown in figure 5 e. The triangle-shaped intruder shows a mostly negative rate of vertical displacement because the V-shaped structure formed by its two edges at the vertices disables it to climb over the bed of particles. The intruder displayed a non-monotonic sink rate dependence on the time period. However, the magnitude of the sink rate is significantly higher than the other shapes considered in our study. The reason for it is due to an almost continual downward stress exerted by the particles on the triangle's slant edges. We have also found that there is an increase in the rate of rise of the intruder with the increase in the intergranular friction of $\mu=0.0$ to 0.5 (figure 6). The same kind of trend has also been reported by \cite{liu2021locomotion} for a 3D system with a circular intruder and also stated that there is a saturation of rise rate when $\mu\geq0.5$.  
\begin{figure}
	\centering
	\includegraphics[width=0.9\linewidth]{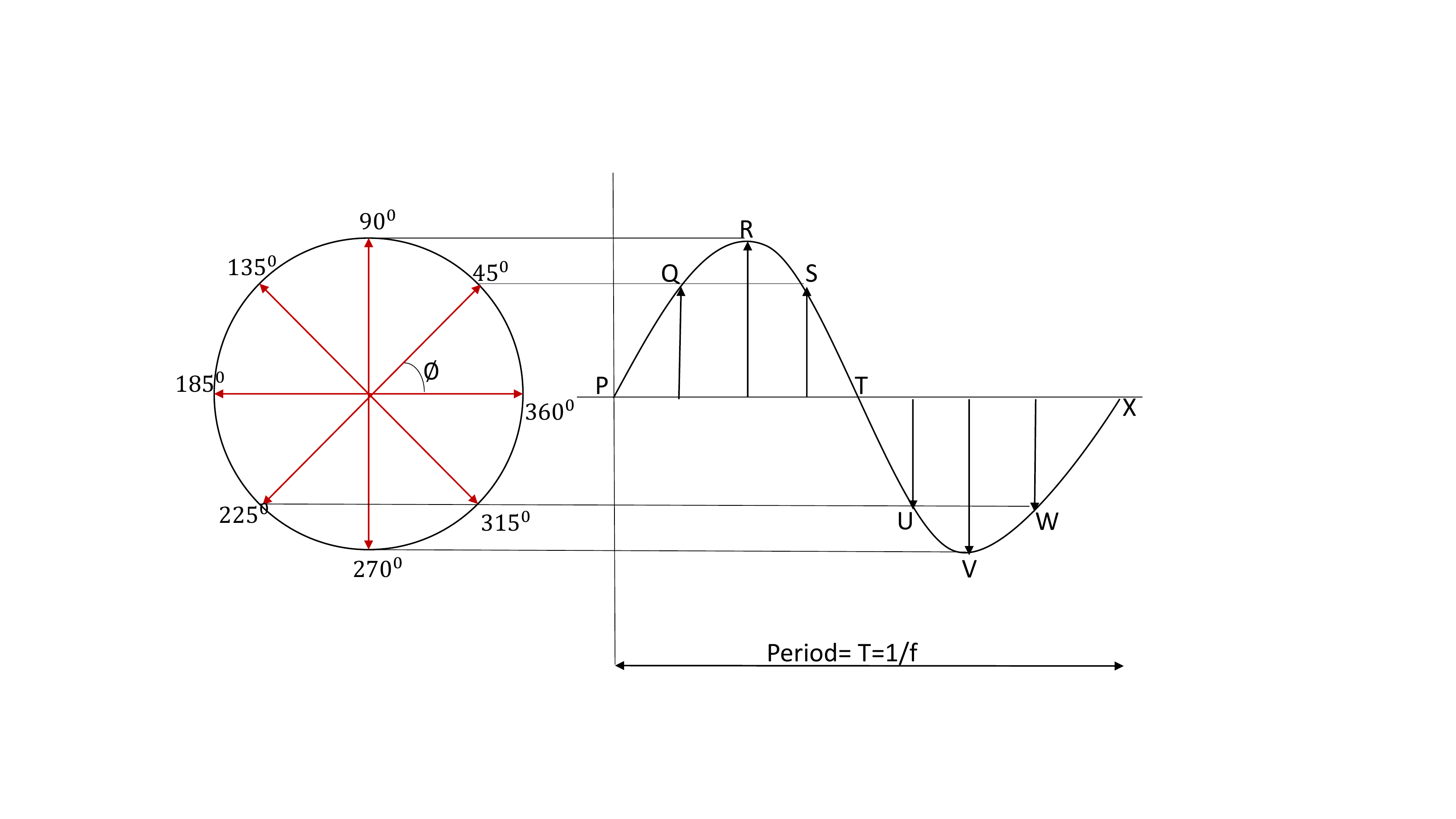}
	\vspace*{-1cm}
	\caption{Oscillatory motion of the intruder as sine wave function representing the phase angle($\phi$)}
\end{figure}

Moreover, due to its unique geometry, each shape has a different angle of contact around its surface with the surrounding particles when oscillating horizontally. With square and rectangle shapes having a flat outer surface, triangular shape with an inclined angle, and disk with a smooth curvature. If we compare the positive $V_{y}$ value which correspond to the rise of the intruder for the different shapes at $A/d=$ 3 and 5, than it is in the order of : $V_{y};OS1 > V_{y};OS3 > V_{y};OS2 > V_{y};OS4 > V_{y};OS5$. Thus confirming that the geometric shape and orientation also affect the intruder's movement and its rate of rise in the system when oscillated horizontally. 

\subsection{Average rate of vertical displacement of an intruder with phase angle}
\begin{figure}
	\centering
	\includegraphics[width=0.75\linewidth]{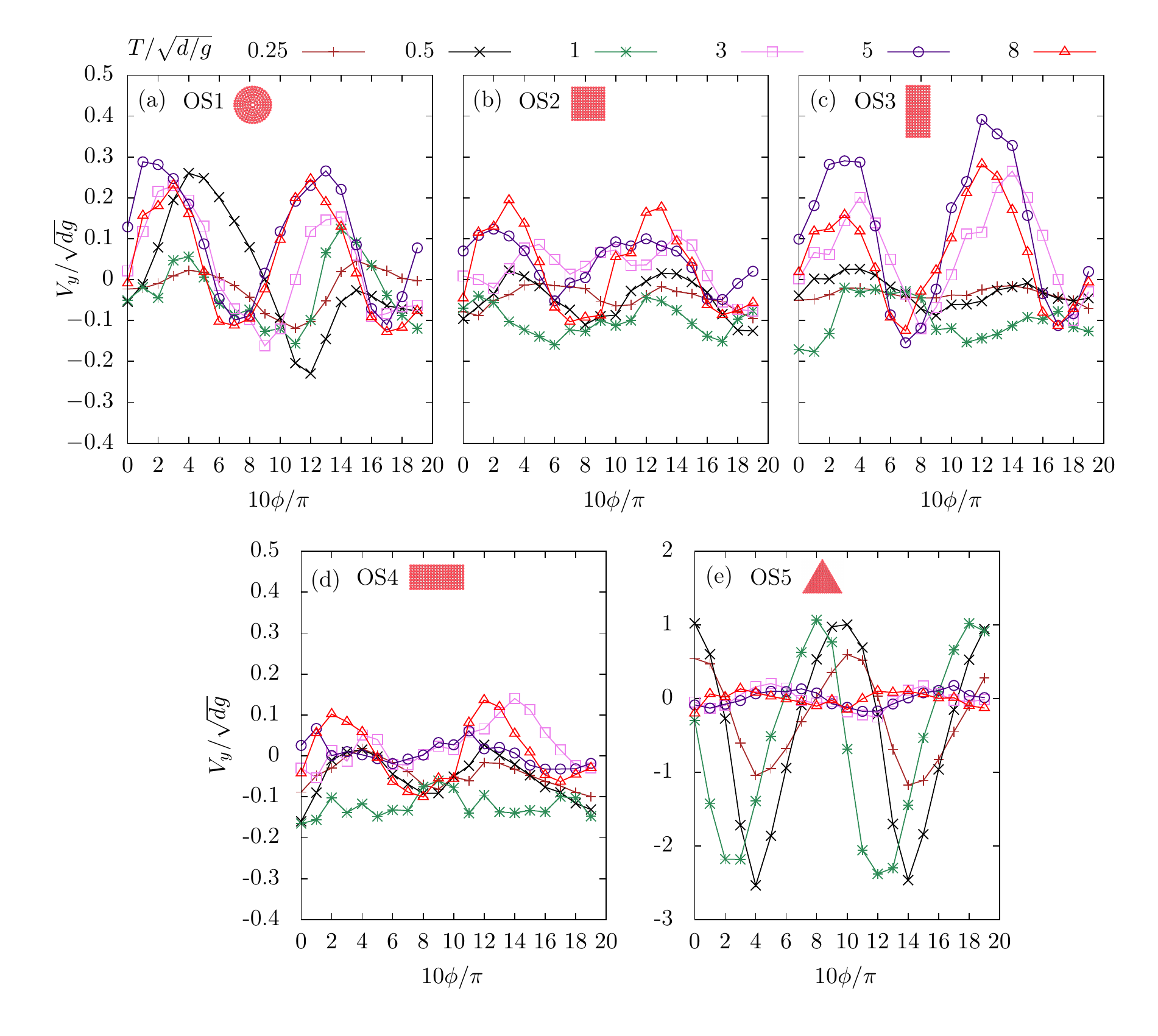}
	\caption{The rise rate(Vy) of the intruder vs the phase angle at $A/d=3$ and different $T/\sqrt{d/g}$ for (a) disk, (b) square, (c) narrow-rectangle, (d) wide-rectangle, and (e) triangle shaped intruder object.}
\end{figure}

In this subsection, we elucidate how the average rate of vertical displacement of the intruder varies with different positions of the oscillation cycle. One oscillation cycle involves the intruder travelling a distance of $``A"$ from the centre to the right, then a distance of $``2A"$ from right to the left, followed by a distance of $``A"$ from left to the centre, where $``A"$ is the amplitude. If we consider one oscillation cycle as a sine wave function, then each phase angle ($\phi$) represents different horizontal positions of the intruder as shown in figure 7. Here, the phase angle $\phi=0$ or $10\phi/\pi=0$ represents the centre point of oscillation when the intruder is moving from left to right, $\phi=\pi/2$ or $10\phi/\pi=5$ indicates the rightmost point of oscillation. In the same way, $\phi=\pi$ or $10\phi/\pi=10$ denotes the centre point of oscillation when the intruder is moving from right to left, $\phi=3\pi/2$ or $10\phi/\pi=15$ indicates the leftmost point of oscillation and $\phi=2\pi$ or $10\phi/\pi=20$ represents the centre point. Figure 8 a displays the average rate of vertical displacement ($V_y$) of the disk-shaped intruder as a function of its phase angle at different time periods and for an amplitude of $A/d=3$. Here, $V_y$ increases with $\phi$ from $10\phi/\pi=0$ to $10\phi/\pi\approx5$ where the intruder gradually climbs over the particles. When the intruder changes its direction of motion ($10\phi/\pi=5$) and moves from right to left, the wake to the left of the intruder gradually decreases. This is associated with an increase in the number of particles interacting with the intruder increasing the resistance to the vertical motion of the intruder. Therefore, $V_y$ decreases from $10\phi/\pi\approx5$ to $10\phi/\pi\approx10$. However, from $10\phi/\pi\approx10$ to $10\phi/\pi\approx15$ the intruder again climbs over the particles yielding an increase in the average rate of vertical displacement $V_y$. When the intruder moves from left to right, another wake evolves to the right of it. At $10\phi/\pi=15$, the intruder changes its direction of motion and moves from right to left which results in a decrease in $V_y$ due to the presence of wake to the right of the intruder. Interestingly, we witnessed both the positive and negative rates of vertical displacement in a single oscillation cycle at all timeperiods and for $A/d=3$ for the disk-shaped intruder. At smaller timeperiods of oscillation, lower $V_y$ is evidenced whereas, at larger timeperiods, greater $V_y$ is noticed. This result is consistent with $V_y$ as a function of time period (figure 5 a). 

The square-shaped intruder (figure 8 b) shows qualitatively similar trends to that of the disk-shaped one. The magnitude of $V_y$ is greater for the disk-shaped intruder ($OS1$) as compared to that of a square-shaped one ($OS2$) although the diameter of $OS1$ ($6.5d$) and the side of $OS2$ ($7.0d$) are almost of the same length. The disk has a curved surface at the top whereas the square has a flat top surface and hence square-shaped intruder experiences greater stress from the particles above it than the disk-shaped one. This results in a smaller $V_y$ for a square-shaped intruder as compared to that of a disk-shaped one. However, for a rectangle-shaped intruder having a shorter top surface ($4.5d$), a greater $V_y$ is evidenced (figure 8 c) as compared to the square one, due to smaller stress experienced from the particles above it. For a rectangle-shaped intruder with a larger base ($9.5d$), we noticed a smaller $V_y$  due to greater stress experienced by the intruder from the particles above it (figure 8 d). For a triangle-shaped intruder, we evidenced a negative rate of vertical displacement $V_y$ for most of the phase angles. This could be due to its asymmetric geometry and either of the two inclining side walls experiencing stress from the particles above it throughout the oscillation cycle.

\subsection{Mean flow fields}

\begin{figure}
	\hspace{-0.9cm}
	\begin{subfigure}[b]{0.25\textwidth}
		\centering
		\includegraphics[width=\textwidth]{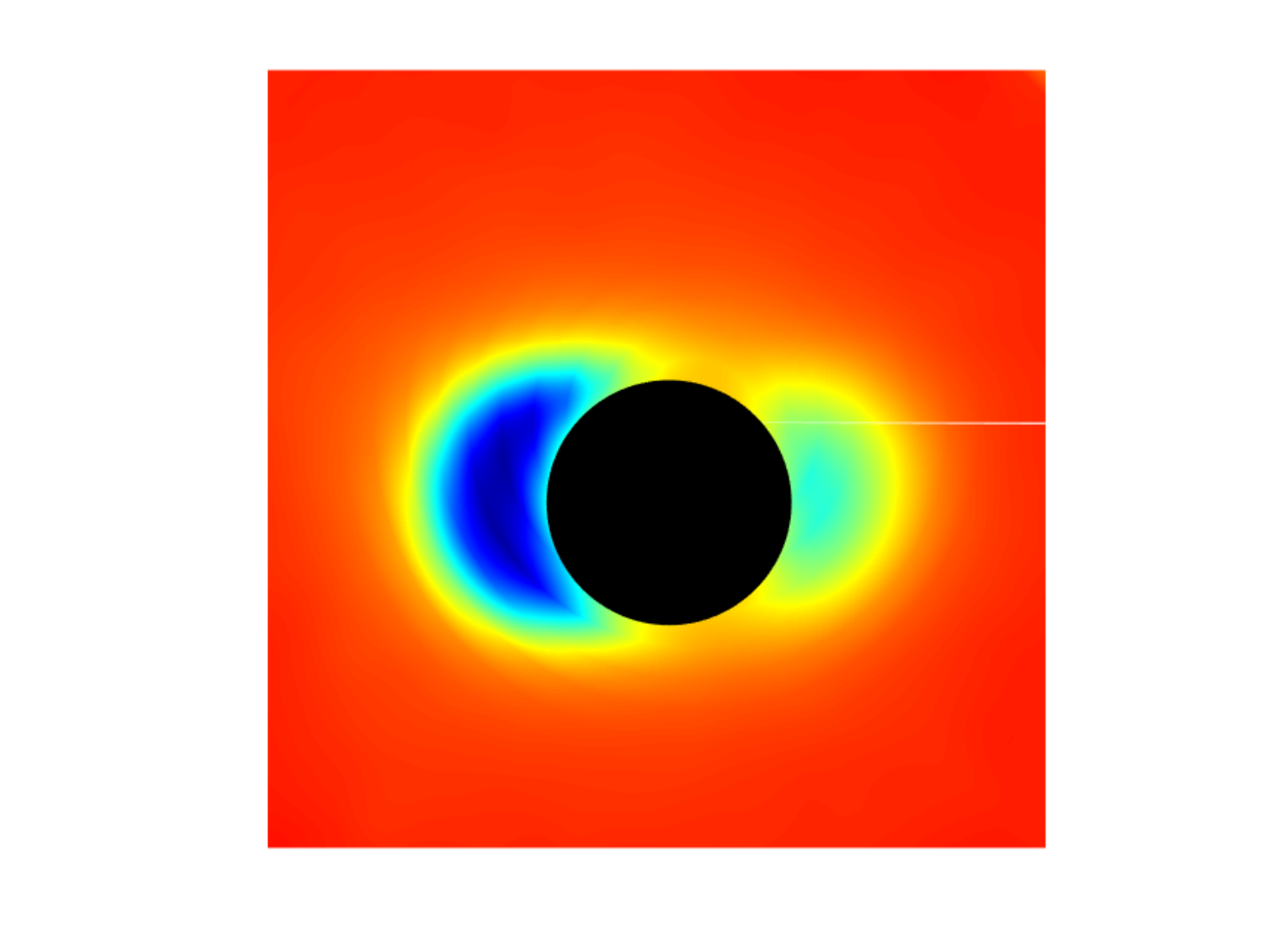}
		\caption{}
		\label{fig:y equals x}
	\end{subfigure}
	\hspace{-1.1cm}
	\hfill
	\begin{subfigure}[b]{0.25\textwidth}
		\centering
		\includegraphics[width=\textwidth]{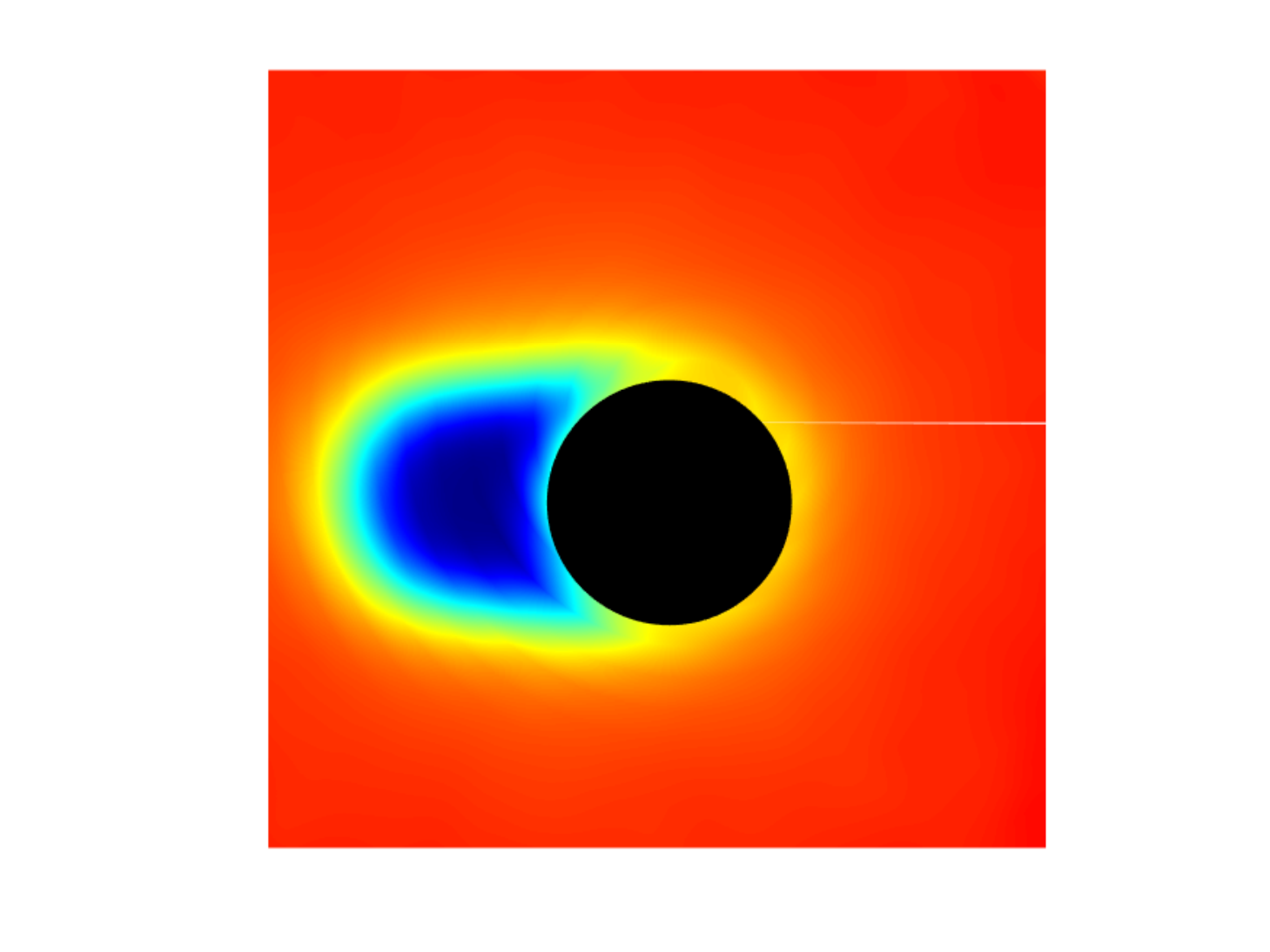}
		\caption{}
		\label{fig:three sin x}
	\end{subfigure}
	\hspace{-1.1cm}
	\hfill
	\begin{subfigure}[b]{0.25\textwidth}
		\centering
		\includegraphics[width=\textwidth]{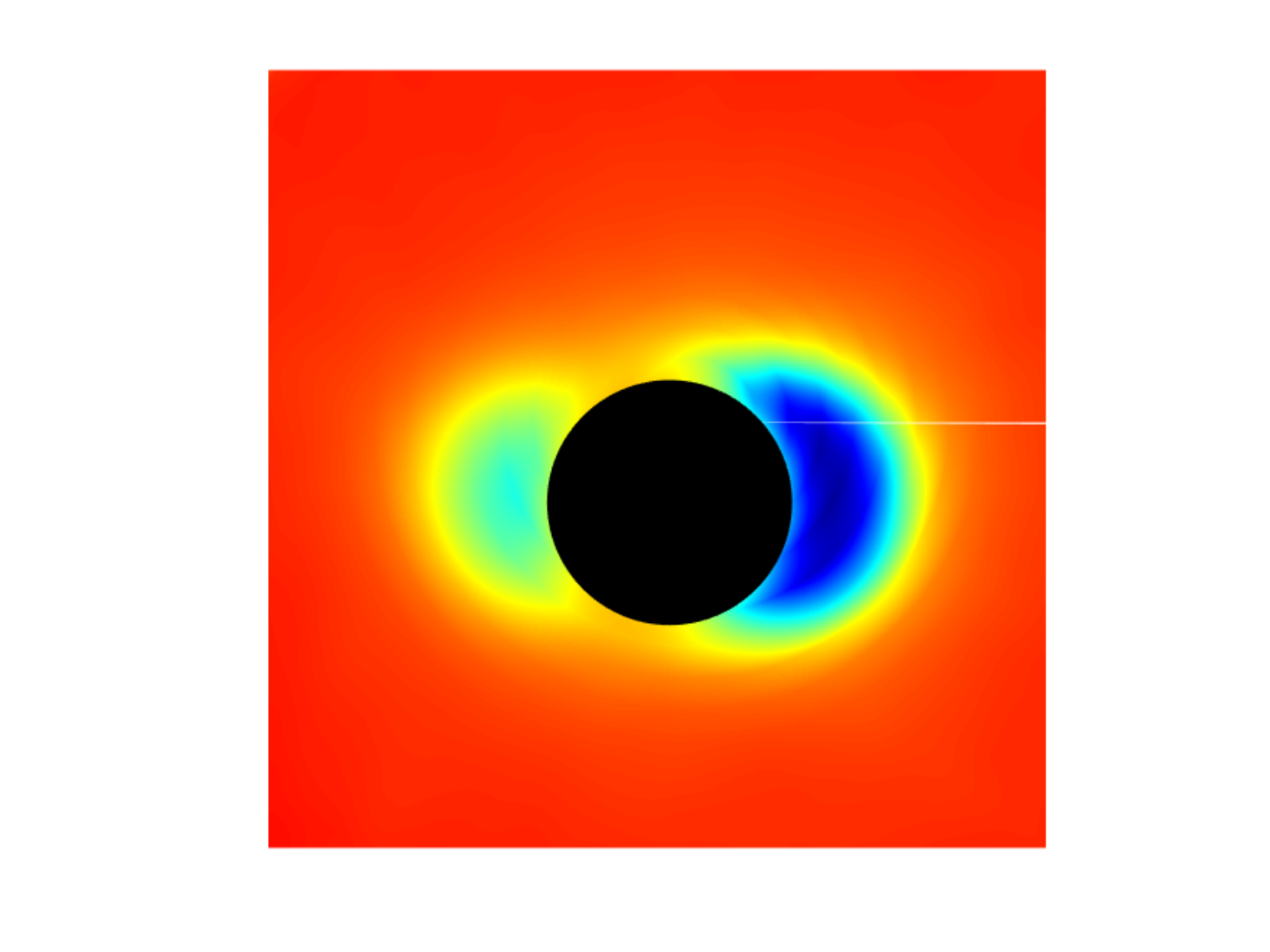}
		\caption{}
		\label{fig:five over x}
	\end{subfigure}
	\hspace{-1.1cm}
	\hfill
	\begin{subfigure}[b]{0.25\textwidth}
		\centering
		\includegraphics[width=\textwidth]{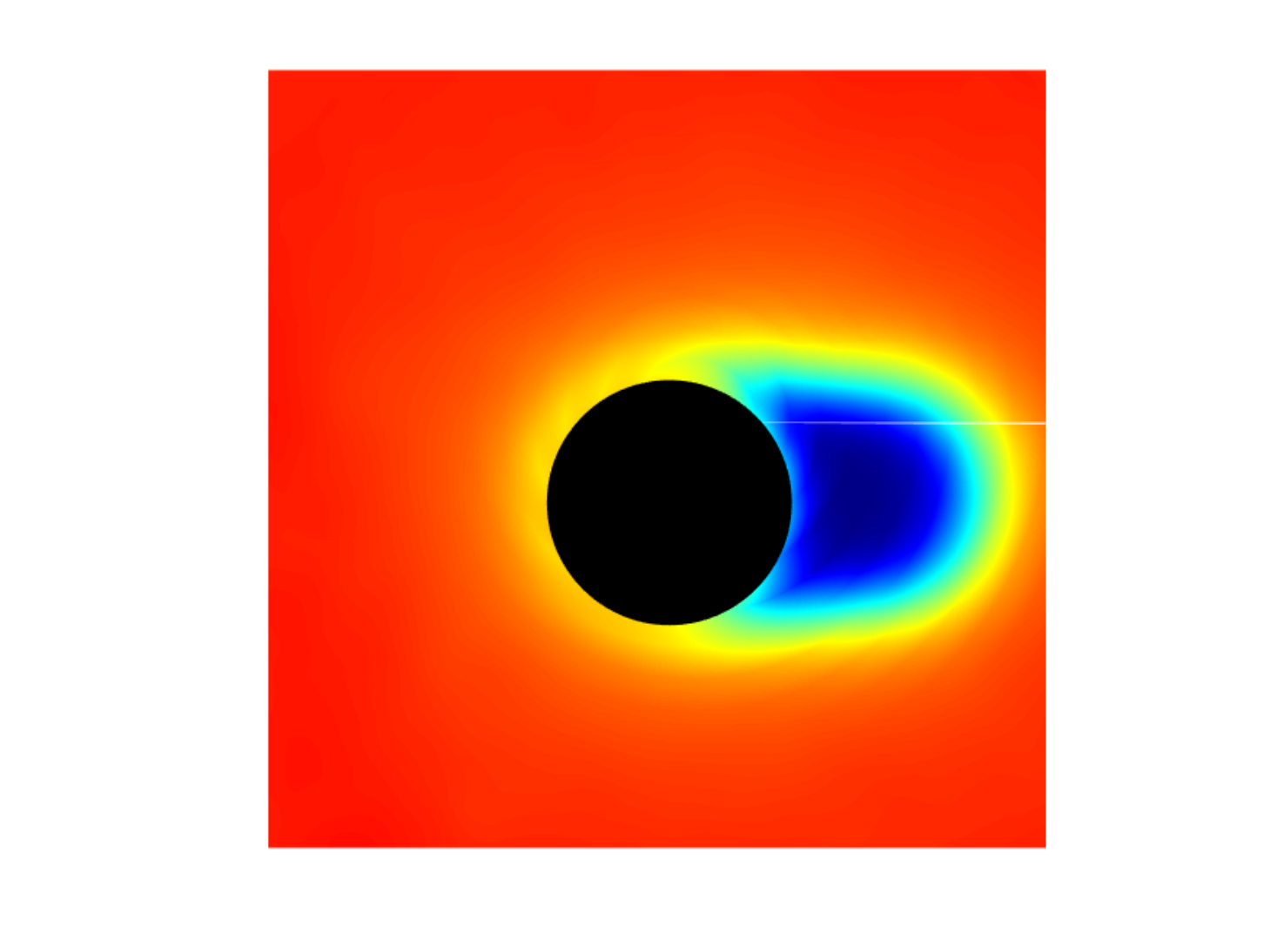}
		\caption{}
		\label{fig:three sin x}
	\end{subfigure}
	\hspace{-1.0cm}
	\hfill
	\begin{subfigure}[b]{0.25\textwidth}
		\centering
		\includegraphics[width=\textwidth]{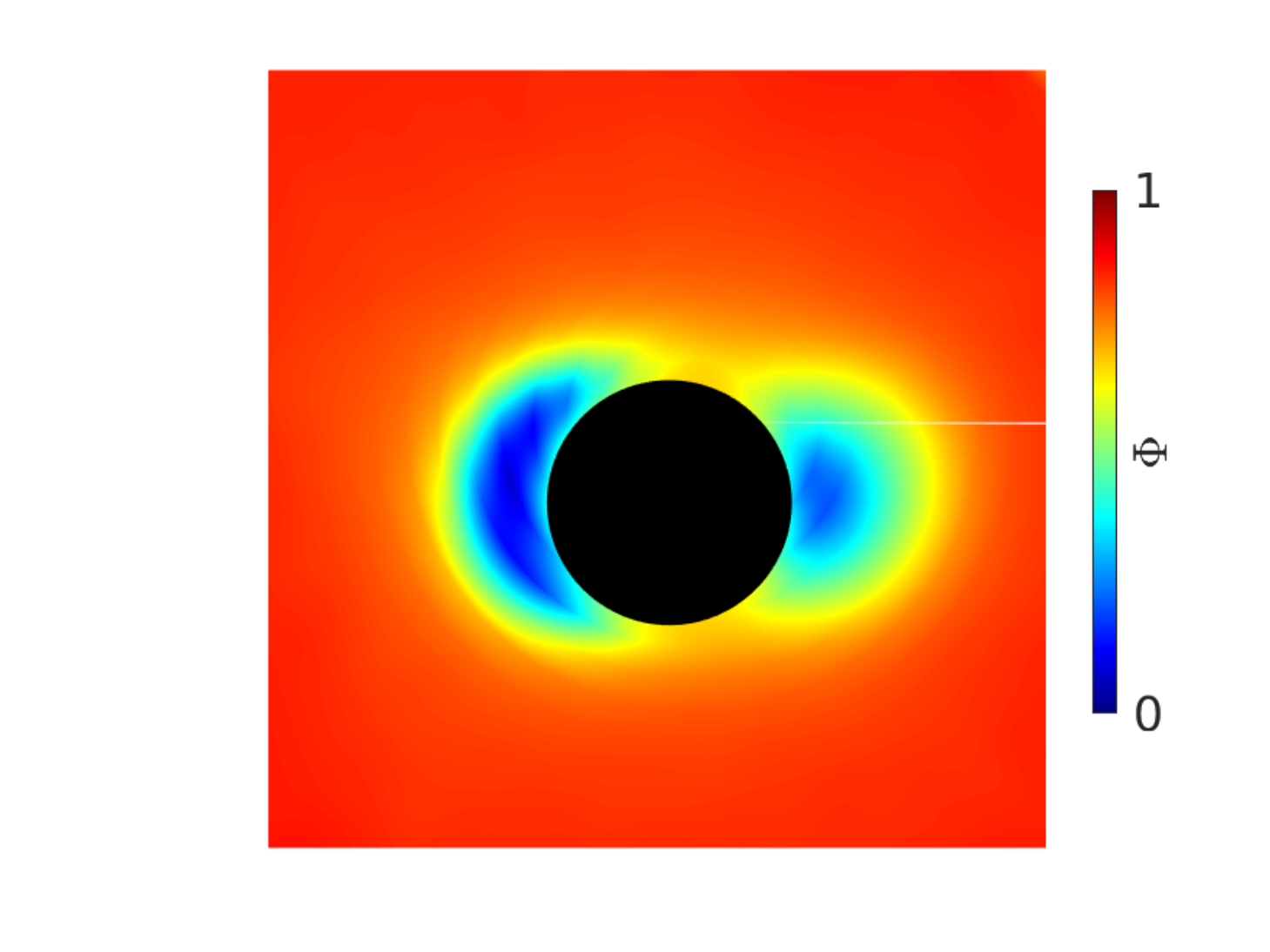}
		\caption{}
		\label{fig:five over x}
	\end{subfigure}
	
	\hspace{-0.9cm}
	\begin{subfigure}[b]{0.25\textwidth}
		\centering
		\includegraphics[width=\textwidth]{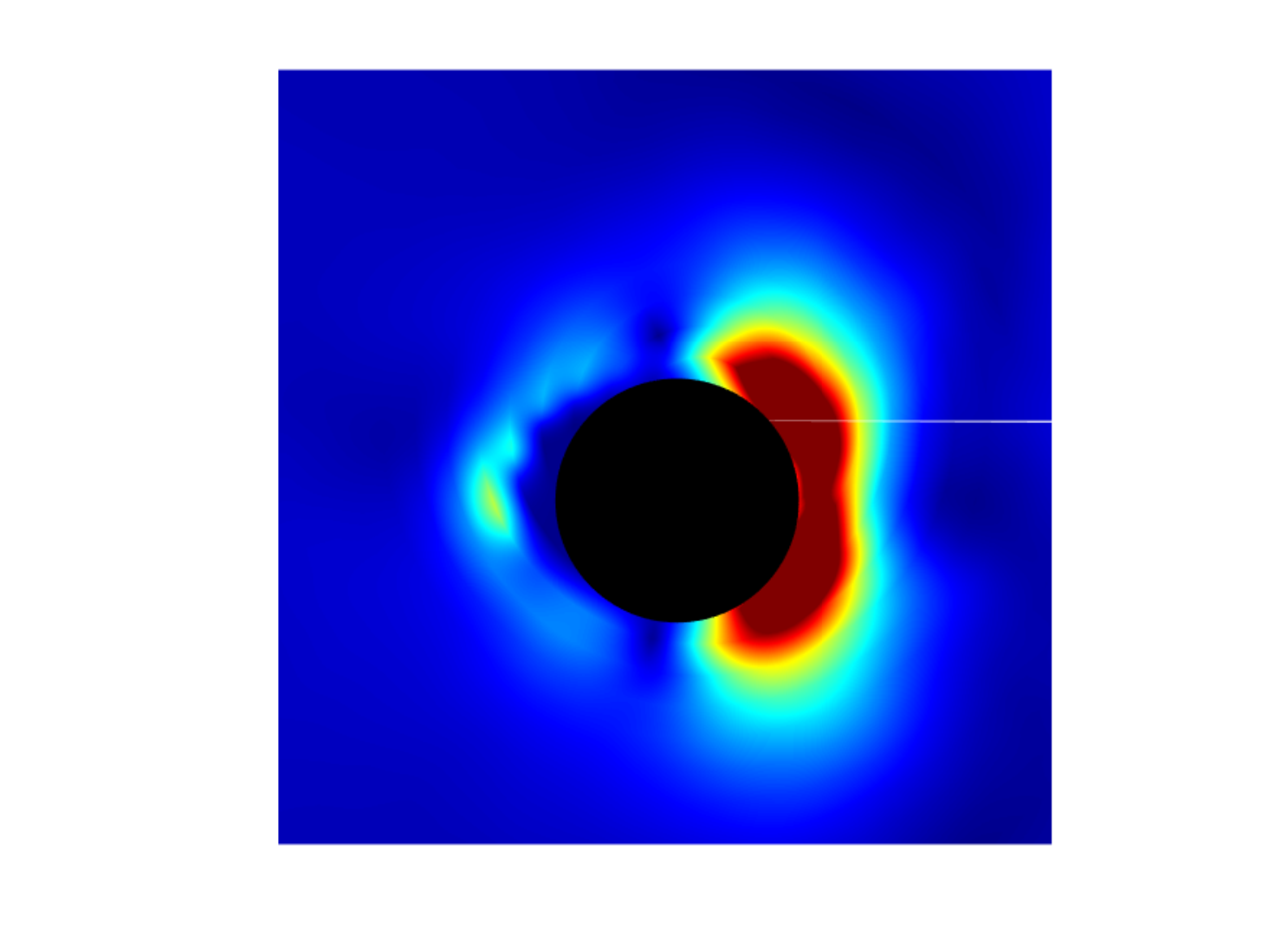}
		\caption{}
		\label{fig:y equals x}
	\end{subfigure}
	\hspace{-1.1cm}
	\hfill
	\begin{subfigure}[b]{0.25\textwidth}
		\centering
		\includegraphics[width=\textwidth]{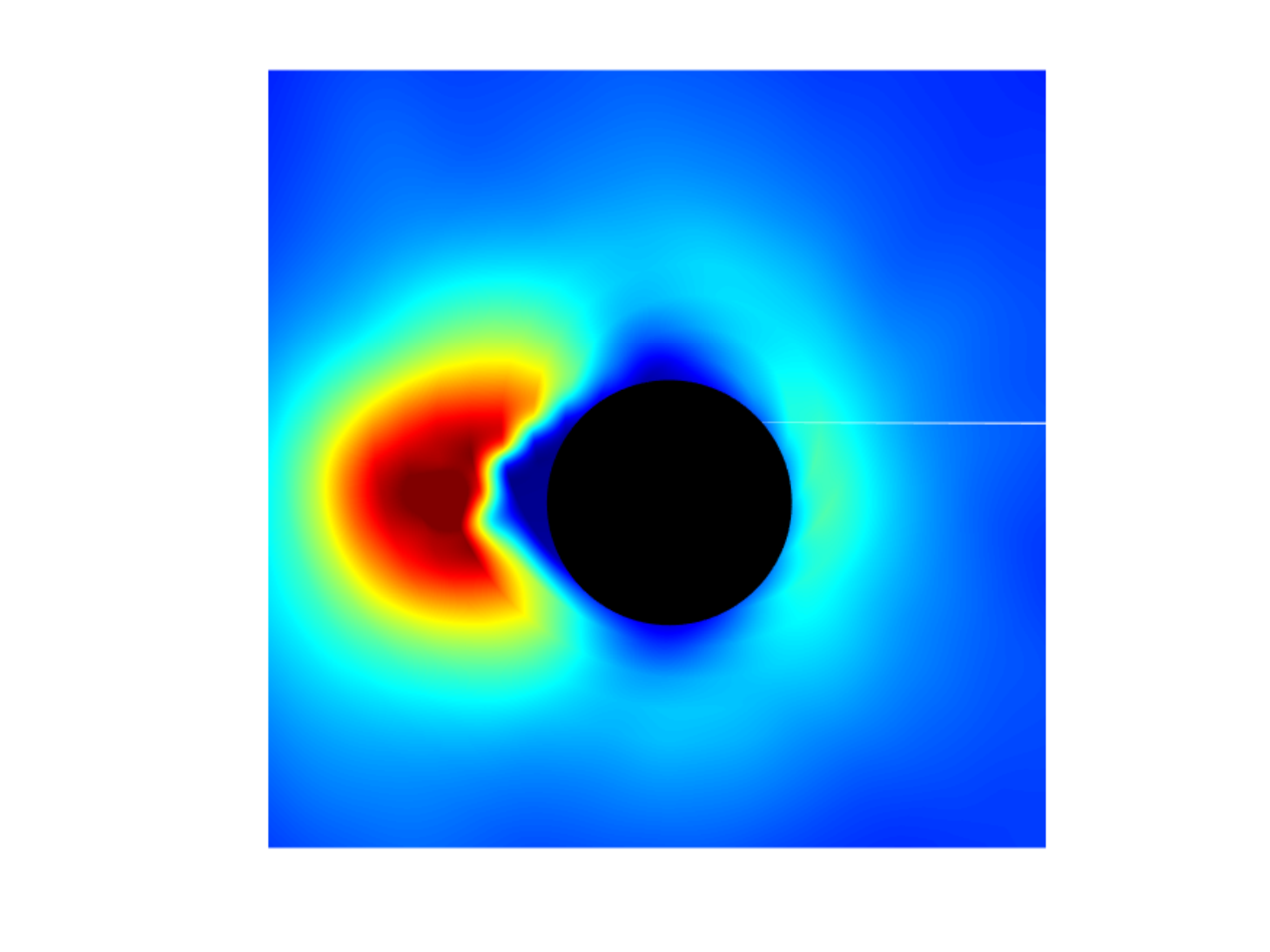}
		\caption{}
		\label{fig:three sin x}
	\end{subfigure}
	\hspace{-1.1cm}
	\hfill
	\begin{subfigure}[b]{0.25\textwidth}
		\centering
		\includegraphics[width=\textwidth]{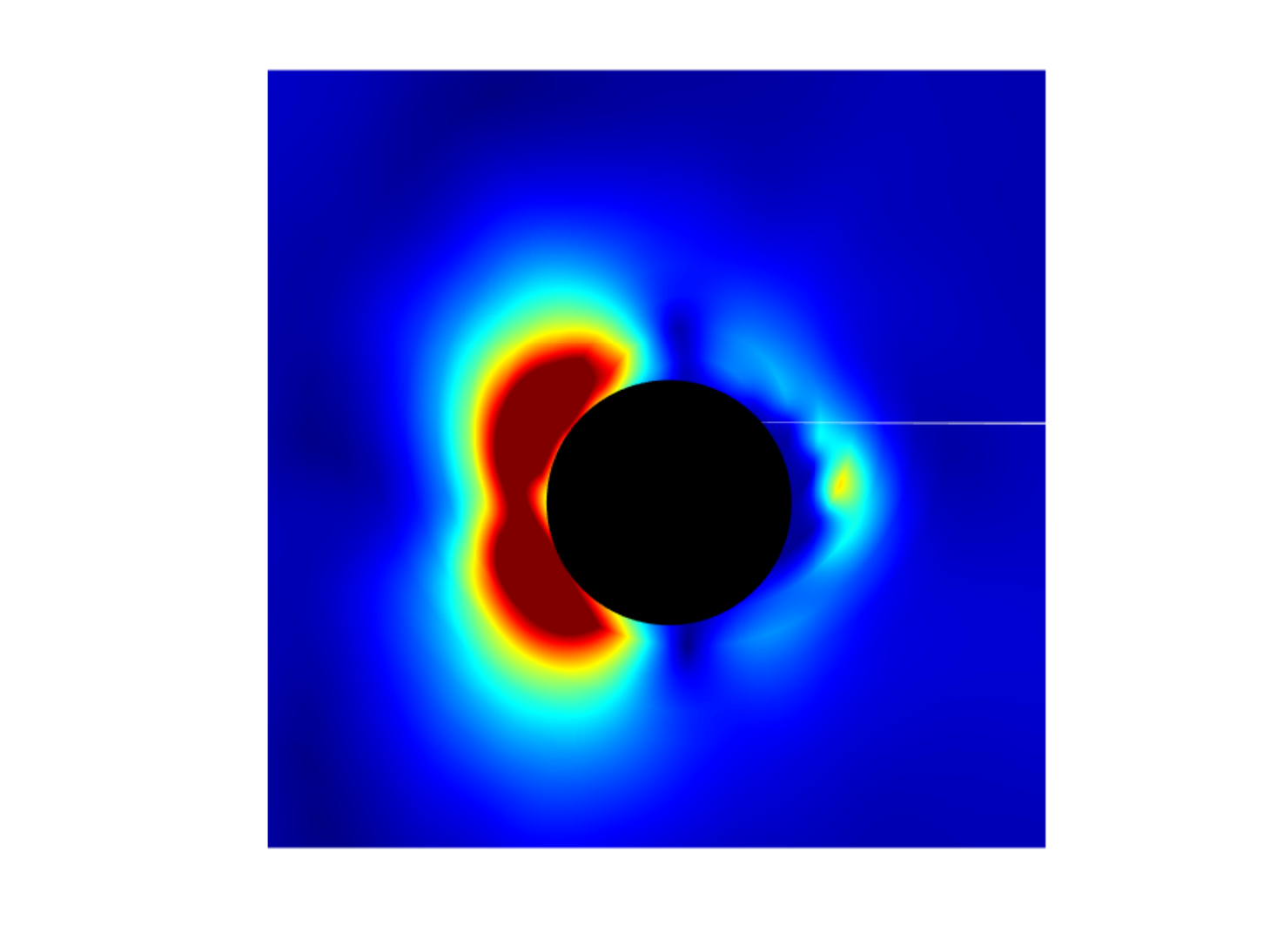}
		\caption{}
		\label{fig:five over x}
	\end{subfigure}
	\hspace{-1.1cm}
	\hfill
	\begin{subfigure}[b]{0.25\textwidth}
		\centering
		\includegraphics[width=\textwidth]{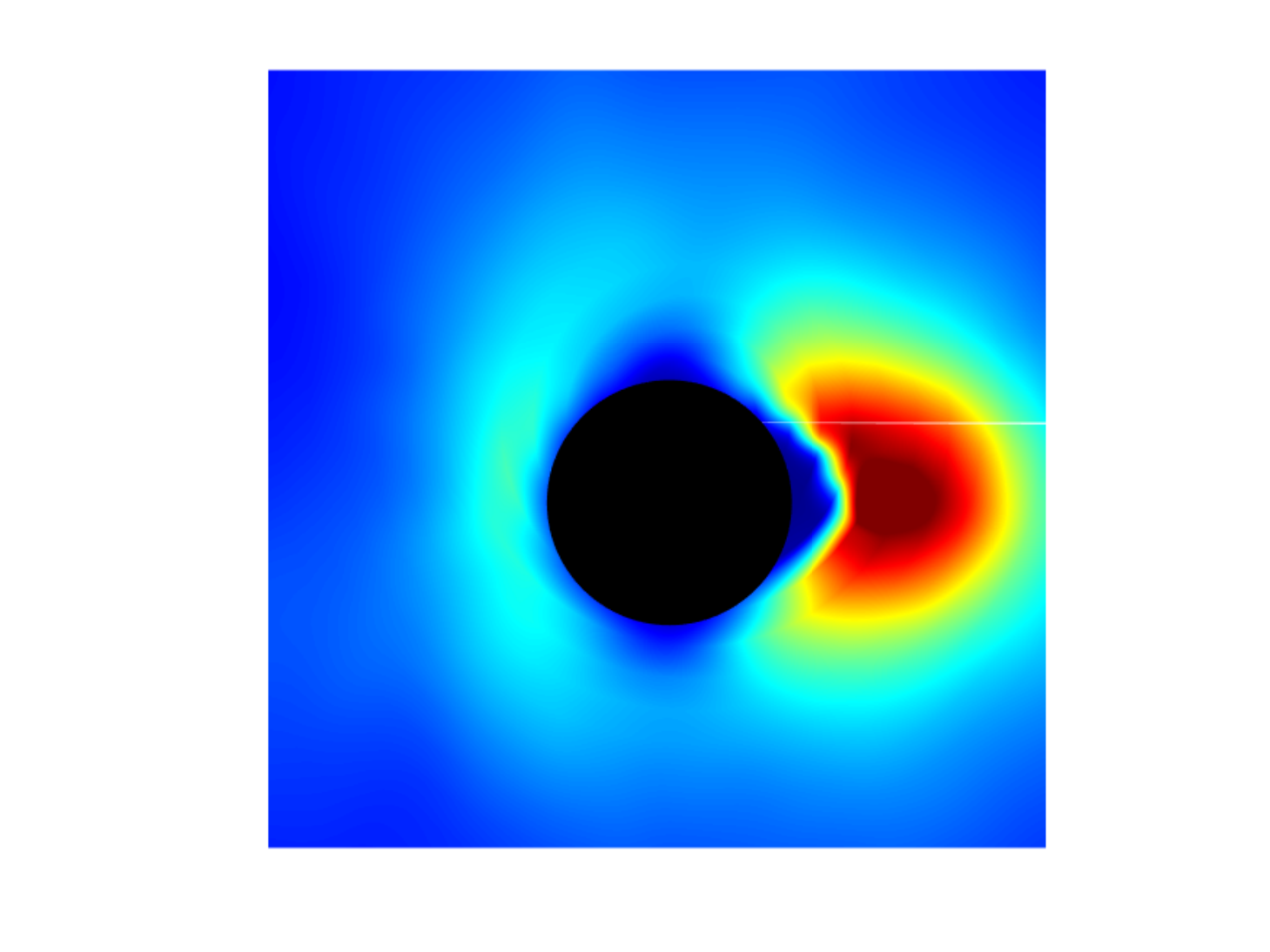}
		\caption{}
		\label{fig:three sin x}
	\end{subfigure}
	\hspace{-1.0cm}
	\hfill
	\begin{subfigure}[b]{0.25\textwidth}
		\centering
		\includegraphics[width=\textwidth]{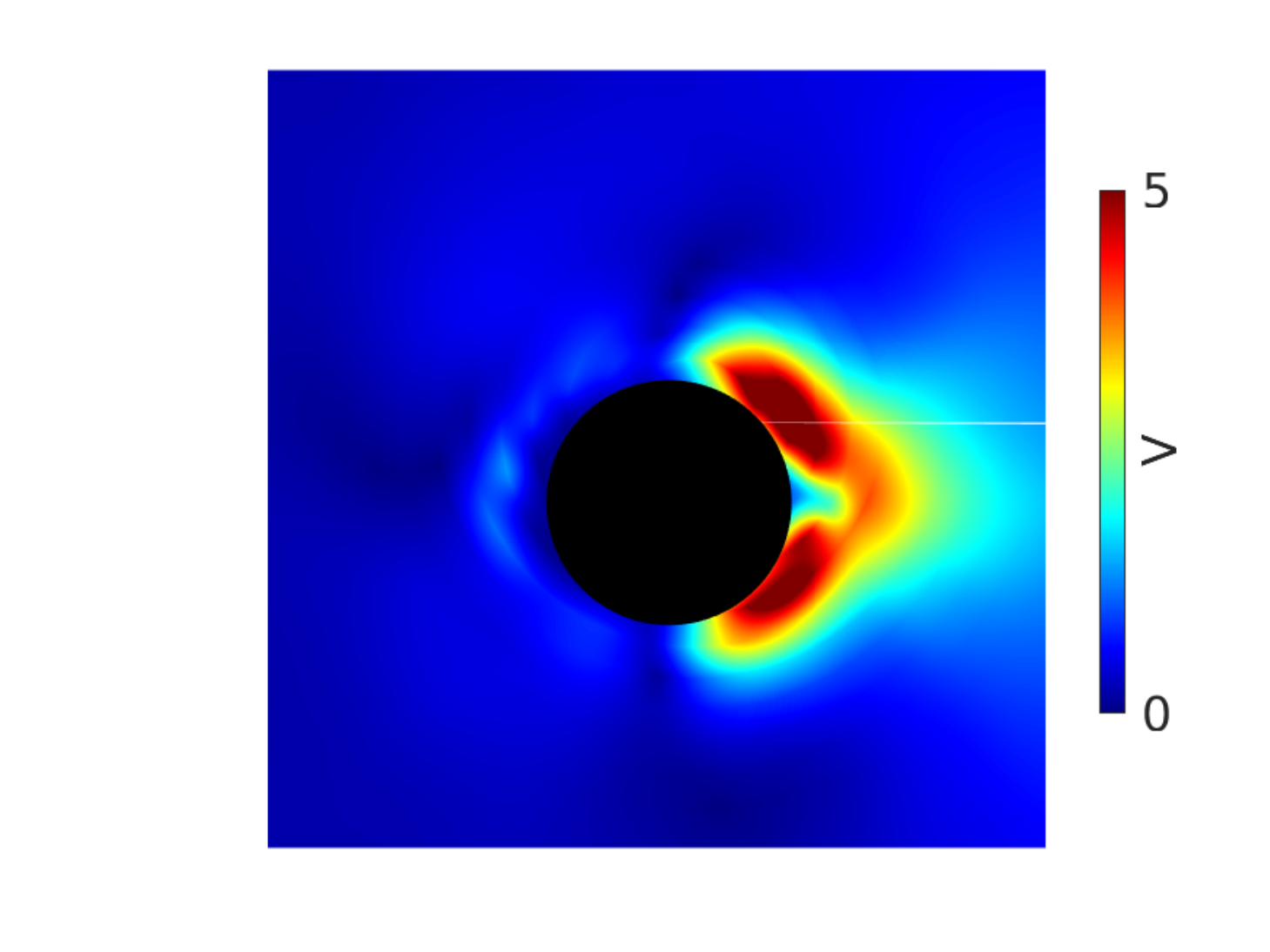}
		\caption{}
		\label{fig:five over x}
	\end{subfigure}
	
	\hspace{-0.9cm}
	\begin{subfigure}[b]{0.25\textwidth}
		\centering
		\includegraphics[width=\textwidth]{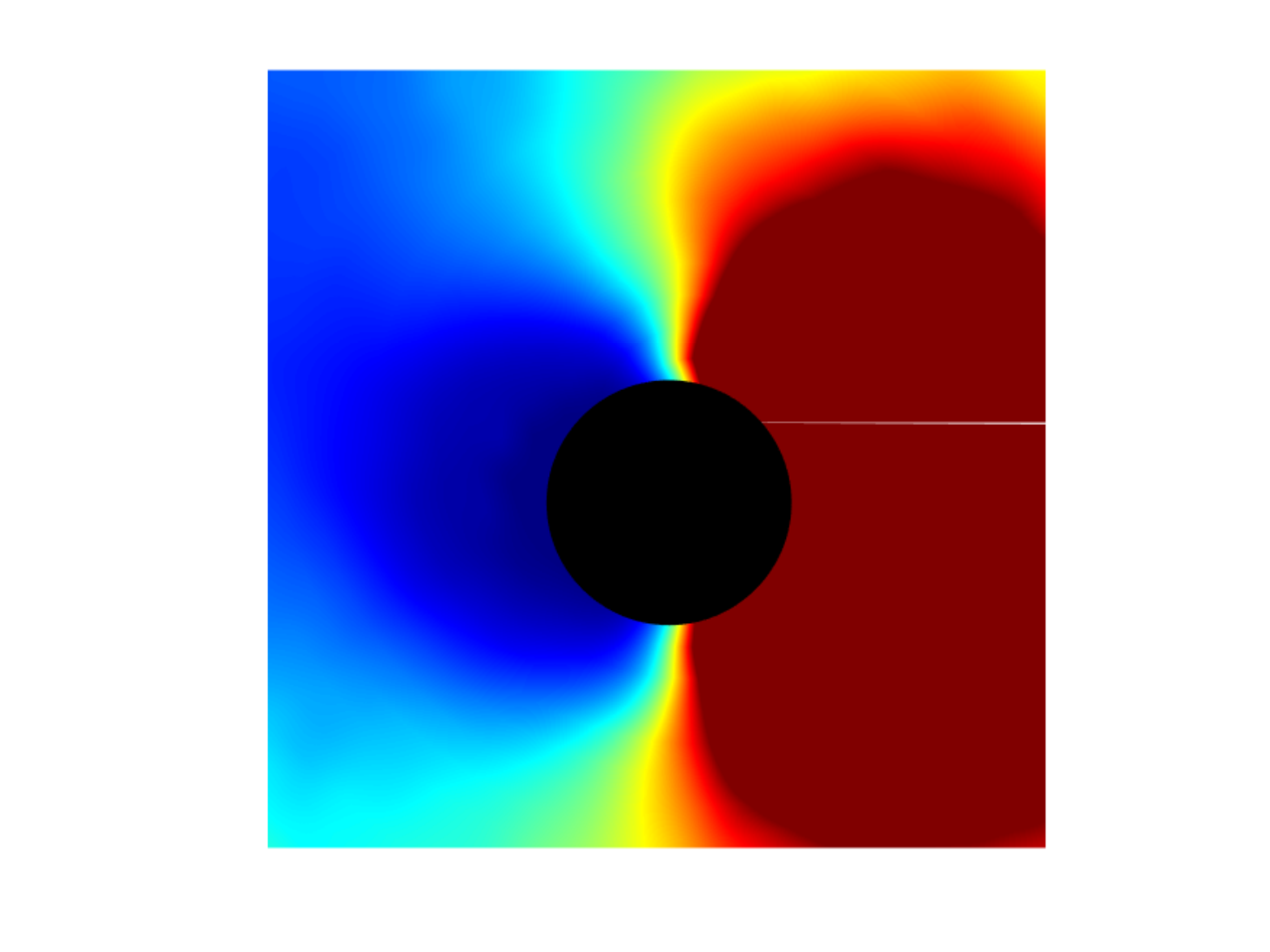}
		\caption{$10\phi/\pi=0$}
		\label{fig:y equals x}
	\end{subfigure}
	\hspace{-1.1cm}
	\hfill
	\begin{subfigure}[b]{0.25\textwidth}
		\centering
		\includegraphics[width=\textwidth]{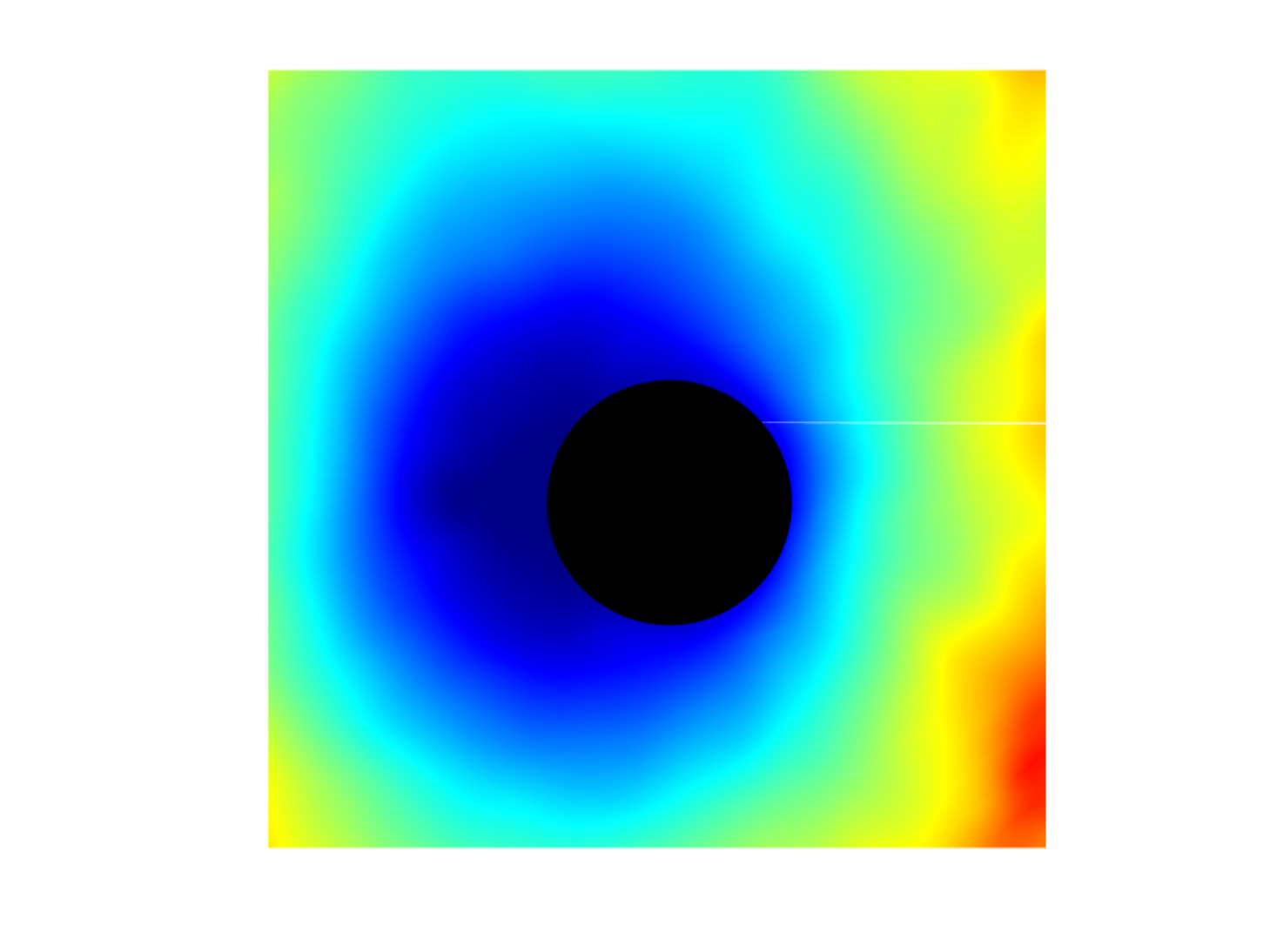}
		\caption{$10\phi/\pi=5$}
		\label{fig:three sin x}
	\end{subfigure}
	\hspace{-1.1cm}
	\hfill
	\begin{subfigure}[b]{0.25\textwidth}
		\centering
		\includegraphics[width=\textwidth]{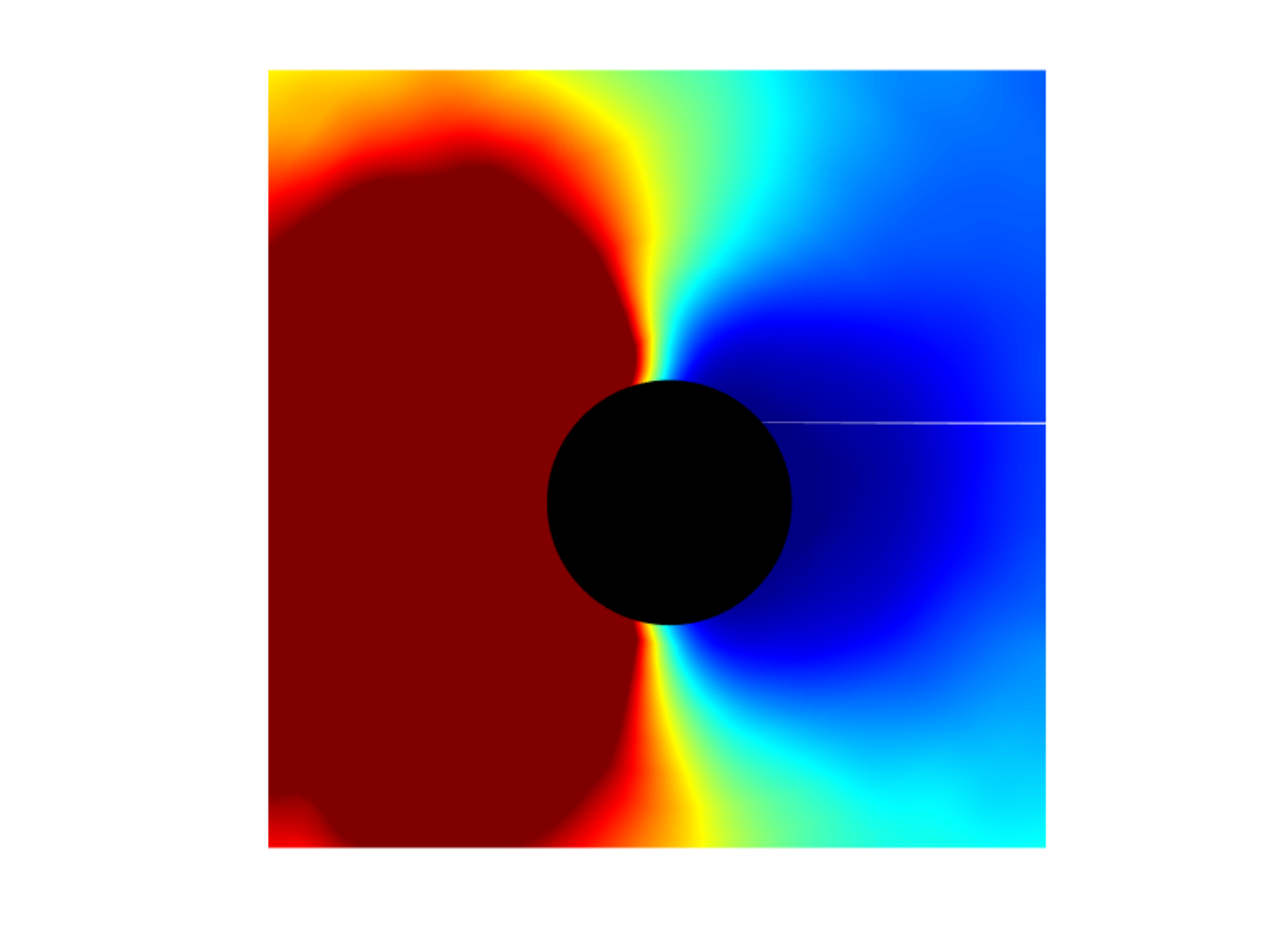}
		\caption{$10\phi/\pi=10$}
		\label{fig:five over x}
	\end{subfigure}
	\hspace{-1.1cm}
	\hfill
	\begin{subfigure}[b]{0.25\textwidth}
		\centering
		\includegraphics[width=\textwidth]{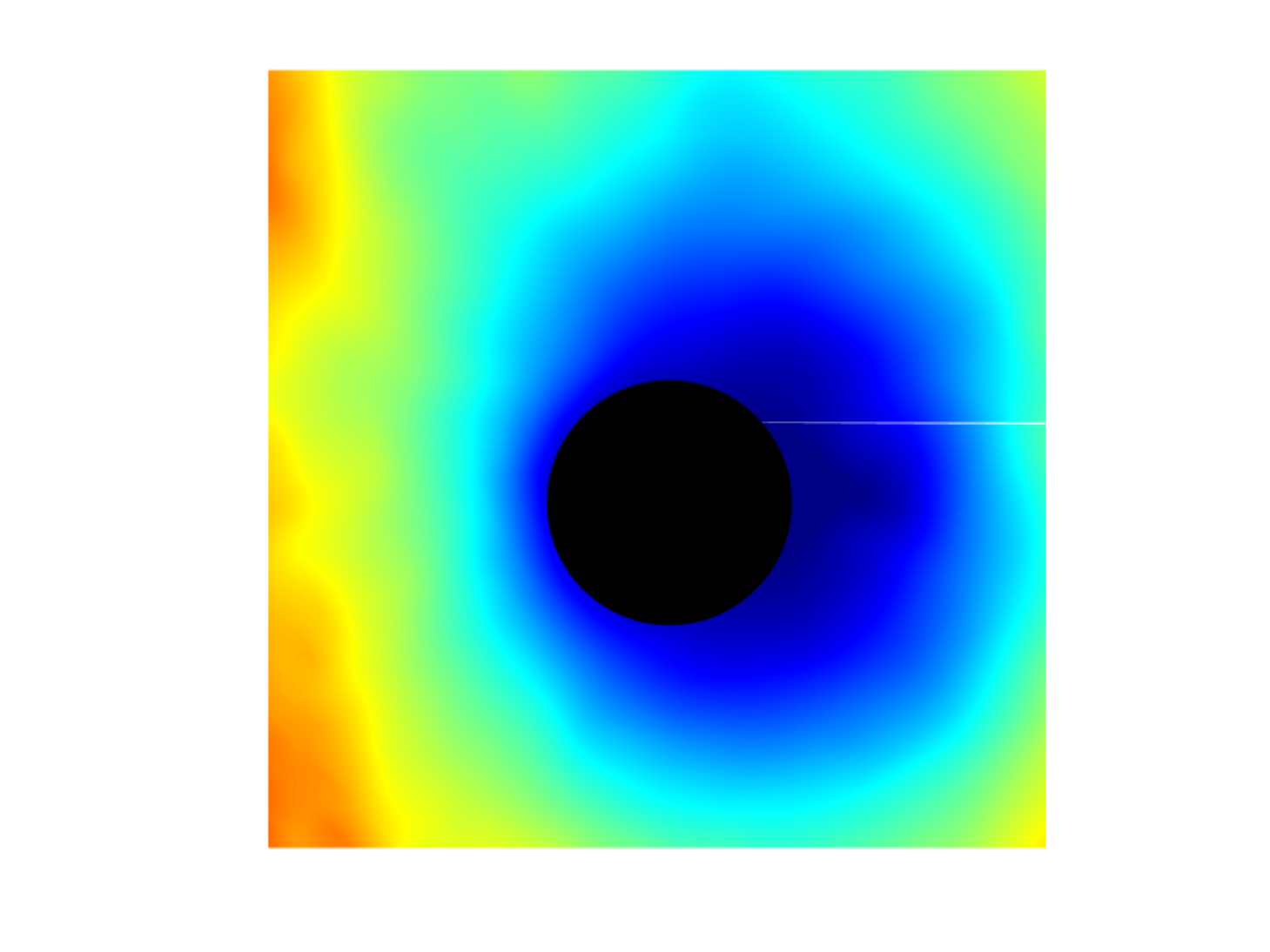}
		\caption{$10\phi/\pi=15$}
		\label{fig:three sin x}
	\end{subfigure}
	\hspace{-1.0cm}
	\hfill
	\begin{subfigure}[b]{0.25\textwidth}
		\centering
		\includegraphics[width=\textwidth]{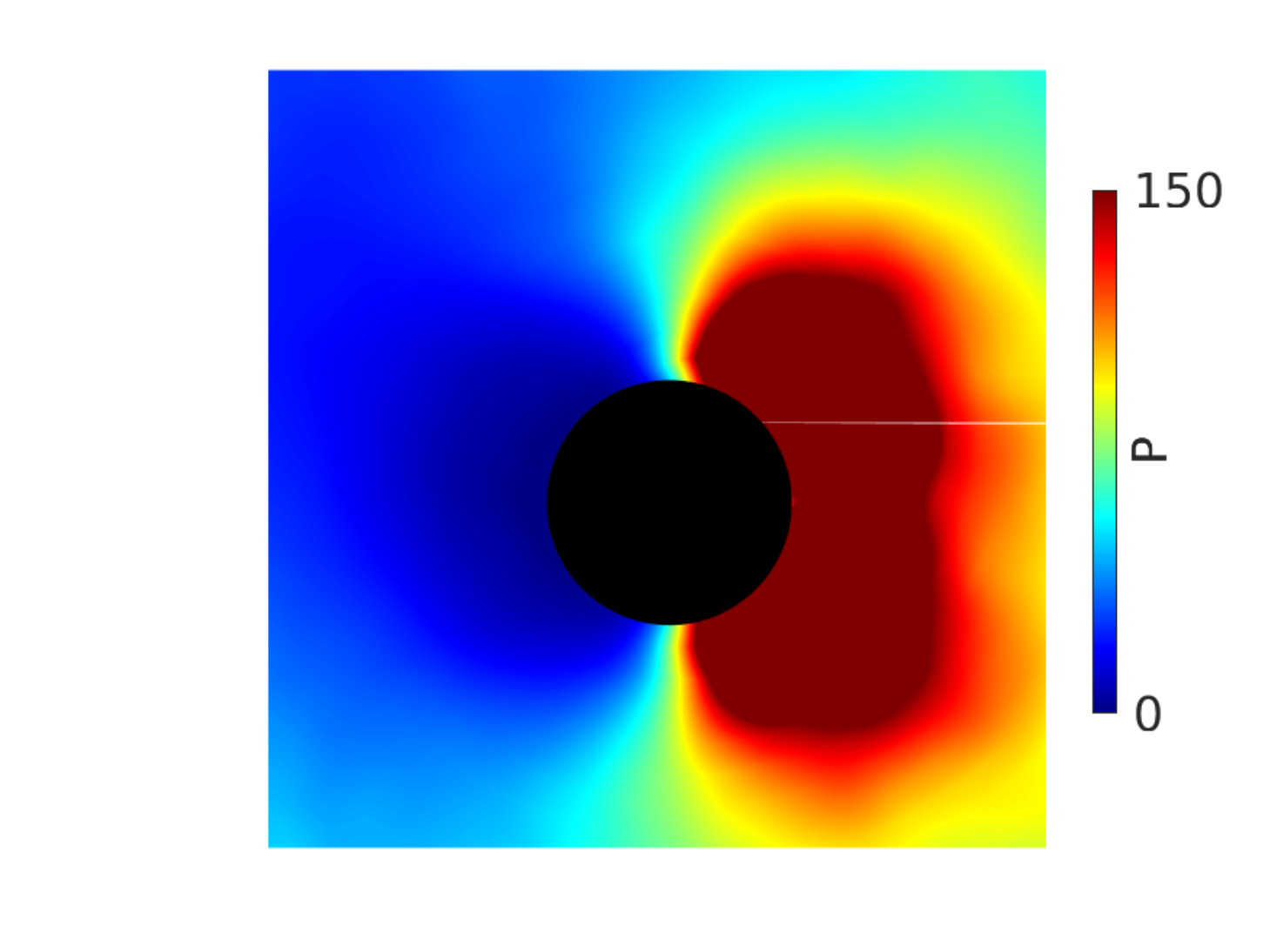}
		\caption{$10\phi/\pi=19$}
		\label{fig:five over x}
	\end{subfigure}
	
	\caption{The time-averaged flow fields for depicting the area fraction $\Phi$, velocity $V$ and pressure $P$ around the intruder $OS1$ (disk) for $T/\sqrt{d/g}=1$ and different phase angle $10\phi/\pi$  : (a), (f), (k) 0.0, (b) (g), (l) 5.0, and (c), (h), (m) 10.0, (d), (i), (n) 15.0 and (e), (j), (o) 19.0. The oscillating amplitude of the intruder is $A/d=3$.\label{fig:mff_disc}}
\end{figure}

\begin{figure}
	\hspace{-0.9cm}
	\begin{subfigure}[b]{0.25\textwidth}
		\centering
		\includegraphics[width=\textwidth]{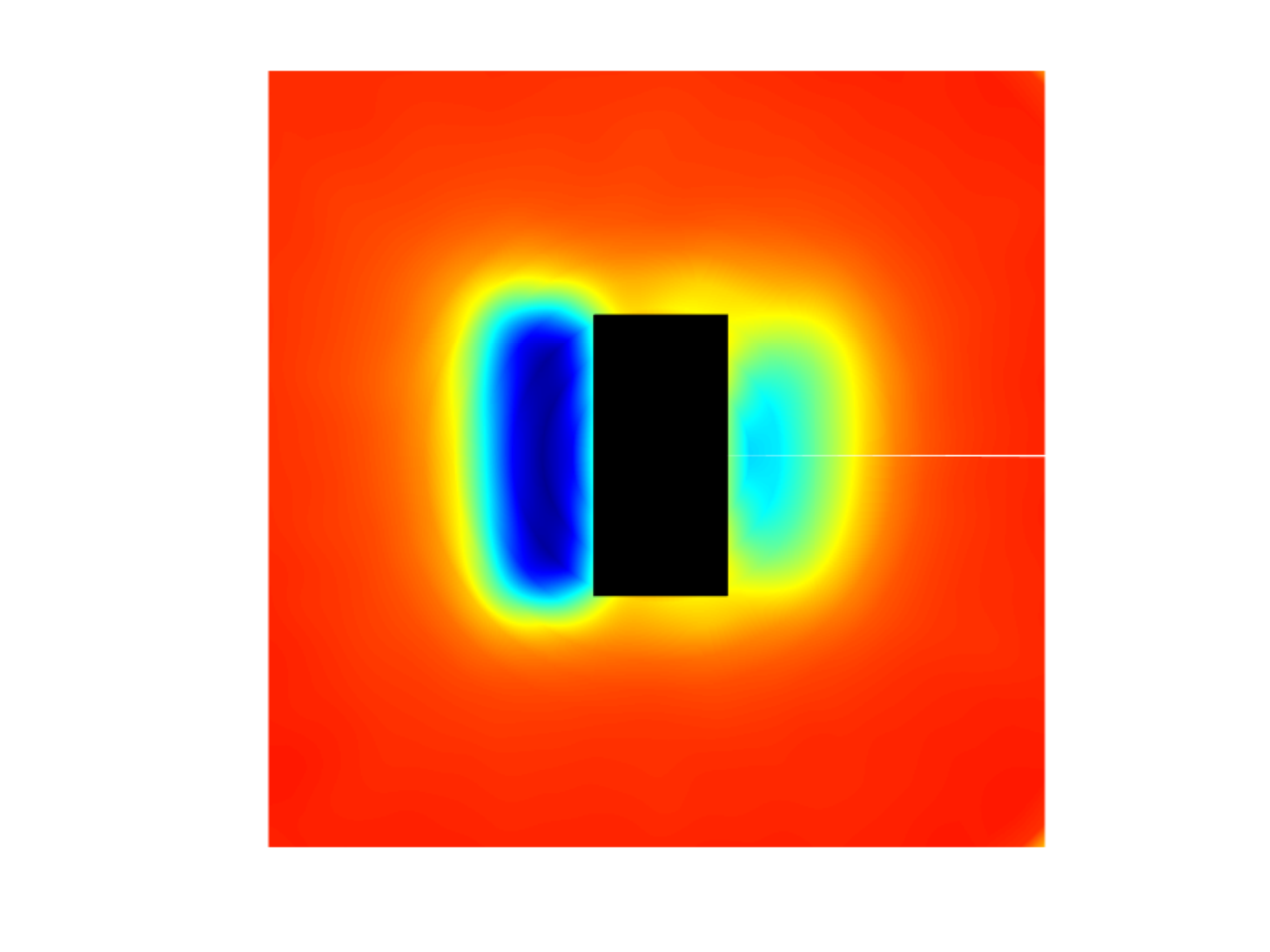}
		\caption{}
		\label{fig:y equals x}
	\end{subfigure}
	\hspace{-1.1cm}
	\hfill
	\begin{subfigure}[b]{0.25\textwidth}
		\centering
		\includegraphics[width=\textwidth]{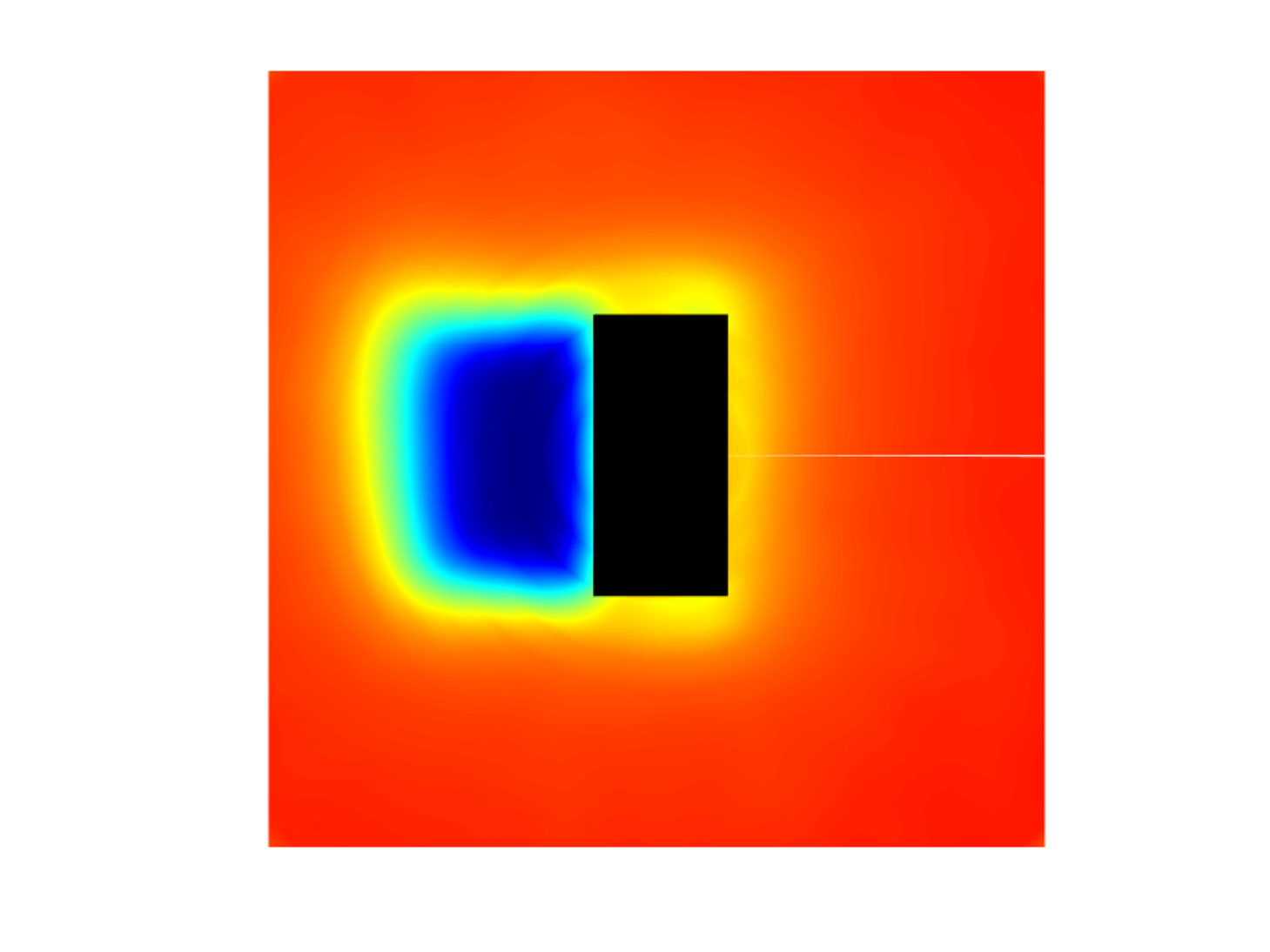}
		\caption{}
		\label{fig:three sin x}
	\end{subfigure}
	\hspace{-1.1cm}
	\hfill
	\begin{subfigure}[b]{0.25\textwidth}
		\centering
		\includegraphics[width=\textwidth]{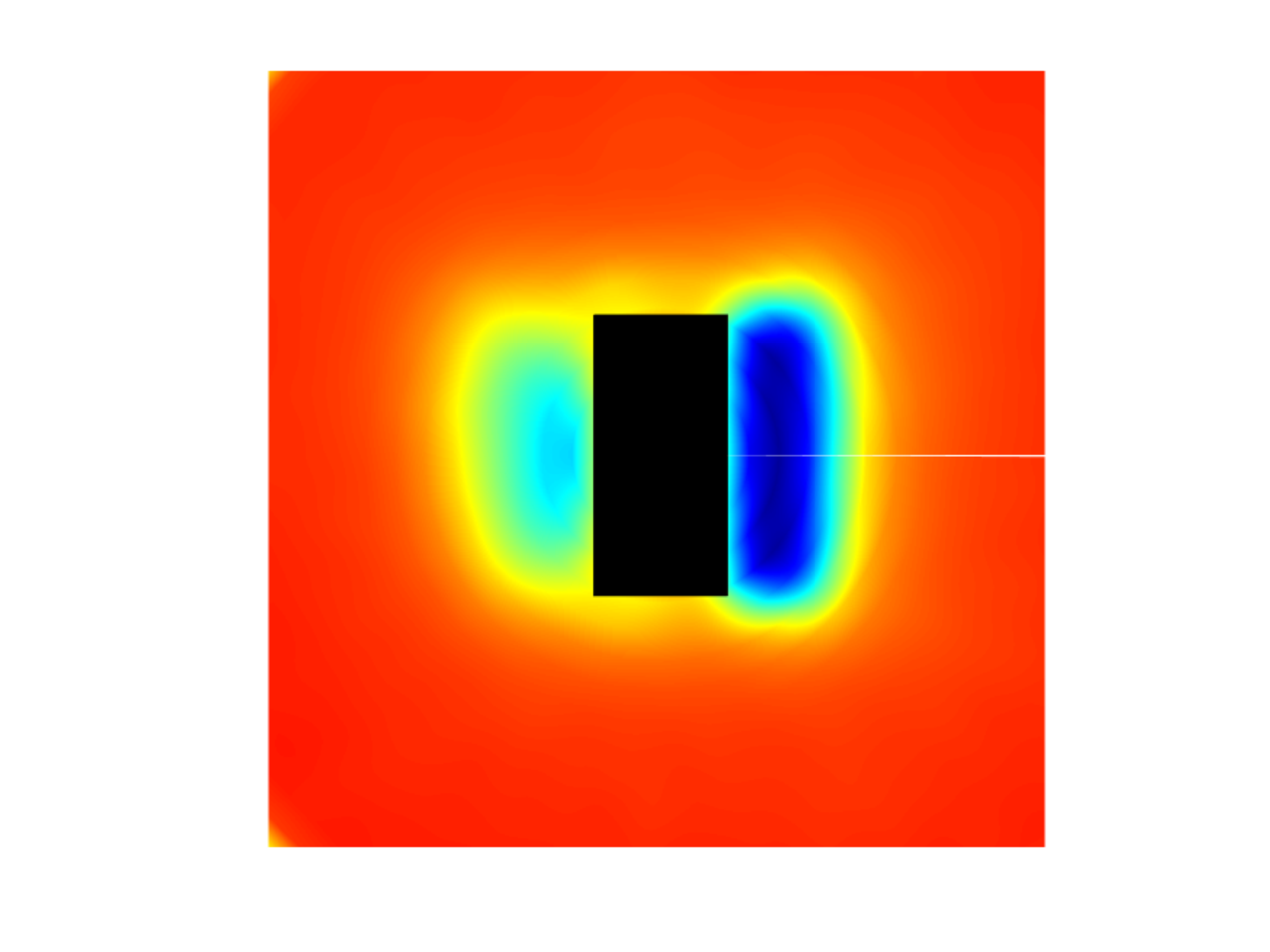}
		\caption{}
		\label{fig:five over x}
	\end{subfigure}
	\hspace{-1.1cm}
	\hfill
	\begin{subfigure}[b]{0.25\textwidth}
		\centering
		\includegraphics[width=\textwidth]{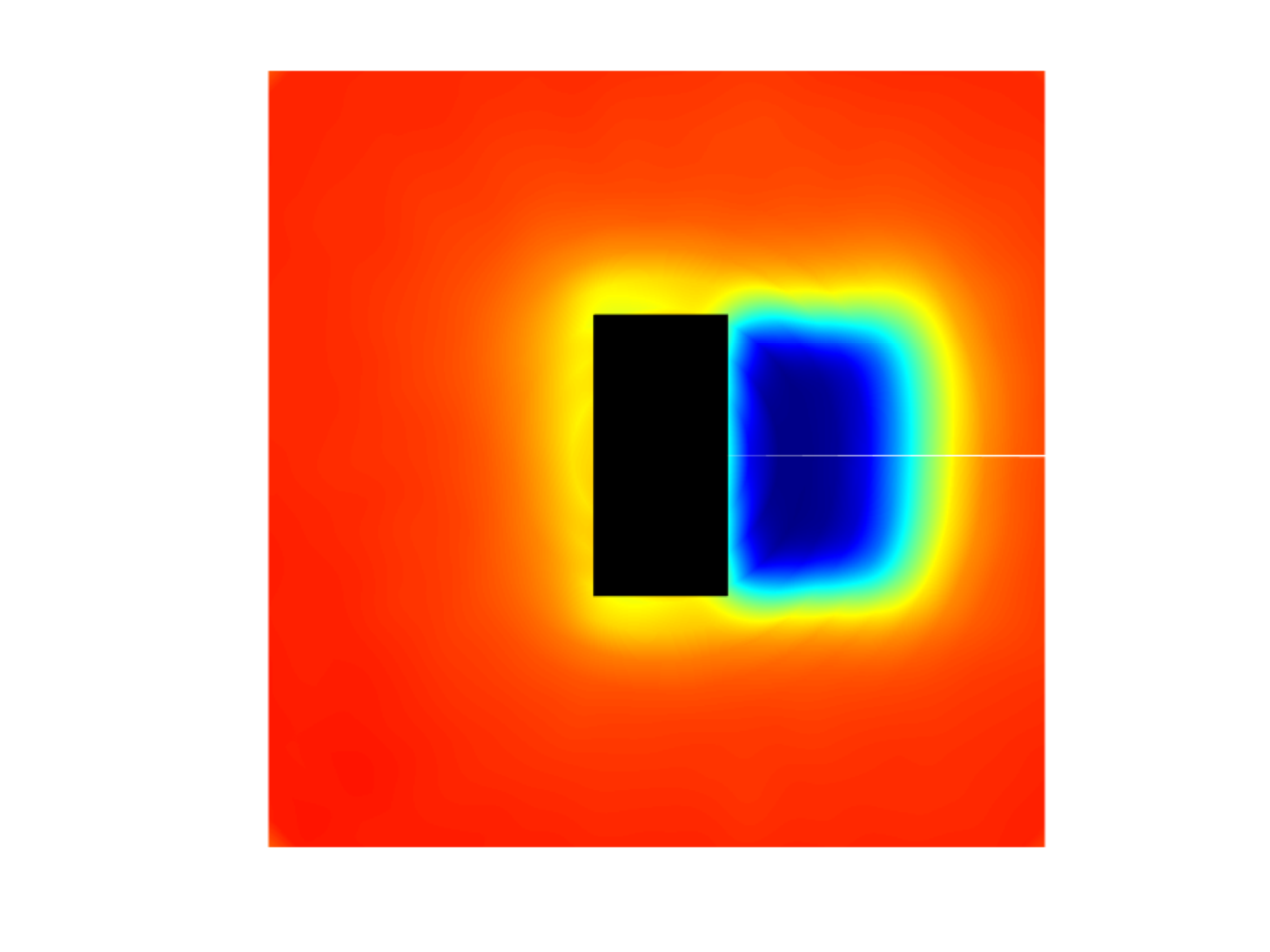}
		\caption{}
		\label{fig:three sin x}
	\end{subfigure}
	\hspace{-1.0cm}
	\hfill
	\begin{subfigure}[b]{0.25\textwidth}
		\centering
		\includegraphics[width=\textwidth]{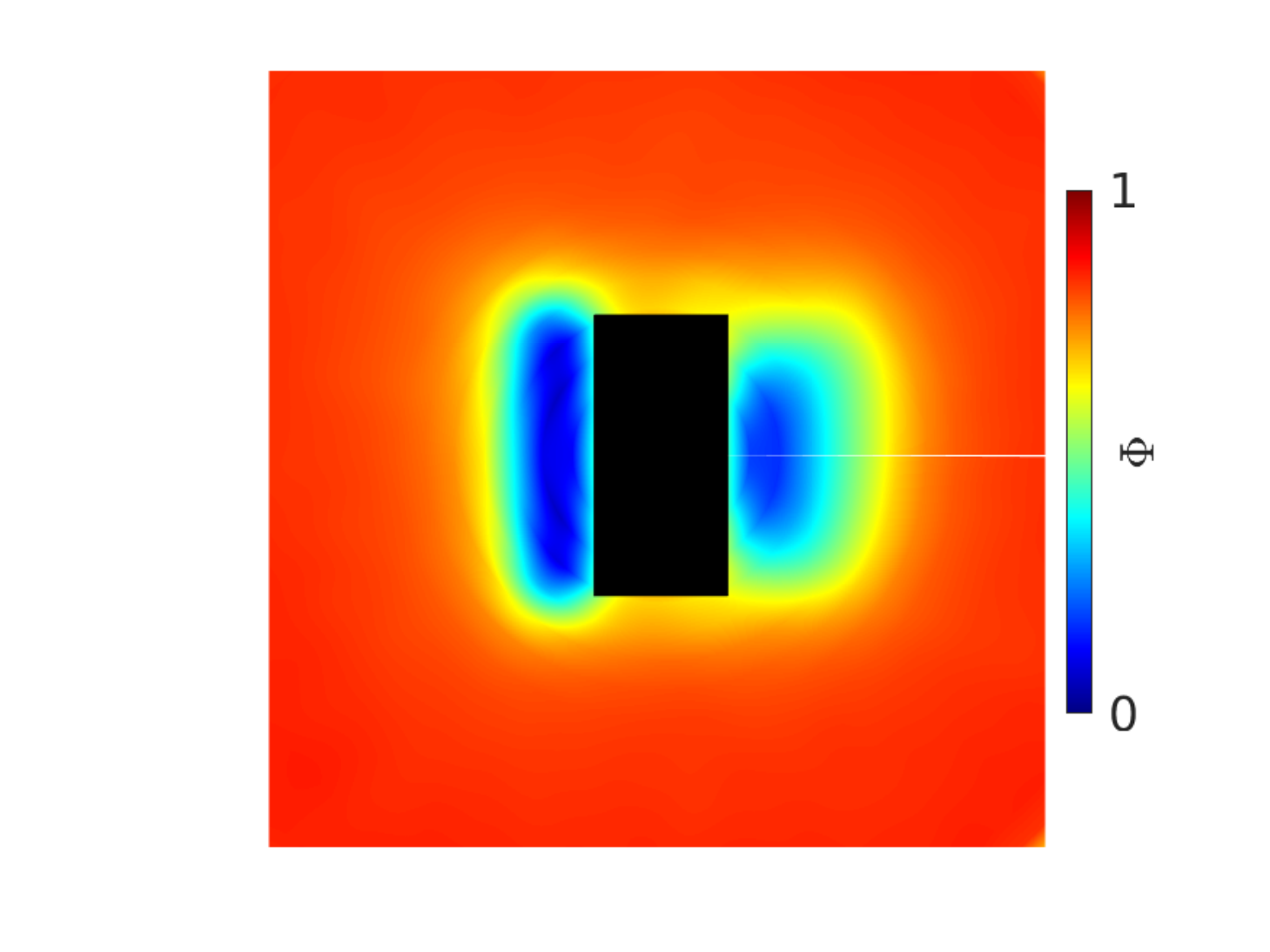}
		\caption{}
		\label{fig:five over x}
	\end{subfigure}
	
	\hspace{-0.9cm}
	\begin{subfigure}[b]{0.25\textwidth}
		\centering
		\includegraphics[width=\textwidth]{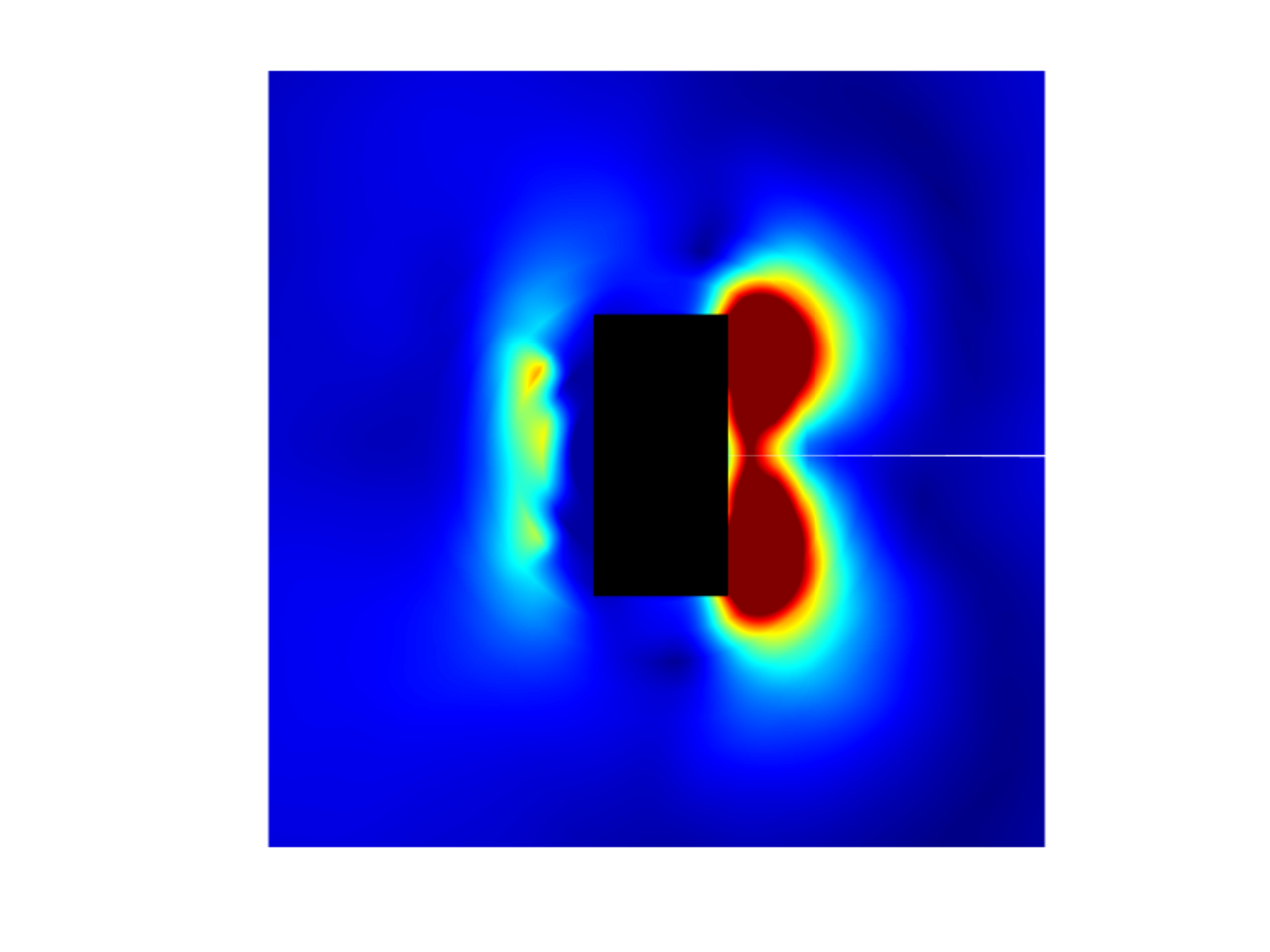}
		\caption{}
		\label{fig:y equals x}
	\end{subfigure}
	\hspace{-1.1cm}
	\hfill
	\begin{subfigure}[b]{0.25\textwidth}
		\centering
		\includegraphics[width=\textwidth]{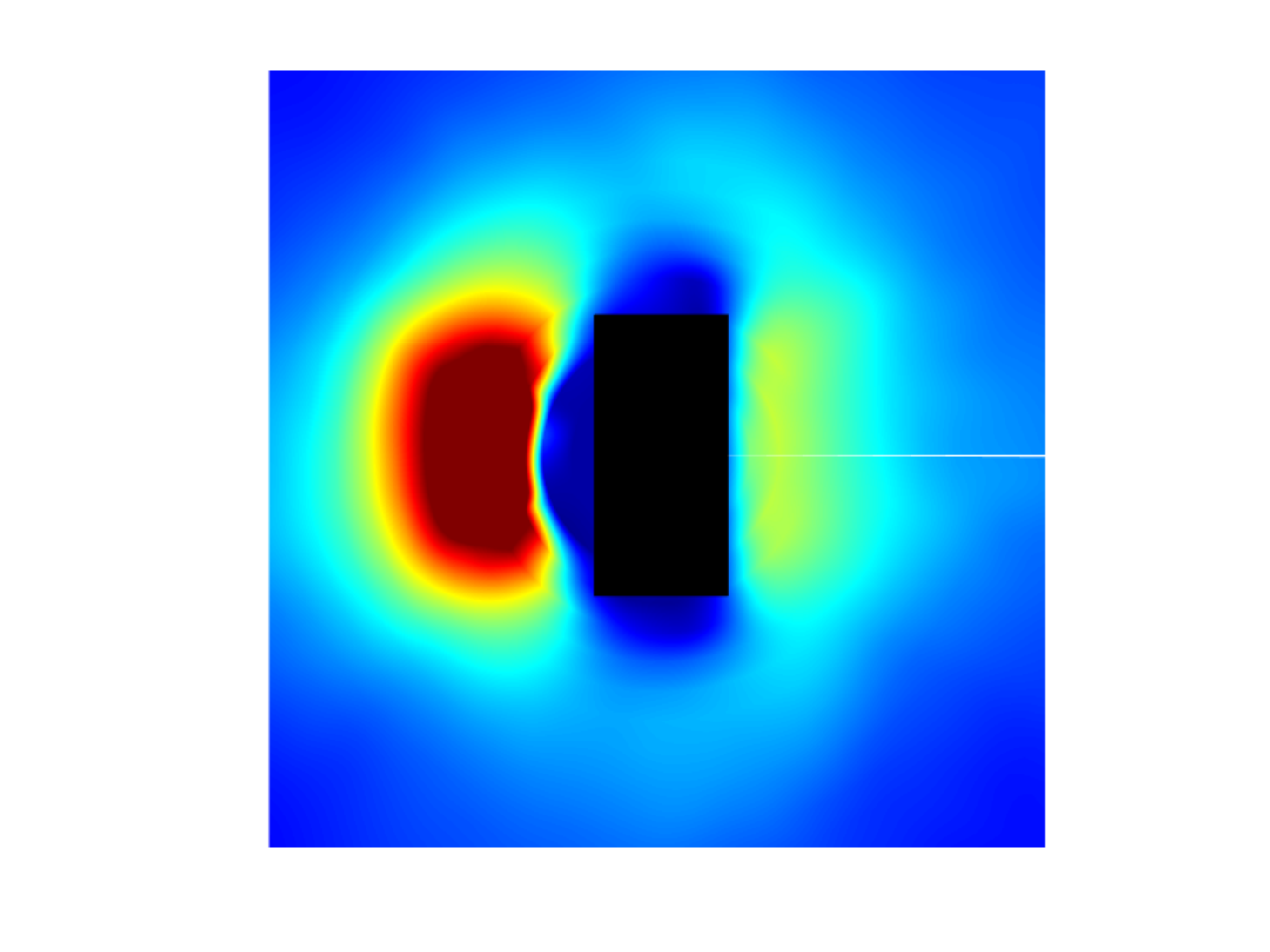}
		\caption{}
		\label{fig:three sin x}
	\end{subfigure}
	\hspace{-1.1cm}
	\hfill
	\begin{subfigure}[b]{0.25\textwidth}
		\centering
		\includegraphics[width=\textwidth]{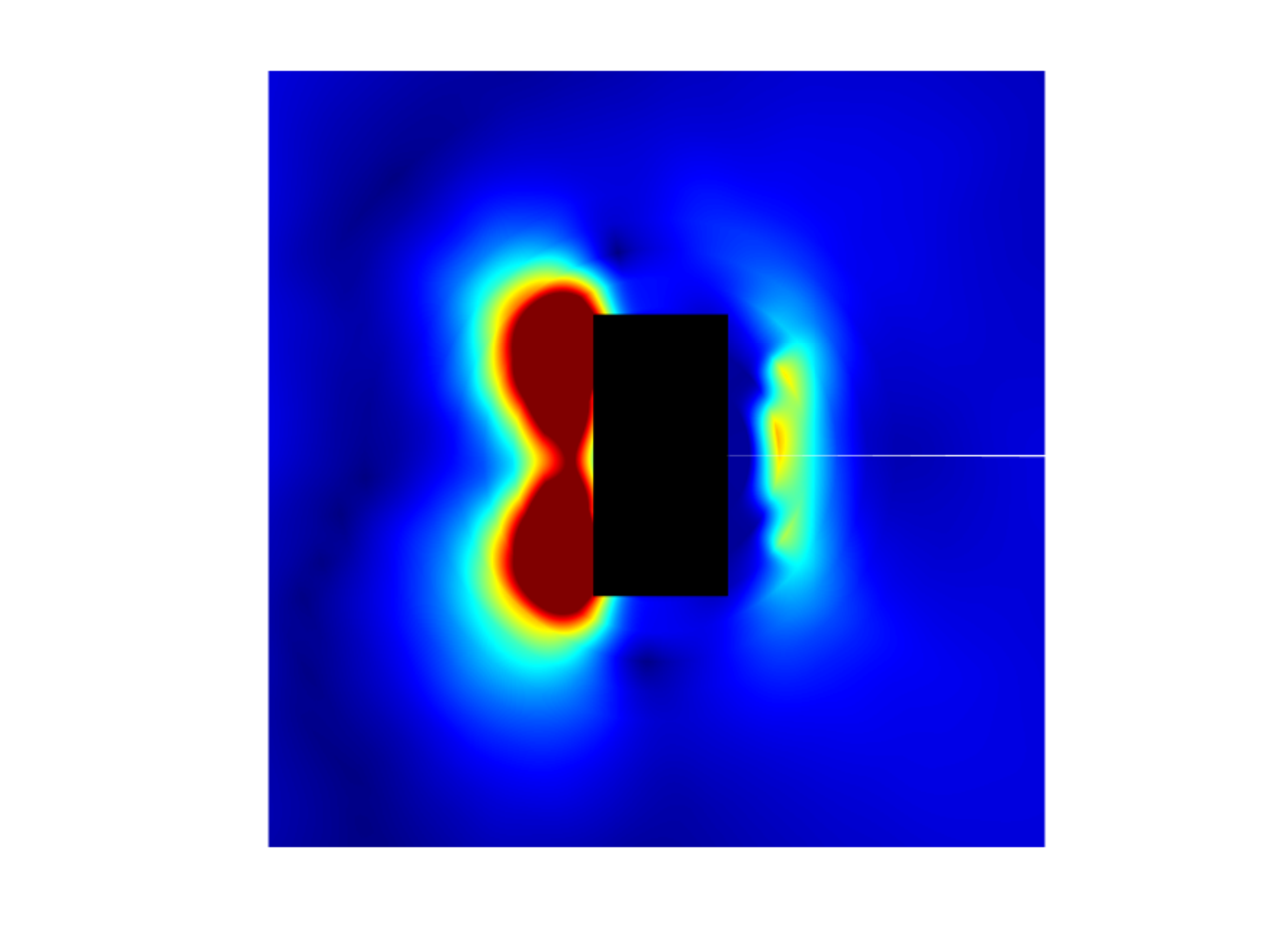}
		\caption{}
		\label{fig:five over x}
	\end{subfigure}
	\hspace{-1.1cm}
	\hfill
	\begin{subfigure}[b]{0.25\textwidth}
		\centering
		\includegraphics[width=\textwidth]{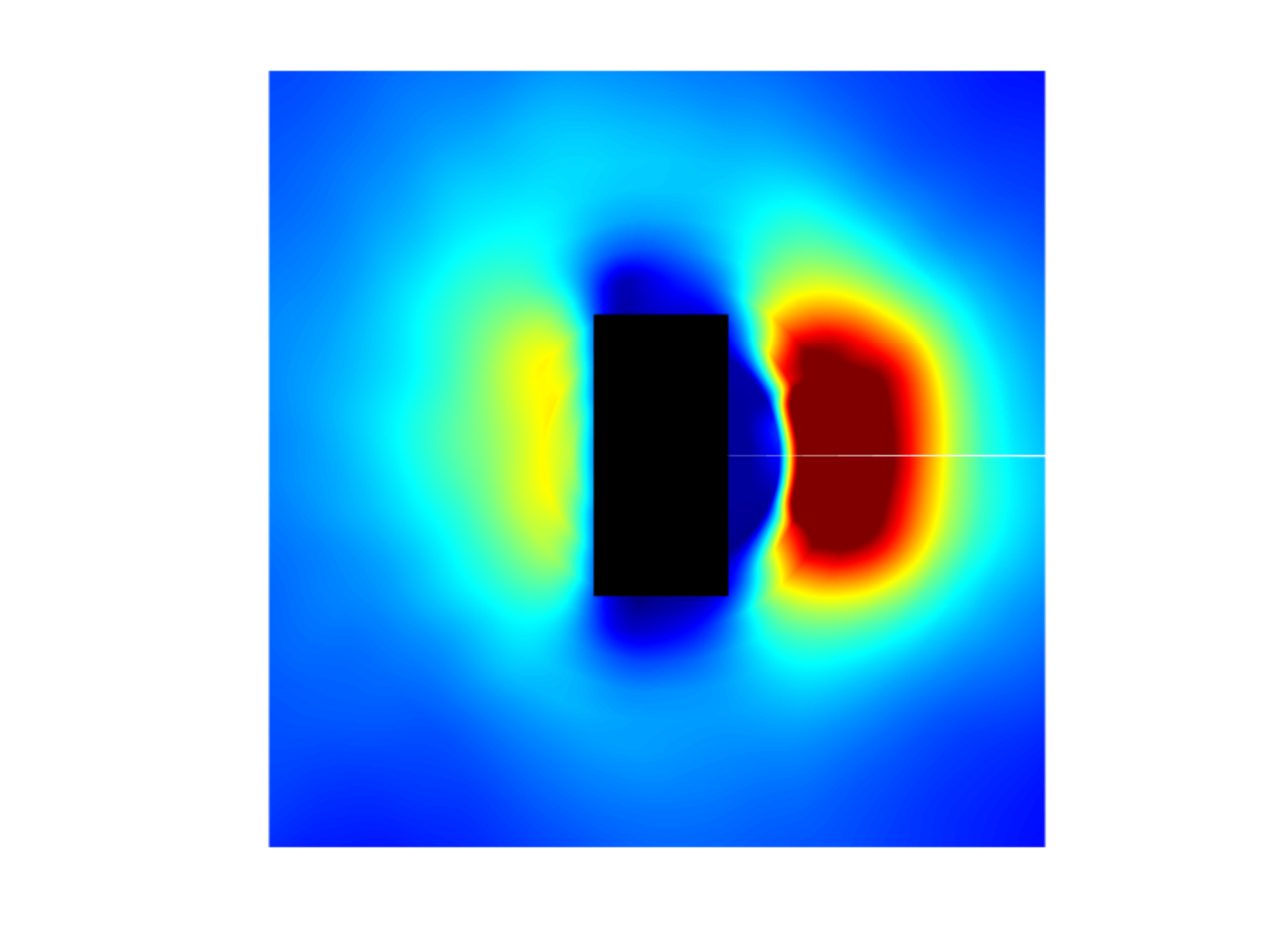}
		\caption{}
		\label{fig:three sin x}
	\end{subfigure}
	\hspace{-1.0cm}
	\hfill
	\begin{subfigure}[b]{0.25\textwidth}
		\centering
		\includegraphics[width=\textwidth]{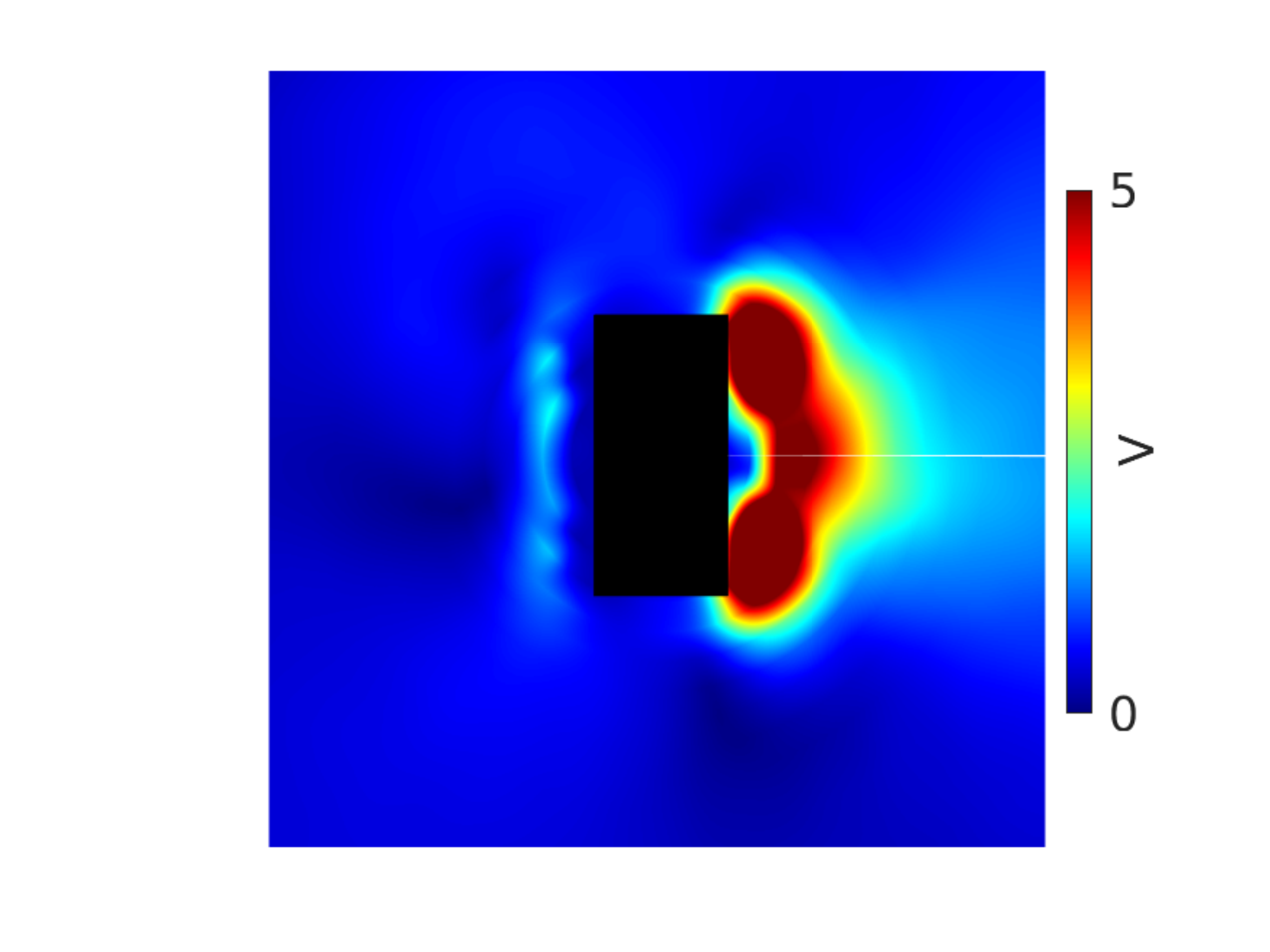}
		\caption{}
		\label{fig:five over x}
	\end{subfigure}
	
	\hspace{-0.9cm}
	\begin{subfigure}[b]{0.25\textwidth}
		\centering
		\includegraphics[width=\textwidth]{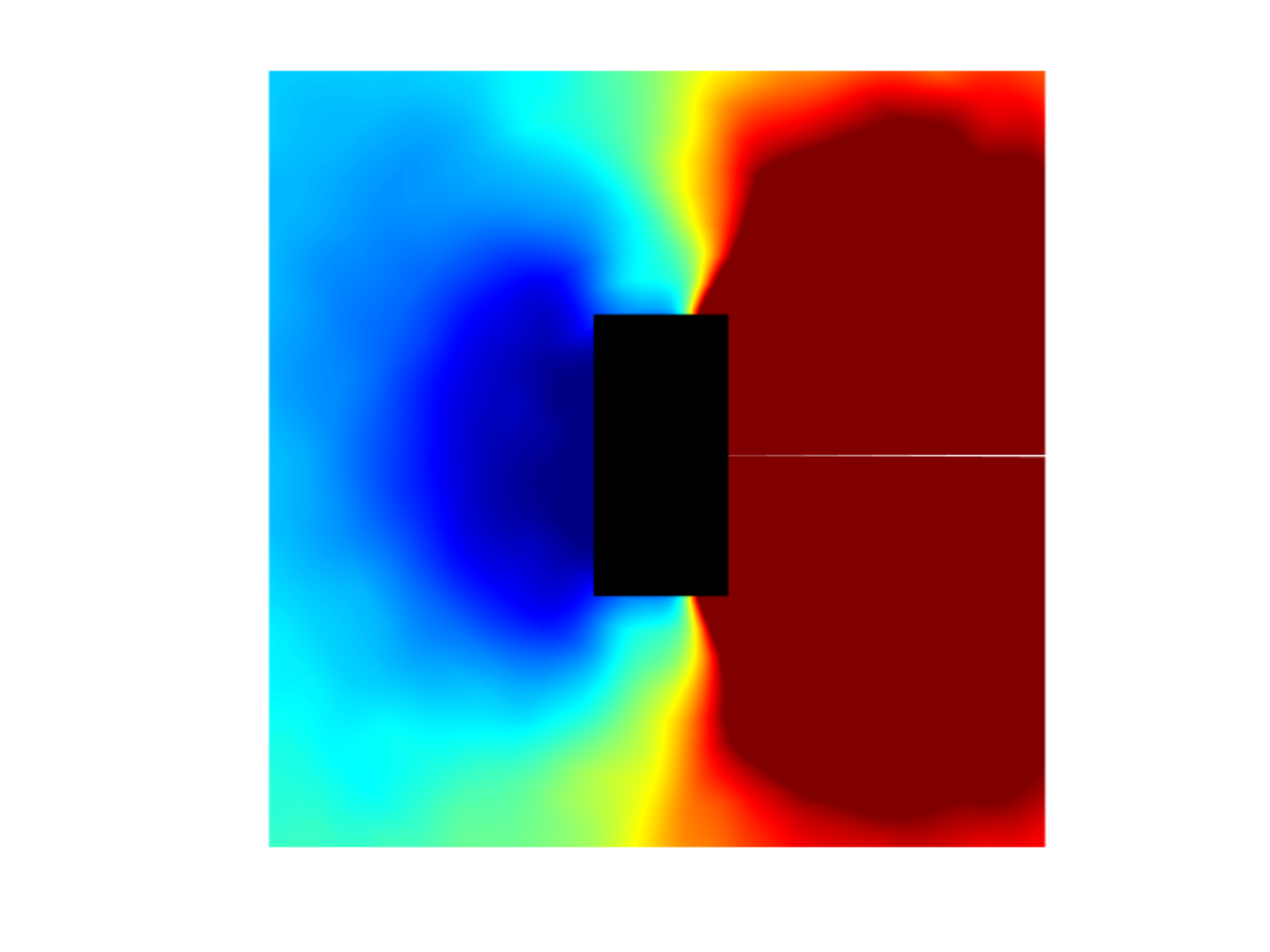}
		\caption{$10\phi/\pi=0$}
		\label{fig:y equals x}
	\end{subfigure}
	\hspace{-1.1cm}
	\hfill
	\begin{subfigure}[b]{0.25\textwidth}
		\centering
		\includegraphics[width=\textwidth]{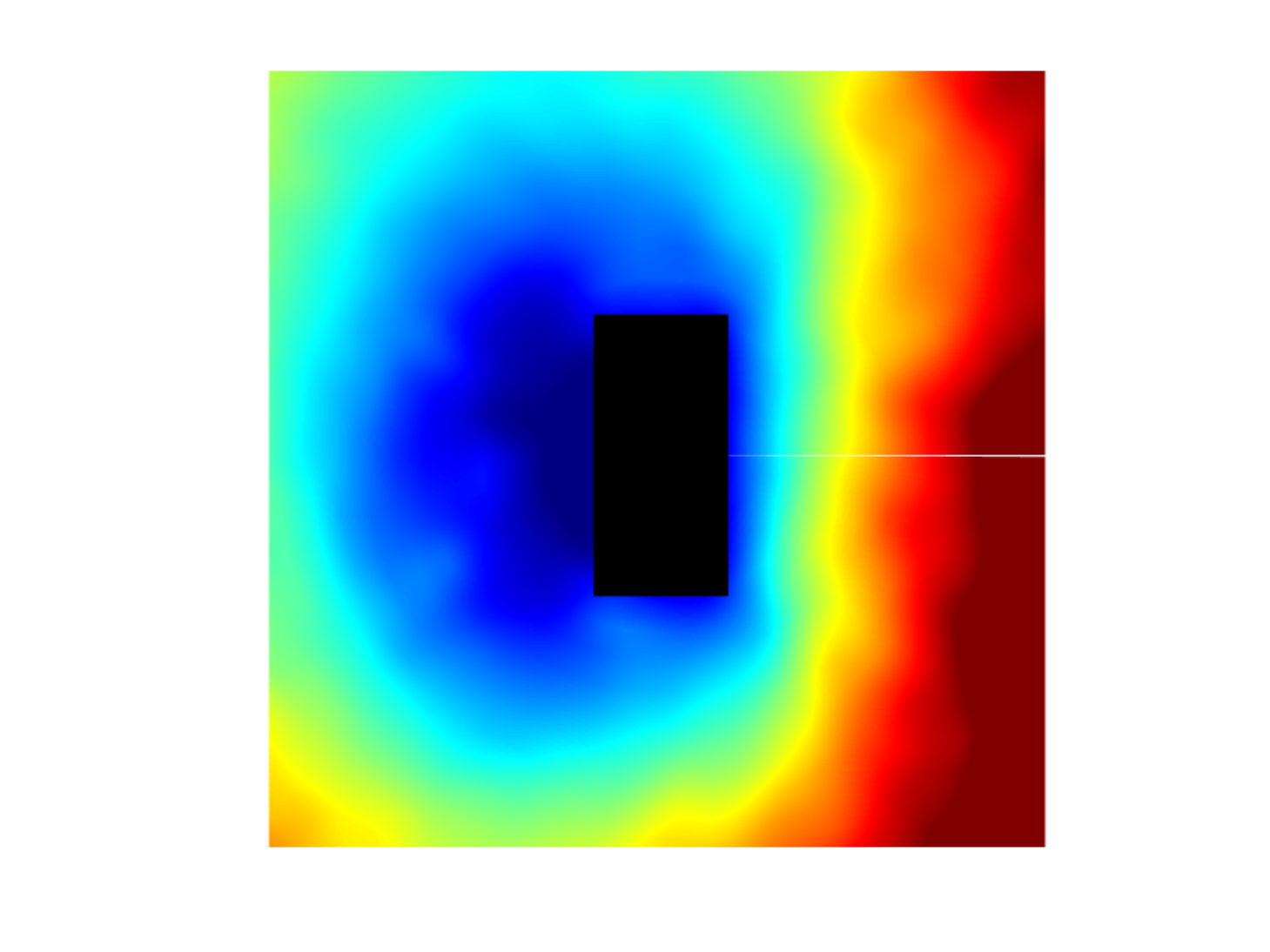}
		\caption{$10\phi/\pi=5$}
		\label{fig:three sin x}
	\end{subfigure}
	\hspace{-1.1cm}
	\hfill
	\begin{subfigure}[b]{0.25\textwidth}
		\centering
		\includegraphics[width=\textwidth]{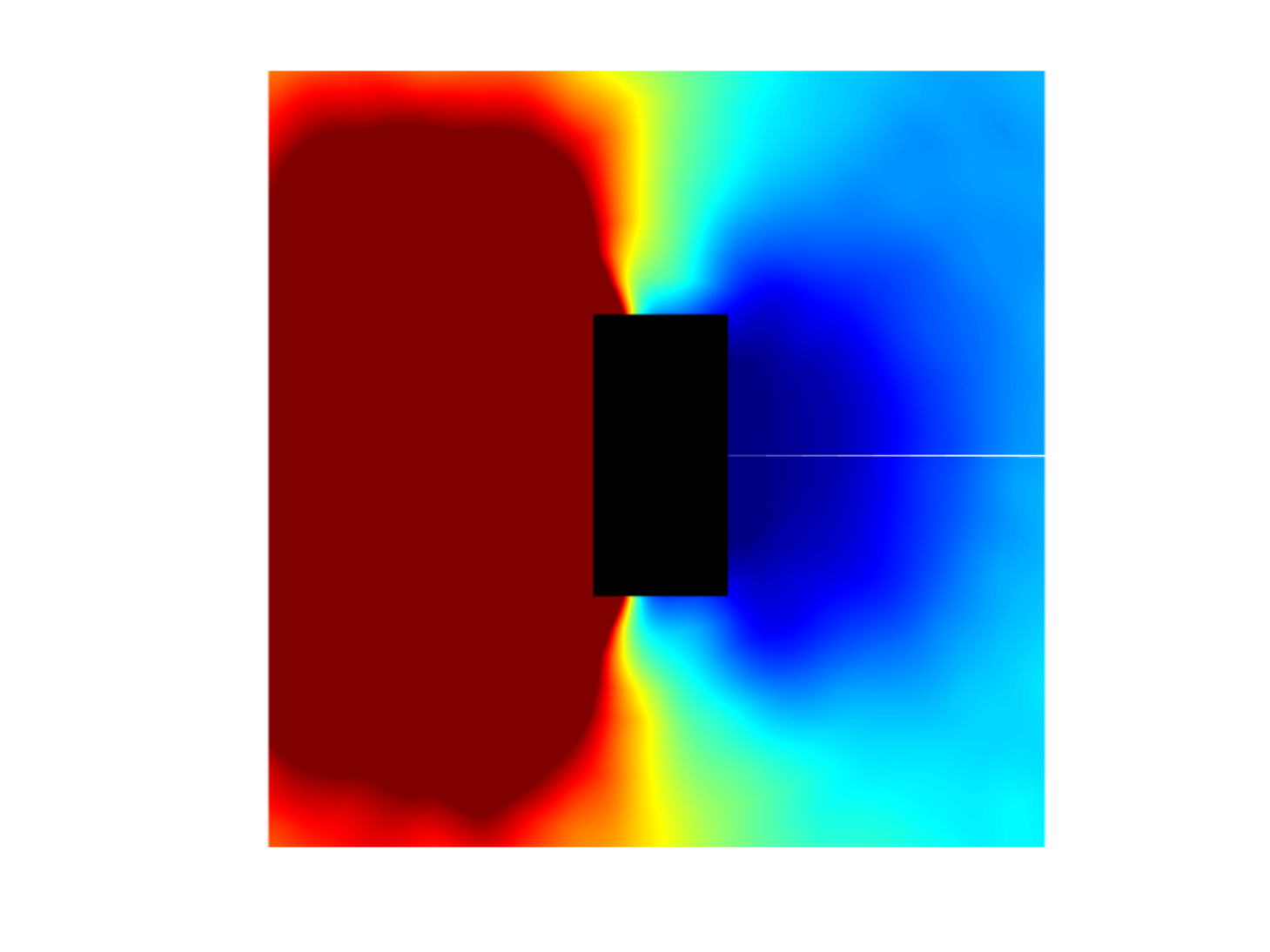}
		\caption{$10\phi/\pi=10$}
		\label{fig:five over x}
	\end{subfigure}
	\hspace{-1.1cm}
	\hfill
	\begin{subfigure}[b]{0.25\textwidth}
		\centering
		\includegraphics[width=\textwidth]{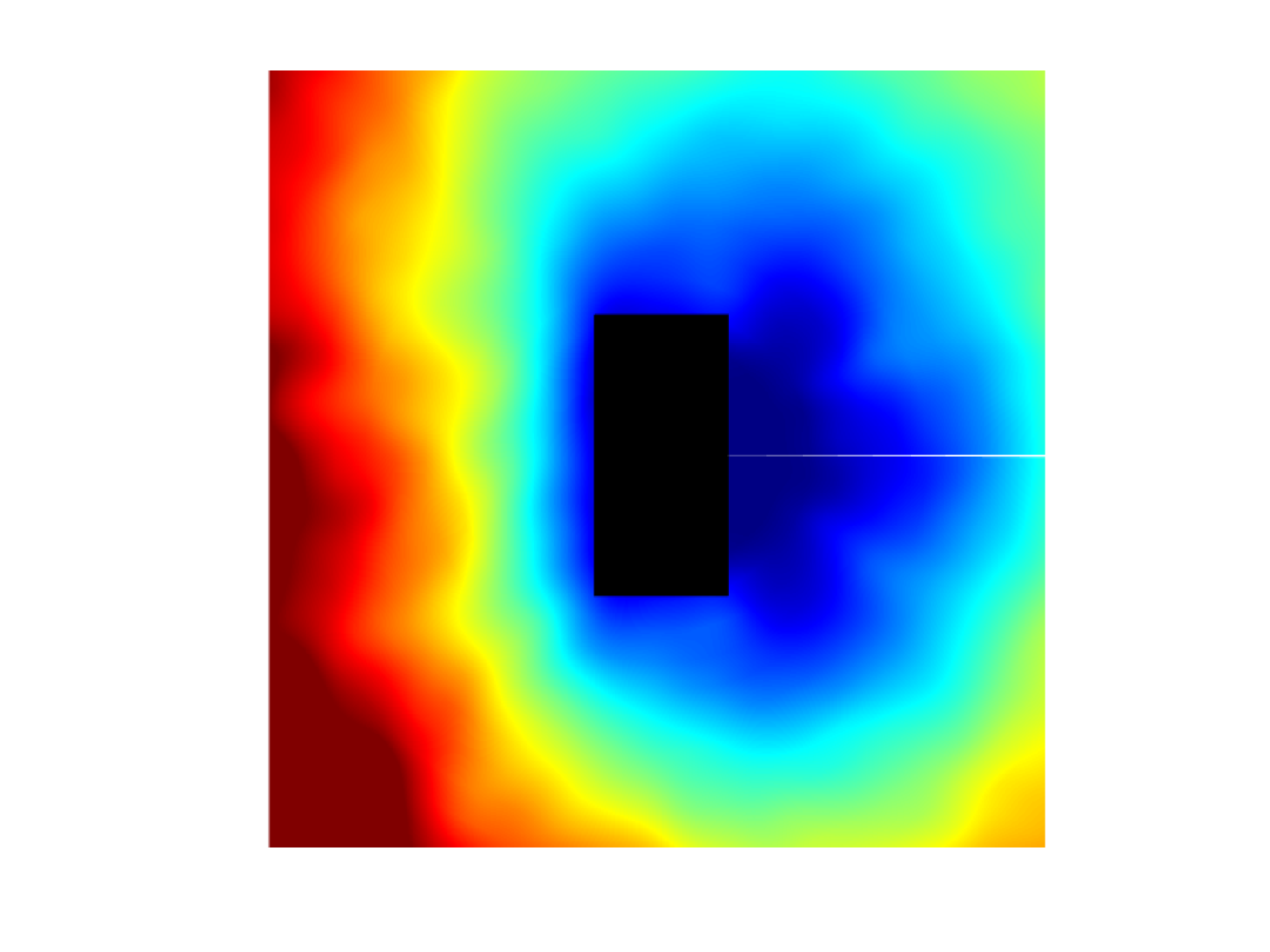}
		\caption{$10\phi/\pi=15$}
		\label{fig:three sin x}
	\end{subfigure}
	\hspace{-1.0cm}
	\hfill
	\begin{subfigure}[b]{0.25\textwidth}
		\centering
		\includegraphics[width=\textwidth]{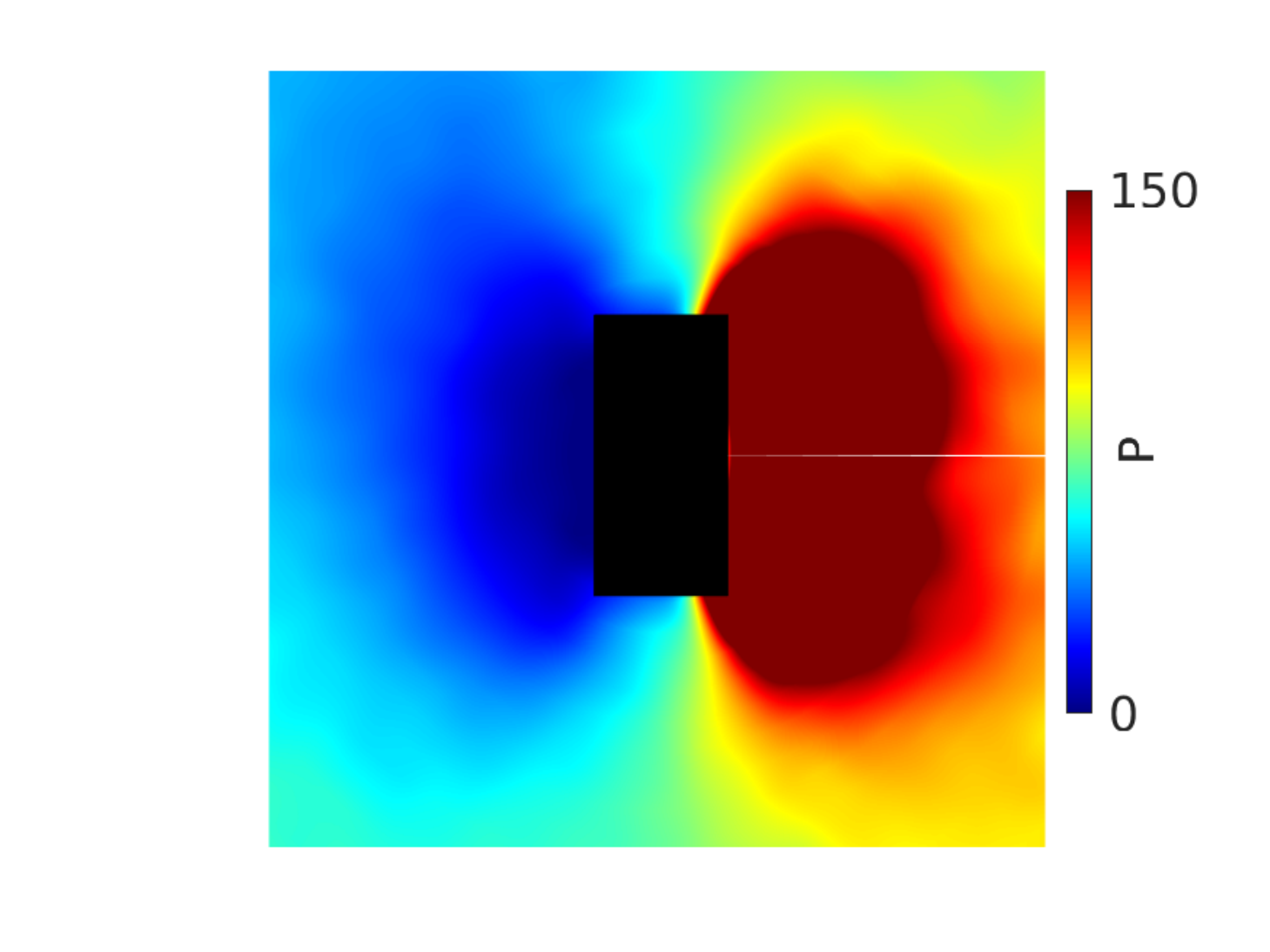}
		\caption{$10\phi/\pi=19$}
		\label{fig:five over x}
	\end{subfigure}
	
	\caption{The time-averaged flow fields for depicting the area fraction $\Phi$, velocity $V$ and pressure $P$ around the intruder $OS3$ (rectangle with its longer dimension along the y-axis) for $T/\sqrt{d/g}=1$ and different phase angle $10\phi/\pi$  : (a), (f), (k) 0.0, (b) (g), (l) 5.0, and (c), (h), (m) 10.0, (d), (i), (n) 15.0 and (e), (j), (o) 19.0. The oscillating amplitude of the intruder is $A/d=3$.\label{fig:mff_rectangle}}
\end{figure}

\begin{figure}
	\hspace{-0.9cm}
	\begin{subfigure}[b]{0.25\textwidth}
		\centering
		\includegraphics[width=\textwidth]{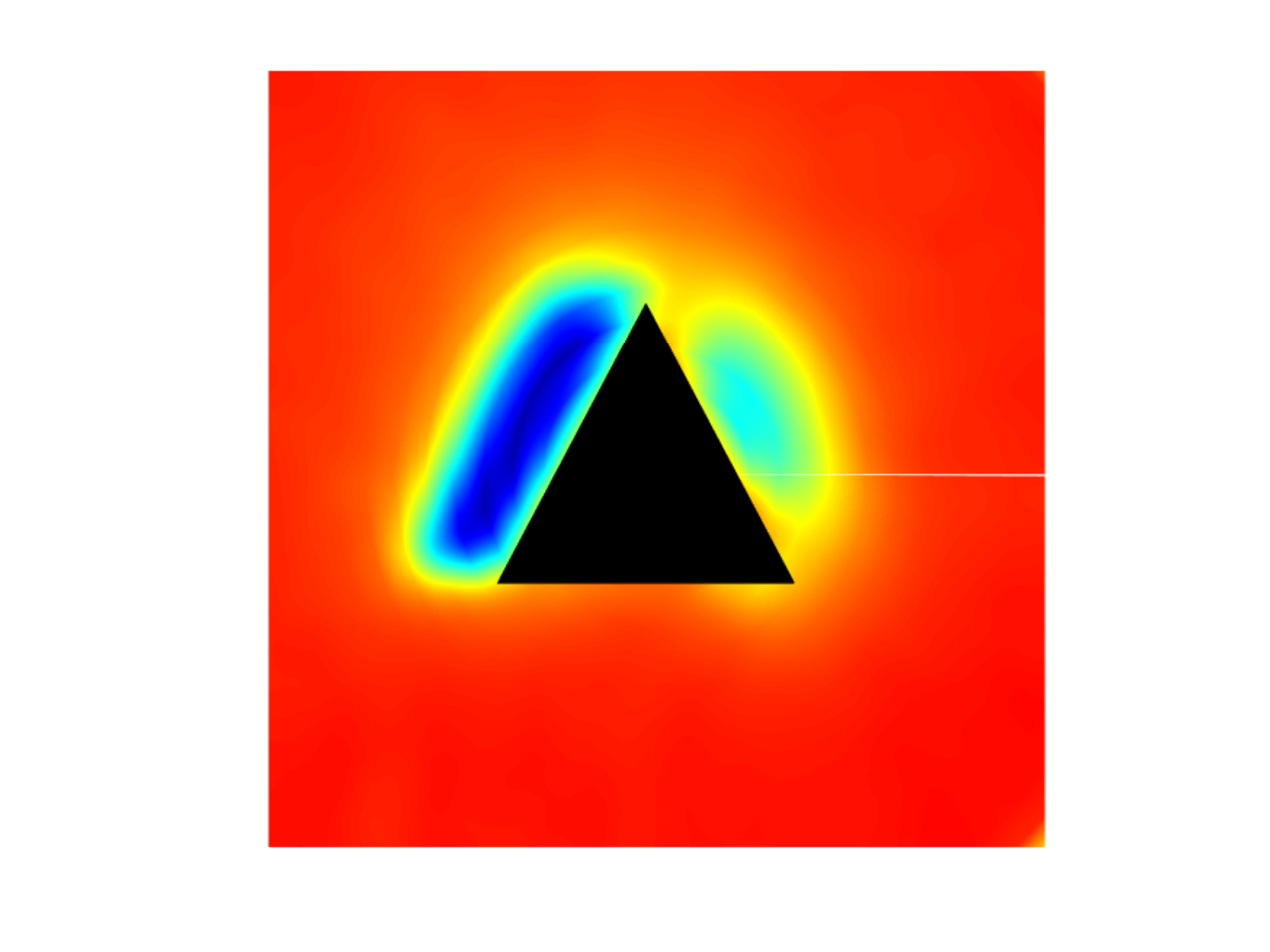}
		\caption{}
		\label{fig:y equals x}
	\end{subfigure}
	\hspace{-1.1cm}
	\hfill
	\begin{subfigure}[b]{0.25\textwidth}
		\centering
		\includegraphics[width=\textwidth]{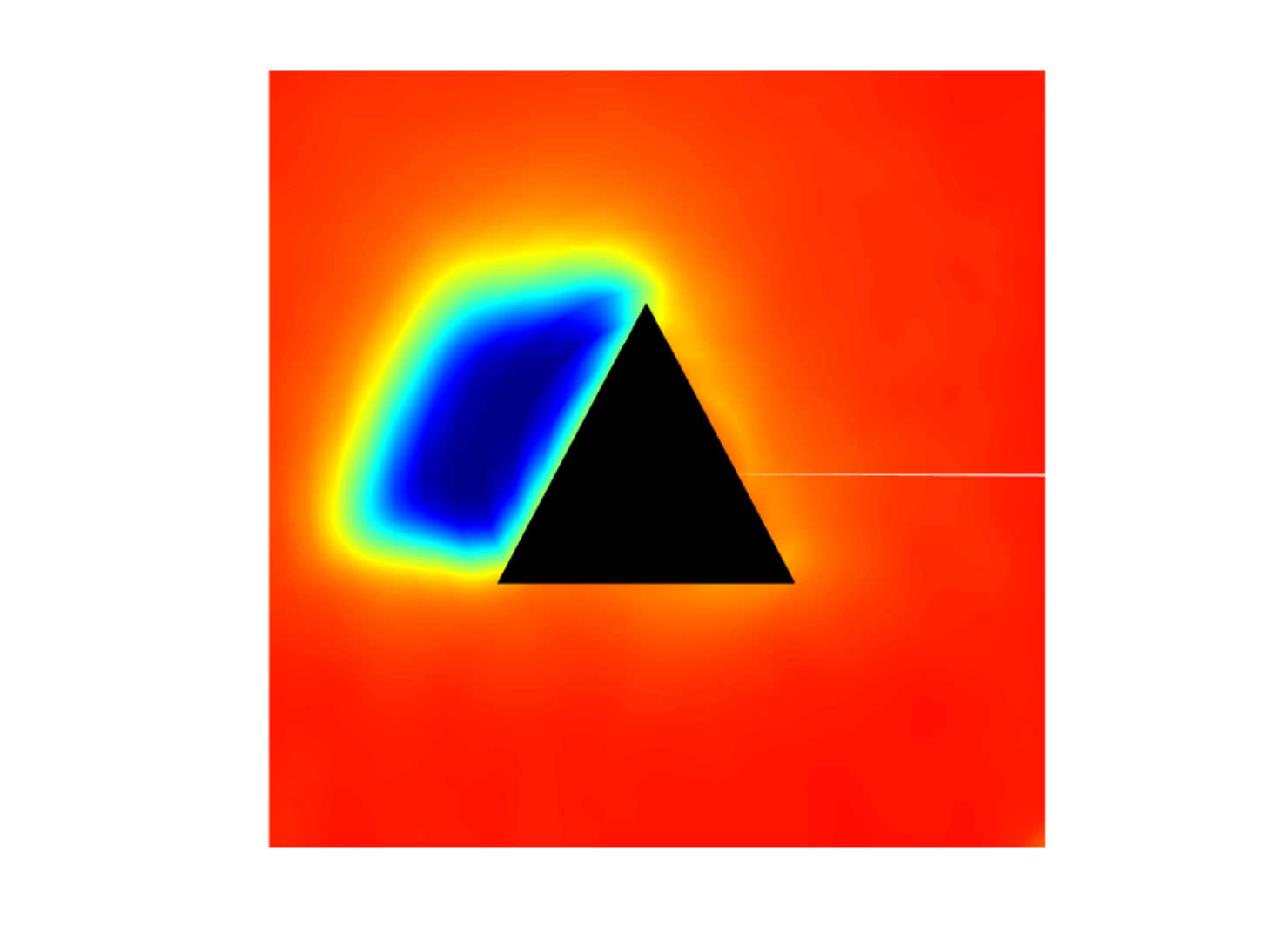}
		\caption{}
		\label{fig:three sin x}
	\end{subfigure}
	\hspace{-1.1cm}
	\hfill
	\begin{subfigure}[b]{0.25\textwidth}
		\centering
		\includegraphics[width=\textwidth]{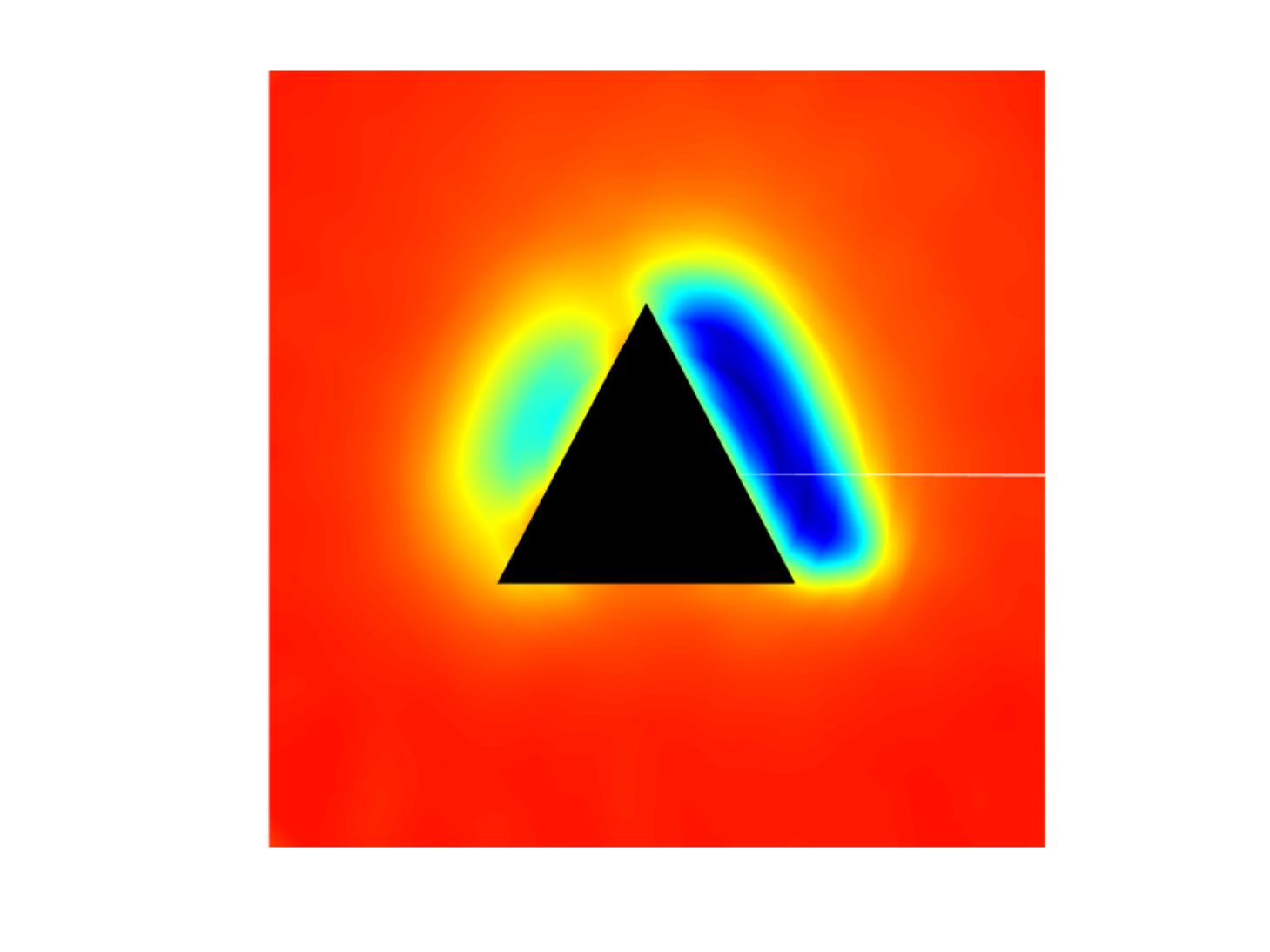}
		\caption{}
		\label{fig:five over x}
	\end{subfigure}
	\hspace{-1.1cm}
	\hfill
	\begin{subfigure}[b]{0.25\textwidth}
		\centering
		\includegraphics[width=\textwidth]{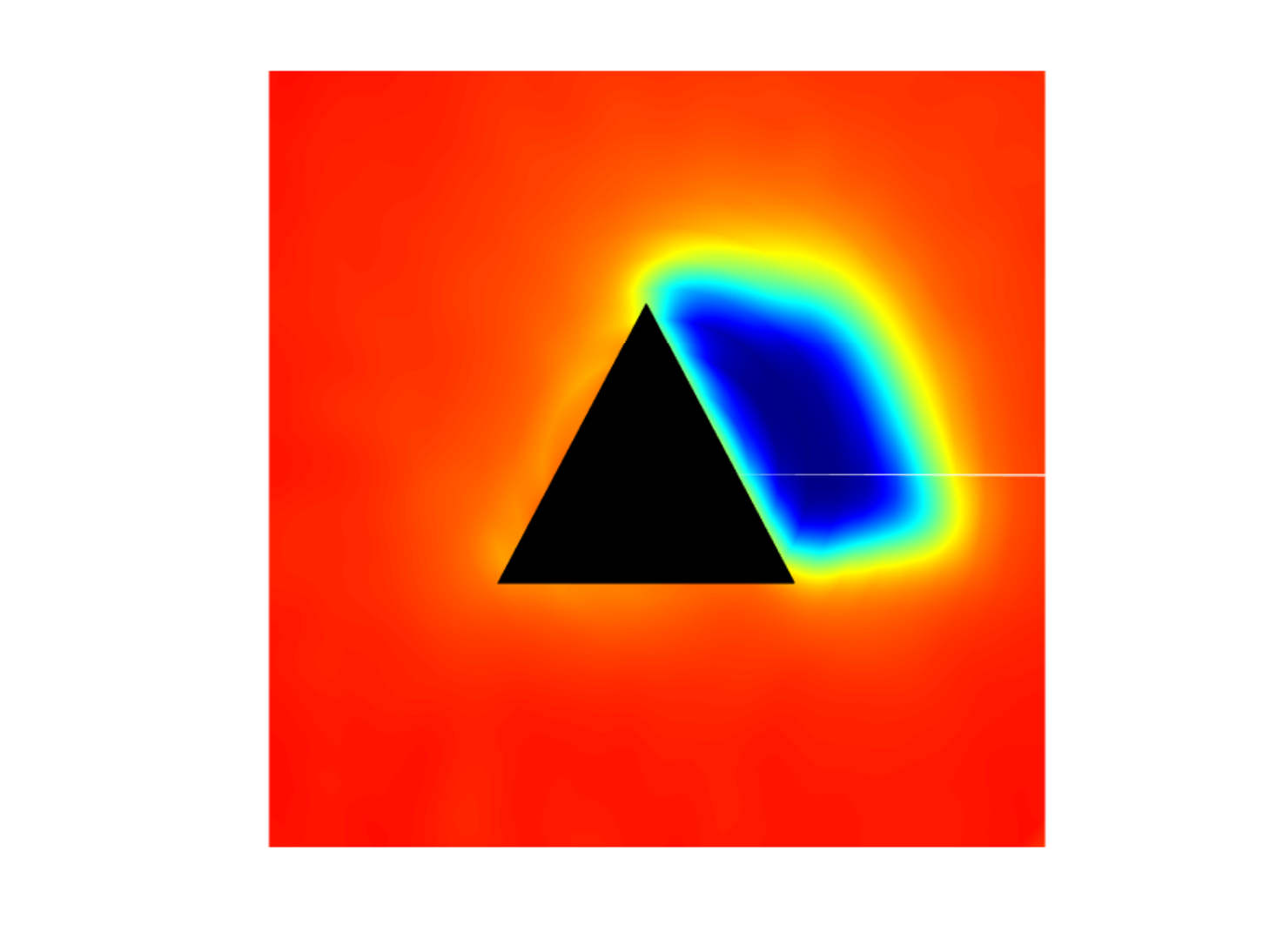}
		\caption{}
		\label{fig:three sin x}
	\end{subfigure}
	\hspace{-1.0cm}
	\hfill
	\begin{subfigure}[b]{0.25\textwidth}
		\centering
		\includegraphics[width=\textwidth]{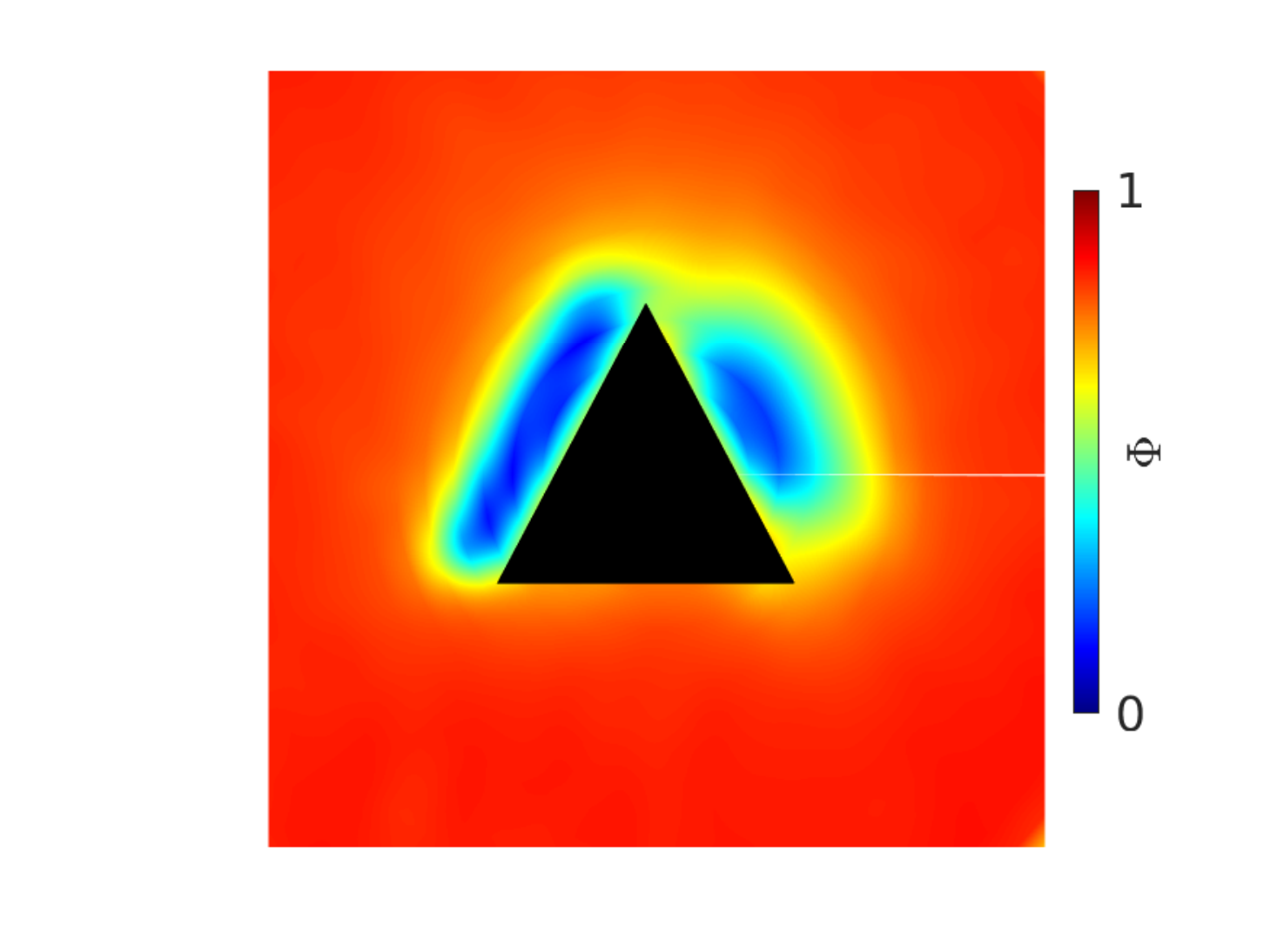}
		\caption{}
		\label{fig:five over x}
	\end{subfigure}

    \hspace{-0.9cm}
    \begin{subfigure}[b]{0.25\textwidth}
    	\centering
    	\includegraphics[width=\textwidth]{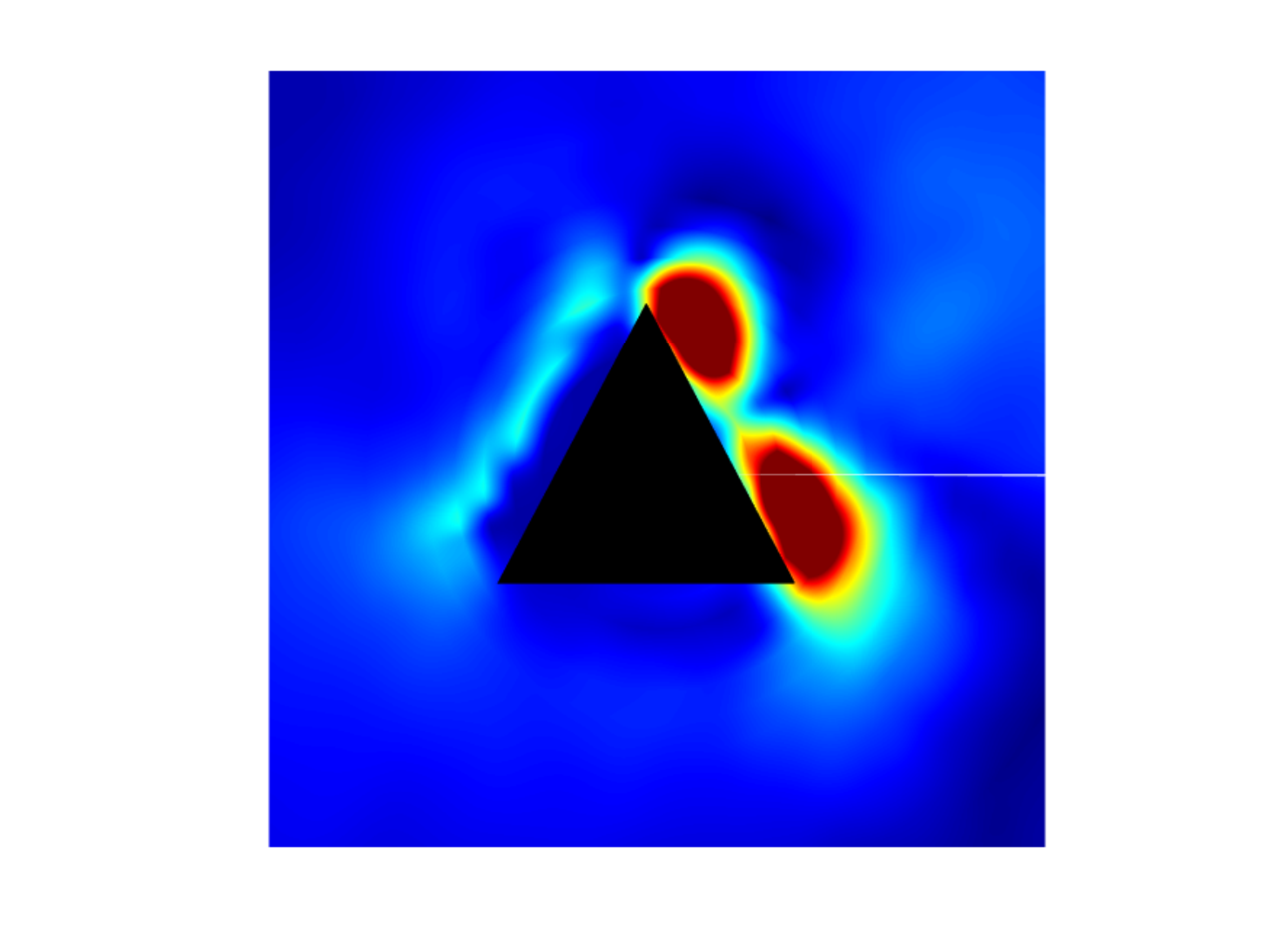}
    	\caption{}
    	\label{fig:y equals x}
    \end{subfigure}
    \hspace{-1.1cm}
    \hfill
    \begin{subfigure}[b]{0.25\textwidth}
    	\centering
    	\includegraphics[width=\textwidth]{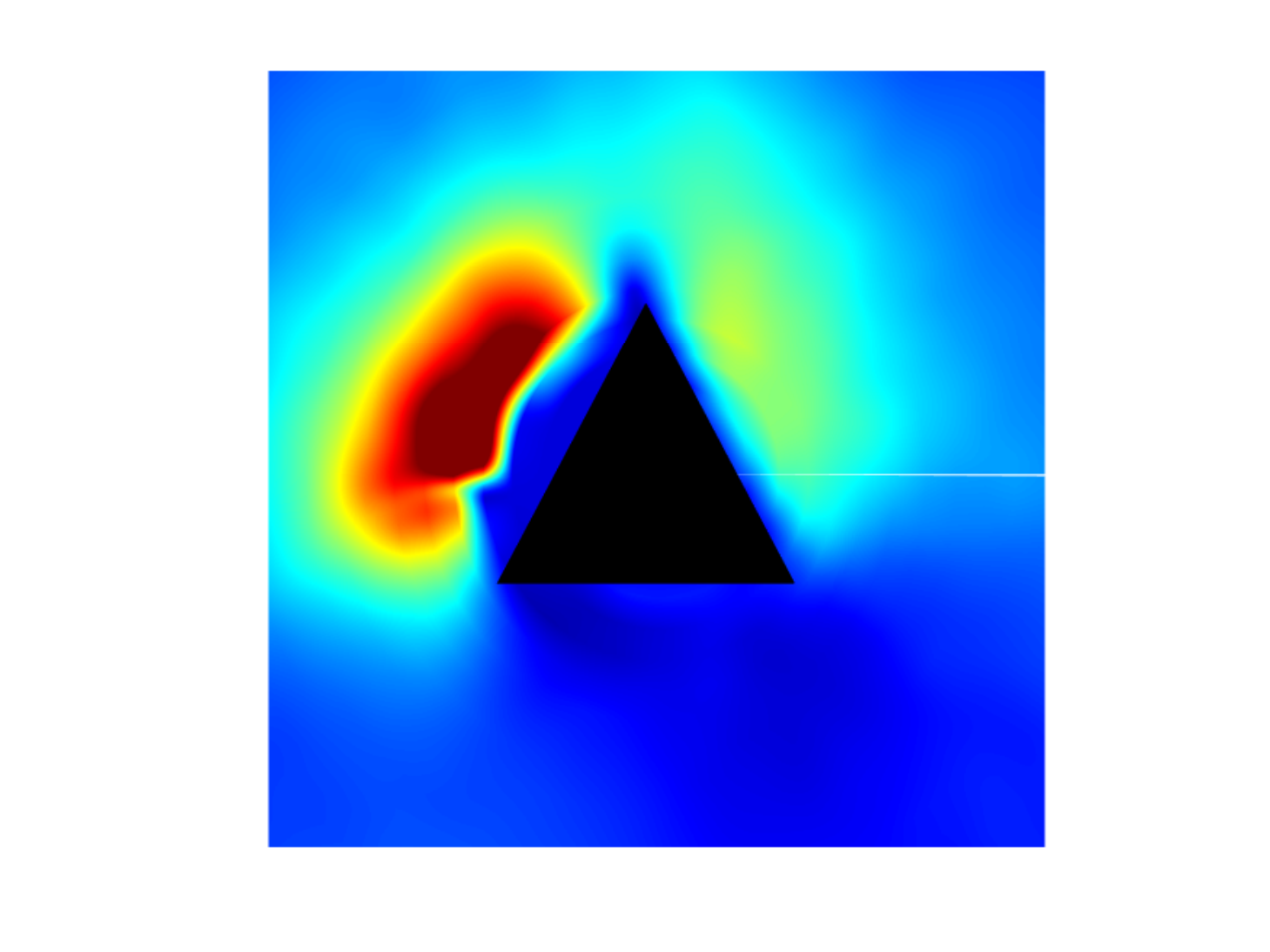}
    	\caption{}
    	\label{fig:three sin x}
    \end{subfigure}
    \hspace{-1.1cm}
    \hfill
    \begin{subfigure}[b]{0.25\textwidth}
    	\centering
    	\includegraphics[width=\textwidth]{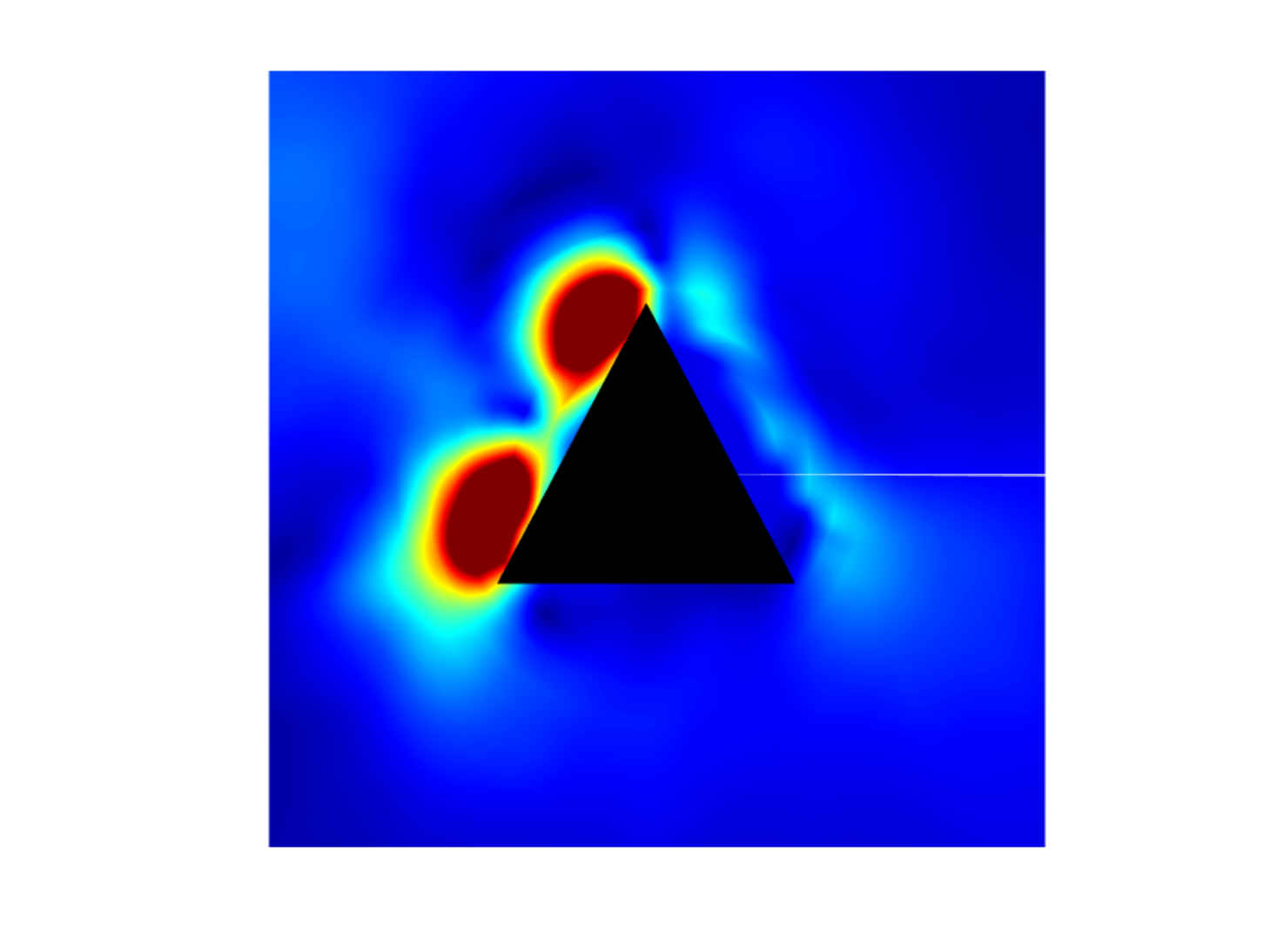}
    	\caption{}
    	\label{fig:five over x}
    \end{subfigure}
    \hspace{-1.1cm}
    \hfill
    \begin{subfigure}[b]{0.25\textwidth}
    	\centering
    	\includegraphics[width=\textwidth]{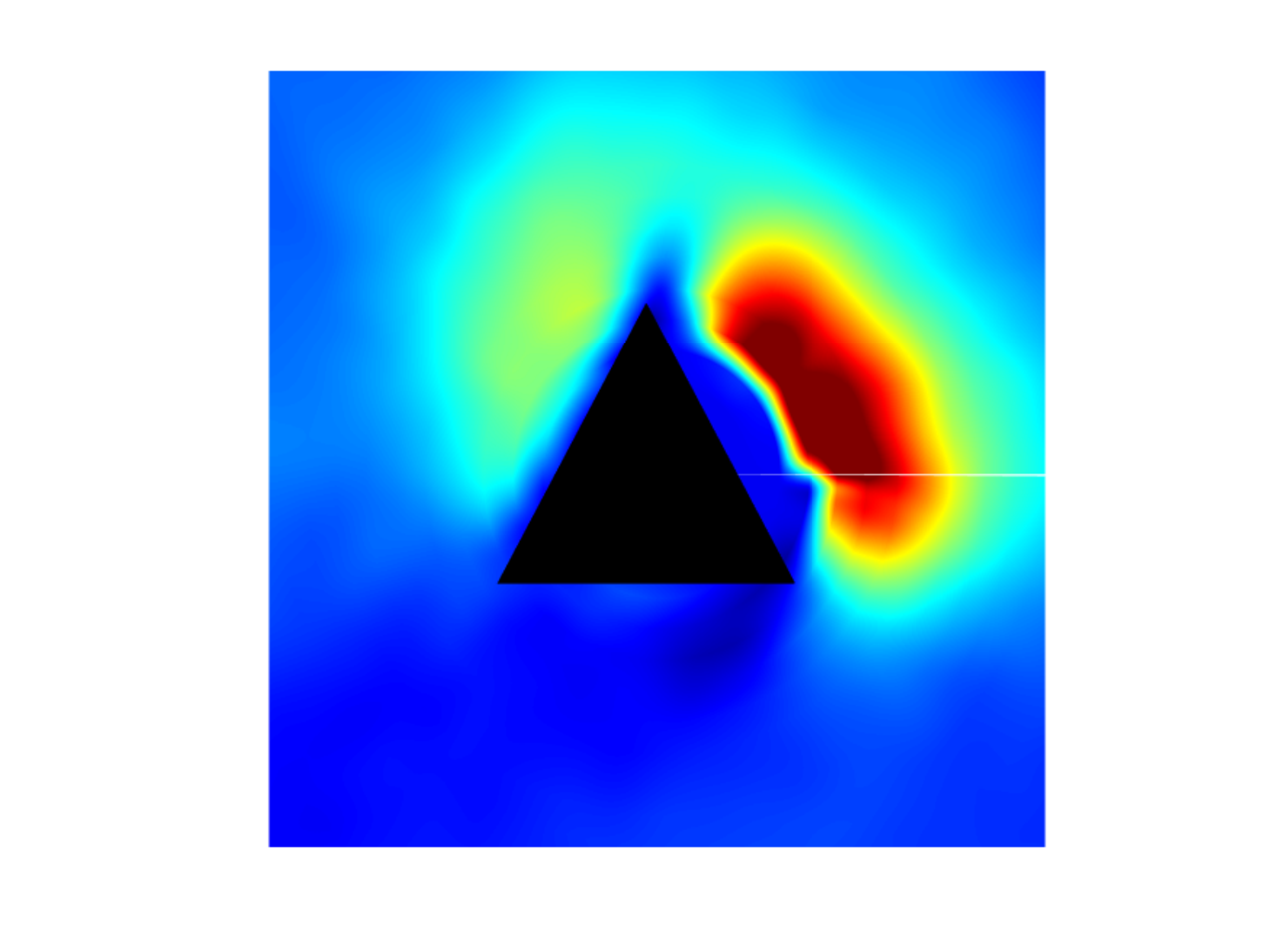}
    	\caption{}
    	\label{fig:three sin x}
    \end{subfigure}
    \hspace{-1.0cm}
    \hfill
    \begin{subfigure}[b]{0.25\textwidth}
    	\centering
    	\includegraphics[width=\textwidth]{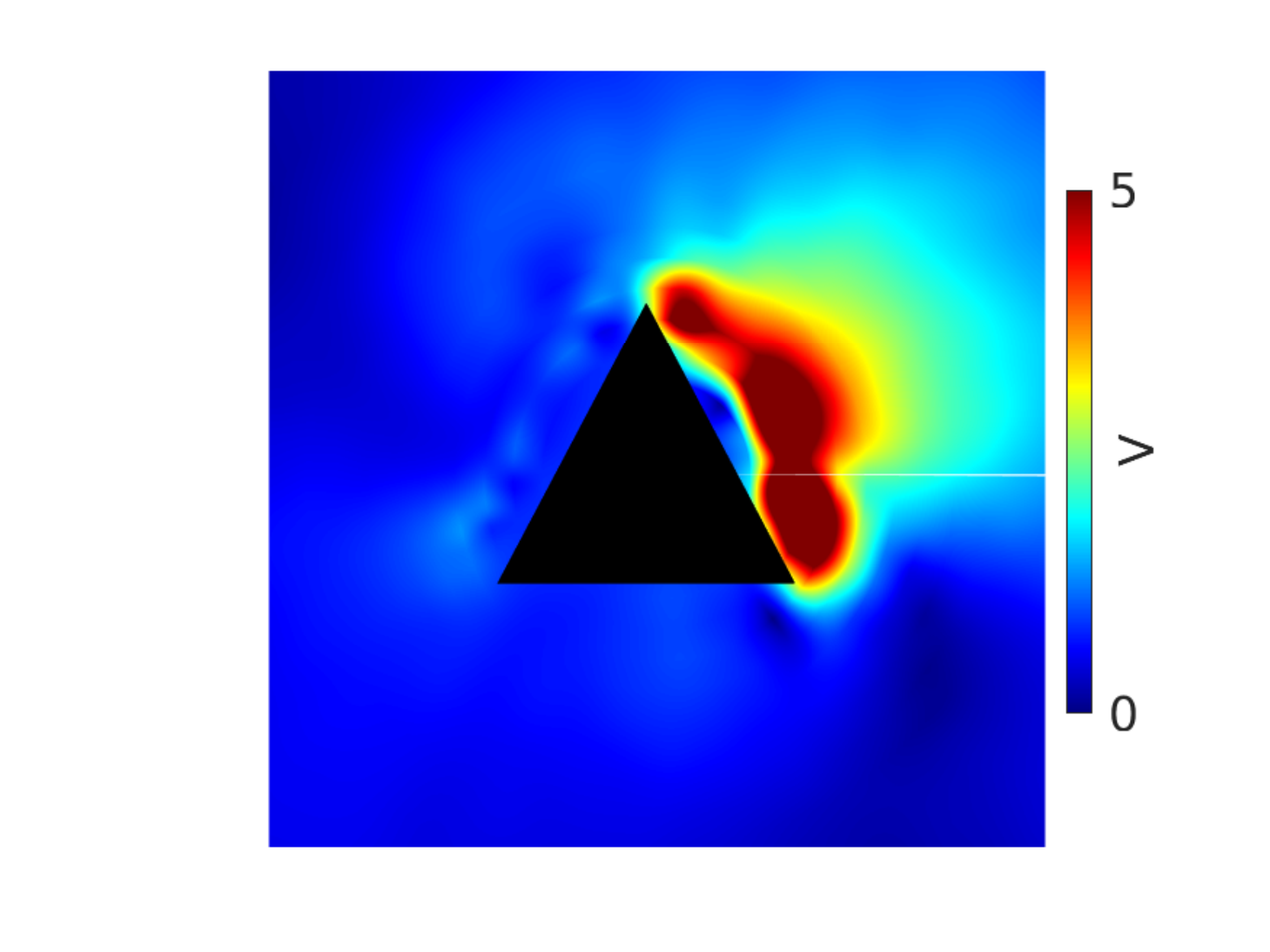}
    	\caption{}
    	\label{fig:five over x}
    \end{subfigure}

    \hspace{-0.9cm}
    \begin{subfigure}[b]{0.25\textwidth}
    	\centering
    	\includegraphics[width=\textwidth]{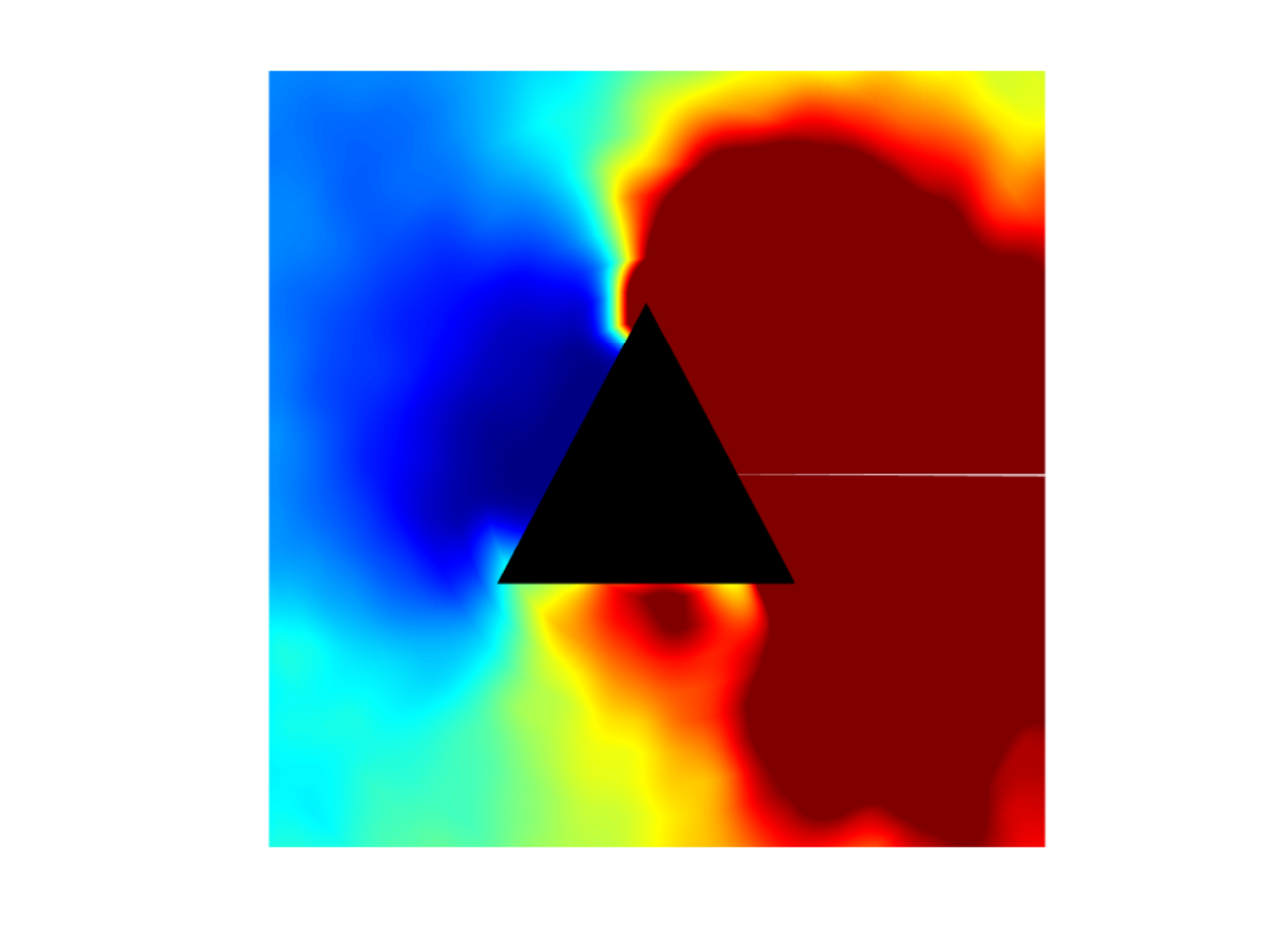}
    	\caption{$10\phi/\pi=0$}
    	\label{fig:y equals x}
    \end{subfigure}
    \hspace{-1.1cm}
    \hfill
    \begin{subfigure}[b]{0.25\textwidth}
    	\centering
    	\includegraphics[width=\textwidth]{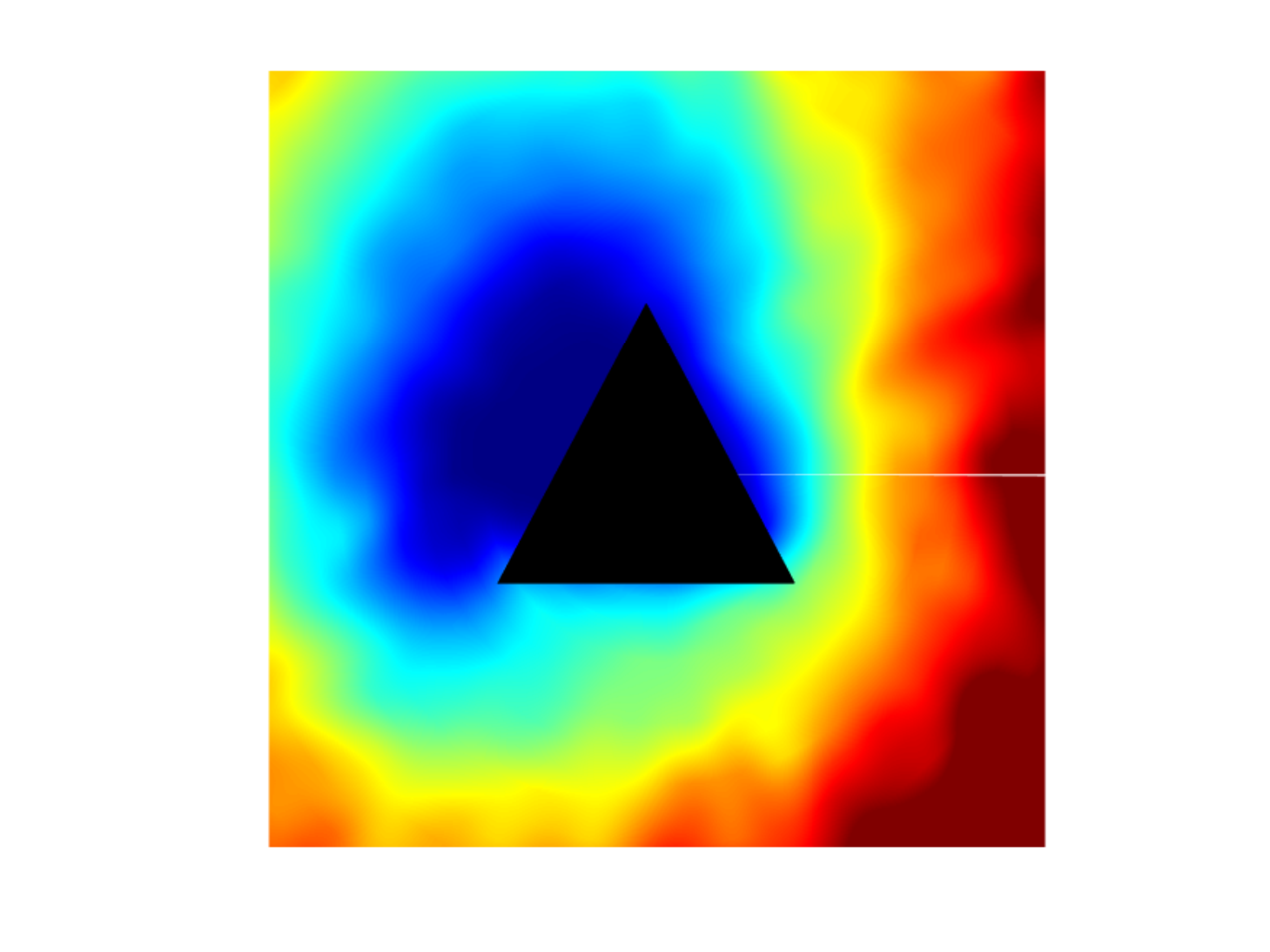}
    	\caption{$10\phi/\pi=5$}
    	\label{fig:three sin x}
    \end{subfigure}
    \hspace{-1.1cm}
    \hfill
    \begin{subfigure}[b]{0.25\textwidth}
    	\centering
    	\includegraphics[width=\textwidth]{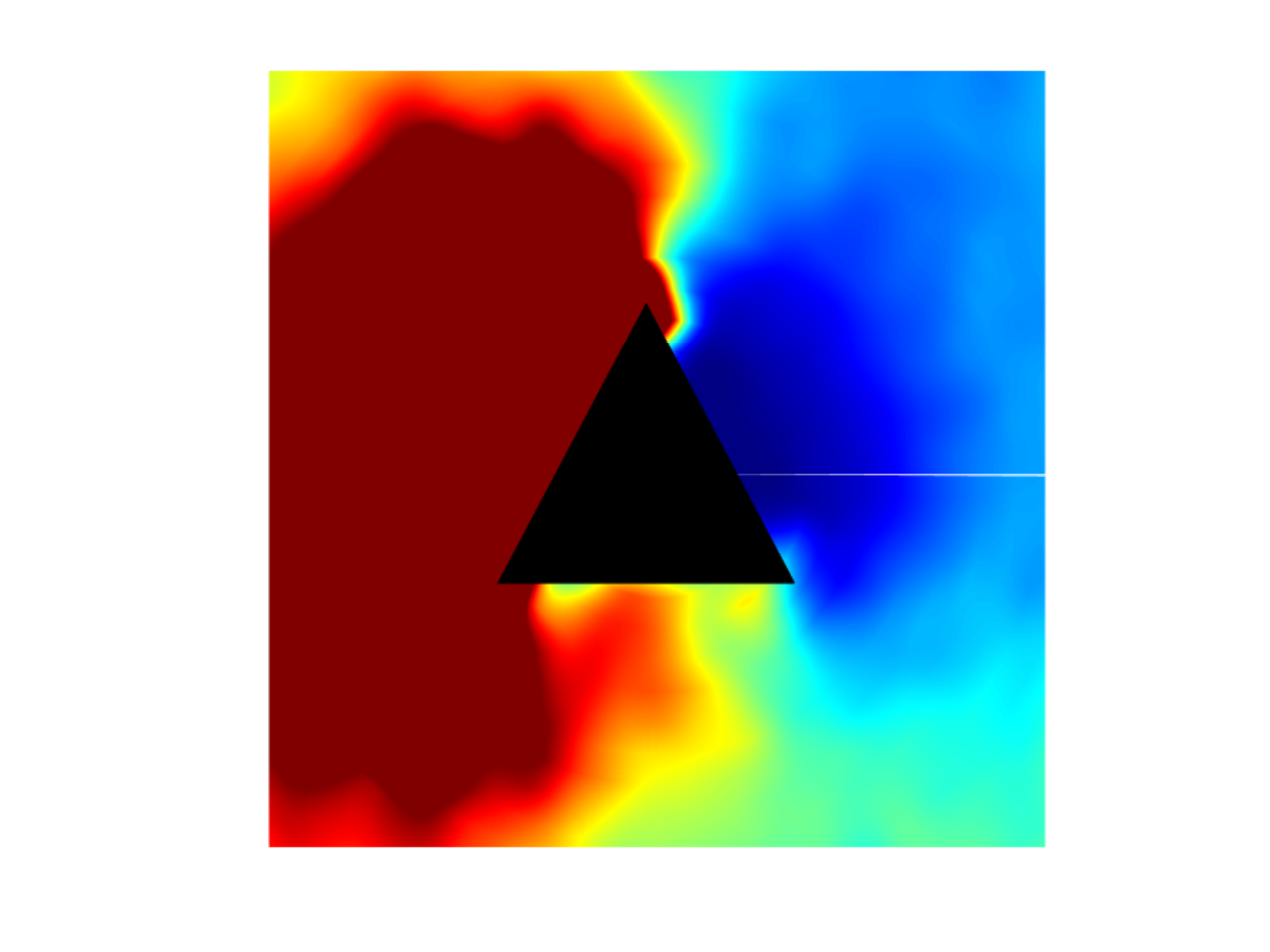}
    	\caption{$10\phi/\pi=10$}
    	\label{fig:five over x}
    \end{subfigure}
    \hspace{-1.1cm}
    \hfill
    \begin{subfigure}[b]{0.25\textwidth}
    	\centering
    	\includegraphics[width=\textwidth]{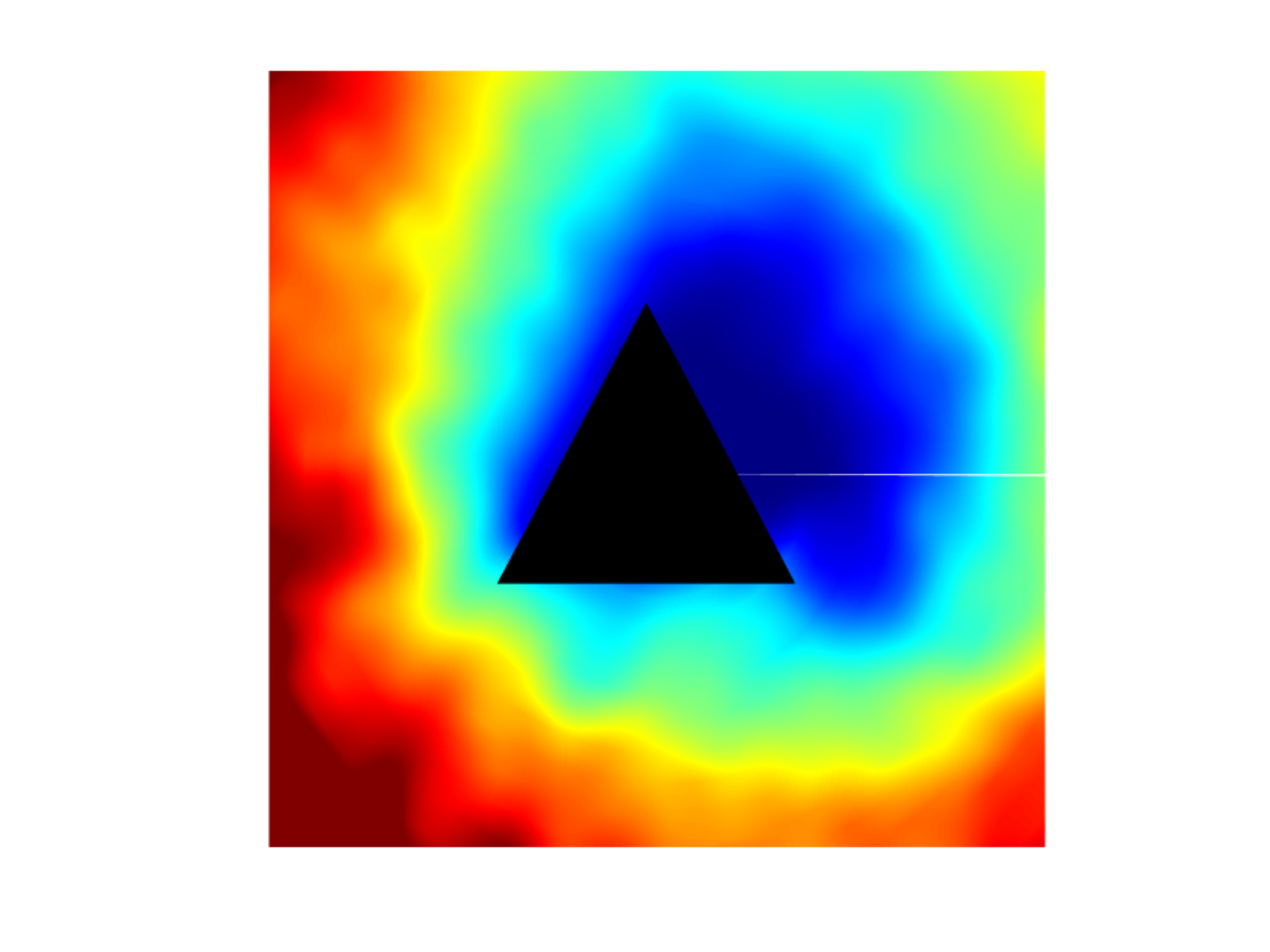}
    	\caption{$10\phi/\pi=15$}
    	\label{fig:three sin x}
    \end{subfigure}
    \hspace{-1.0cm}
    \hfill
    \begin{subfigure}[b]{0.25\textwidth}
    	\centering
    	\includegraphics[width=\textwidth]{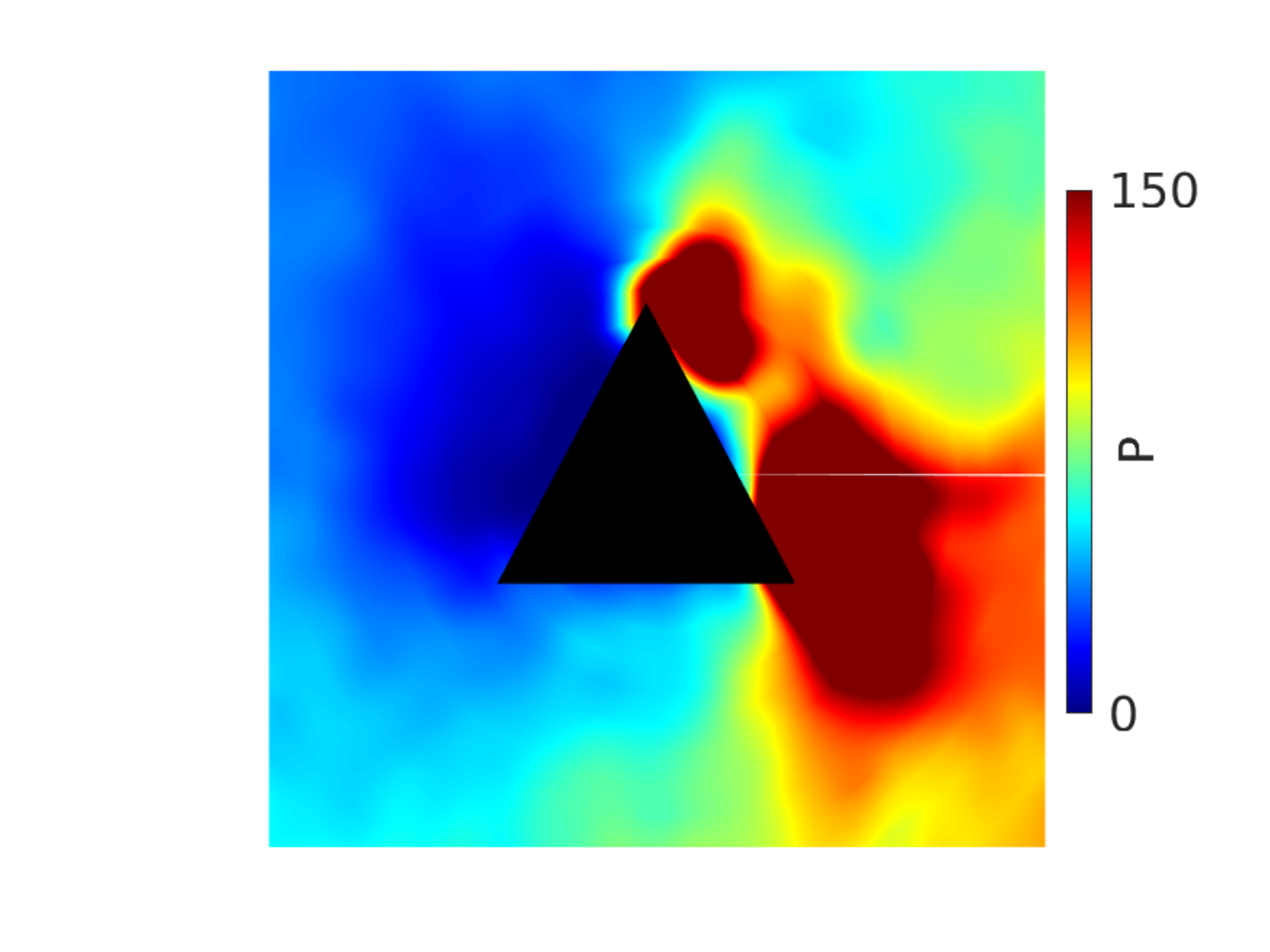}
    	\caption{$10\phi/\pi=19$}
    	\label{fig:five over x}
    \end{subfigure}
	
	\caption{The time-averaged flow fields for depicting the area fraction $\Phi$, velocity $V$ and pressure $P$ around the intruder $OS5$ (triangle) for $T/\sqrt{d/g}=1$ and different phase angle $10\phi/\pi$  : (a), (f), (k) 0.0, (b) (g), (l) 5.0, and (c), (h), (m) 10.0, (d), (i), (n) 15.0 and (e), (j), (o) 19.0. The oscillating amplitude of the intruder is $A/d=3$.\label{fig:mff_triangle}}
\end{figure}

In this subsection, we have shown the time-averaged flow fields by using discrete microscopic data like velocities, positions, stress, etc., of each particle around the designated intruder region. The considered parameters such as the area fraction $\Phi(t)$, velocity $v(t)$, stress tensor $\boldsymbol{\sigma_{ij(t)}}$ and the pressure $P(t)$ are calculated at any position $p$ with position vector $\boldsymbol{r_p}$ and at an arbitrary time $t$ as described in Glasser and Goldhirsch's implementation of the Gaussian coarse-graining function \citep{glasser2001scale, dhiman2020origin, reddy2021granular} : 

\begin{equation}
\Phi(t) =  \sum_{i=1}^{n} \frac{\pi d_i^2}{4} \mathcal{W}(\bm{r_p}-\bm{r_i(t)}) 
\end{equation}

\begin{equation}
\bm{v}(t) = \left[\sum_{i=1}^{n} \frac{\pi d^2_i}{4} \bm{v_i} \mathcal{W}(\bm{r_p}-\bm{r_i(t)}) \right]/\Phi
\end{equation}

\begin{equation}
\bm{\sigma_{ij}}(t)= \sum_{i=1}^{n}\sum_{j=i+1}^{n}(\bm{F^{ij} r_{ij}}) \int_{s=0}^{1} \mathcal{W}(\bm{r_p}-\bm{r_i(t)}+s\bm{r_{ij}})ds
\end{equation}

\begin{equation}
P(t) =\frac{-tr(\sigma_{ij}(t))}{2}
\end{equation}

\begin{equation}
\mathcal{W}(\bm{r})=\frac{1}{\pi w^2}e^{-\bm{r}^2/w^2} 
\end{equation}

Here, $\mathcal{W}(\bm{r})$ is the coarse-graining function, $\rho$, $d_i$ and $\bm{r_i(t)}$ are the density, diameter and the position vector of the particle $i$. $\bm{F^{ij}}$ is the force on the particle $i$ due to particle $j$ and $\bm{r_{ij}}$ is a vector in the direction of line joining the centres of two particles $i$ and $j$. Moreover, $w=1.0$ and $\mathcal{W}(\bm{r_p- r_i})=0$ if $|\bm{r_p- r_i}|>3w$ and $\phi,\bm{v}$ and $P$ are the time-averaged area fraction, velocity and pressure. The flow fields presented are the region around the $15d$ radius from the centre of the intruder ($D_i=5d$) and they are averaged over a minimum of $5000$ frames.

Figure \ref{fig:mff_disc} displays the mean flow field of area fraction when the intruder is at the centre point of oscillation ($10\phi/\pi=0$) while it is moving from left to right. The area fraction is minimum in a small region on the left side of the intruder due to the development of wake. However, in a small region to the right side of the intruder, the area fraction is witnessed to be slightly lower than that of the bulk. In the above and below the regions of the intruder, the area fraction is almost the same as that of the bulk as the intruder can affect only those particles that are close to it. Figure \ref{fig:mff_disc} b shows the area fraction at the rightmost point of oscillation ($10\phi/\pi=5$) when the intruder has changed its direction of motion and has just started to move from right to left. A larger wake is witnessed to the left side of the intruder in this case as compared to that of the one in figure \ref{fig:mff_disc} a. However, this wake gradually disappears as the intruder moves from right to left as the particles surrounding the wake try to fill it due to the action of gravity. Consequently, by the time the intruder reaches the centre point of oscillation ($10\phi/\pi=10$), this wake is replaced by a mildly fluidized set of particles or in other words, a slightly smaller area fraction to the left of the intruder (figure \ref{fig:mff_disc} b). Moreover, another wake begins to develop to the right side of the intruder during this time. As the intruder reaches the rightmost point of oscillation ($10\phi/\pi=15$), a large trailing wake is evidenced to the right side of the intruder (figure \ref{fig:mff_disc} d). Moreover, the mildly fluidized bed to the left side of the intruder is filled by particles above it and hence the area fraction becomes close to that of the bulk. Again as the intruder moves from right to left and reaches close ($10\phi/\pi=19$) to the centre point of oscillation, the wake to its right is slightly filled by particles and generating a fluidized bed(figure \ref{fig:mff_disc} e). A small region of dark blue to the left side of the intruder signifies the presence of another wake.

Figure \ref{fig:mff_disc} f illustrates velocity ($V$) fields when the intruder is at the centre point of oscillation ($10\phi/\pi=0$). The velocity is maximum on the right side of the intruder as the particles move along with the intruder while it is moving from left to right. The velocity is minimum in the regions above and below the intruder as the particles in this region are negligibly affected by the intruder's movement. The velocity of the particles on the left side of the intruder is significantly less than that on the right side. The reason for this might be due to a mildly fluidized bed of particles as noticed in figure \ref{fig:mff_disc} a. As the intruder moves from the centre point of oscillation ($10\phi/\pi=0$) to the rightmost  ($10\phi/\pi=15$), a large wake is developed to the left side of the intruder as shown in figure \ref{fig:mff_disc} b. The particles surrounding the wake try to fill it with larger velocities which are represented by a region of red to the left of the intruder(figure \ref{fig:mff_disc} g). When the intruder moves from right to left, it pushes the particles to move along with it. This can be evidenced in figure \ref{fig:mff_disc} h: the zones of red colour to the left of the intruder. Figure \ref{fig:mff_disc} i shows the velocity field when the intruder is at the leftmost point of oscillation. The particle velocities on the left side of the intruder are slightly lower because the intruder has started to move towards the right. The regions of red colour at a certain distance from the right surface of the intruder represents the particles that are trying to fill the wake that is formed between them and the intruder. As the intruder moves from left to right, the intruder pushes the particles lying to its left and hence a region of red colour is seen in figure \ref{fig:mff_disc} j ($10\phi/\pi=19$). 

Figure \ref{fig:mff_disc} k shows the pressure fields around the intruder when it is at the centre position of oscillation ($10\phi/\pi=0$). As the intruder is moving towards its right, greater pressure is noticed on the particles to the right. The zones of blue colour to the left of the intruder denotes the presence of wake. As the intruder starts to move towards its left after reaching the rightmost point ($10\phi/\pi=5$) of oscillation, the pressure on the particles to its right is small (figure \ref{fig:mff_disc} l). To the left of the intruder, pressure is very less due to the presence of wake as noticed in figure \ref{fig:mff_disc} b. When the intruder reaches the centre position ($10\phi/\pi=10$) from the right, the pressure is more to the left of the intruder (figure \ref{fig:mff_disc} m) as the intruder pushes the particles to its left while it is traversing. Figure \ref{fig:mff_disc} n shows the pressure when the intruder is at the leftmost point of oscillation. Very low pressure to the right of the intruder further confirms the presence of wake. Figure \ref{fig:mff_disc} o shows a greater pressure to the right of it as the intruder is moving towards its right.

Figure \ref{fig:mff_rectangle} shows the mean flow fields around a rectangle-shaped intruder with a small base length ($4.5d$) at different positions of oscillation. The wake region behind the rectangle-shaped intruder (figure \ref{fig:mff_rectangle} b and d) is larger than the disk-one (figure \ref{fig:mff_disc} b and d). The reason for this is the rectangle has a flat surface that pushes more particles while it is oscillating horizontally whereas, the disk with a curved surface can push fewer particles. Moreover, the $y$ dimension length of the rectangle ($9.5d$) is larger than the diameter of the disk ($7.0$d). Due to the above-mentioned reasons, the regions of high velocities beside the wake are wider for the rectangle-shaped one (figures \ref{fig:mff_rectangle} g and i). The pressure fields are observed to be almost the same for both the disk-shaped and the rectangle-shaped intruders. 

Figure \ref{fig:mff_triangle} shows the mean flow fields of area fraction, velocity and pressure around an equilateral triangle-shaped intruder at different positions of oscillation. The pressure fields answer the question: why the triangle-shaped intruder has mostly a negative vertical displacement. Either of the two slant faces of the intruder experience a vertical component of stress pointing downwards during its entire oscillation except when it changes its direction of motion. This can be witnessed in figures \ref{fig:mff_triangle} k, \ref{fig:mff_triangle} m and \ref{fig:mff_triangle} o. In supplementary information we have also shown the mean flow fields around a square (Figure 1 in supplementary information) and a rectangle-shaped intruder (Figure 2 in supplementary information) with a large base length ($9.5d$) at different positions of oscillation. 

\subsection{\label{sec:phase angle}Cavity model}
 
\begin{figure}
	\centering 
	\includegraphics[width=0.75\linewidth]{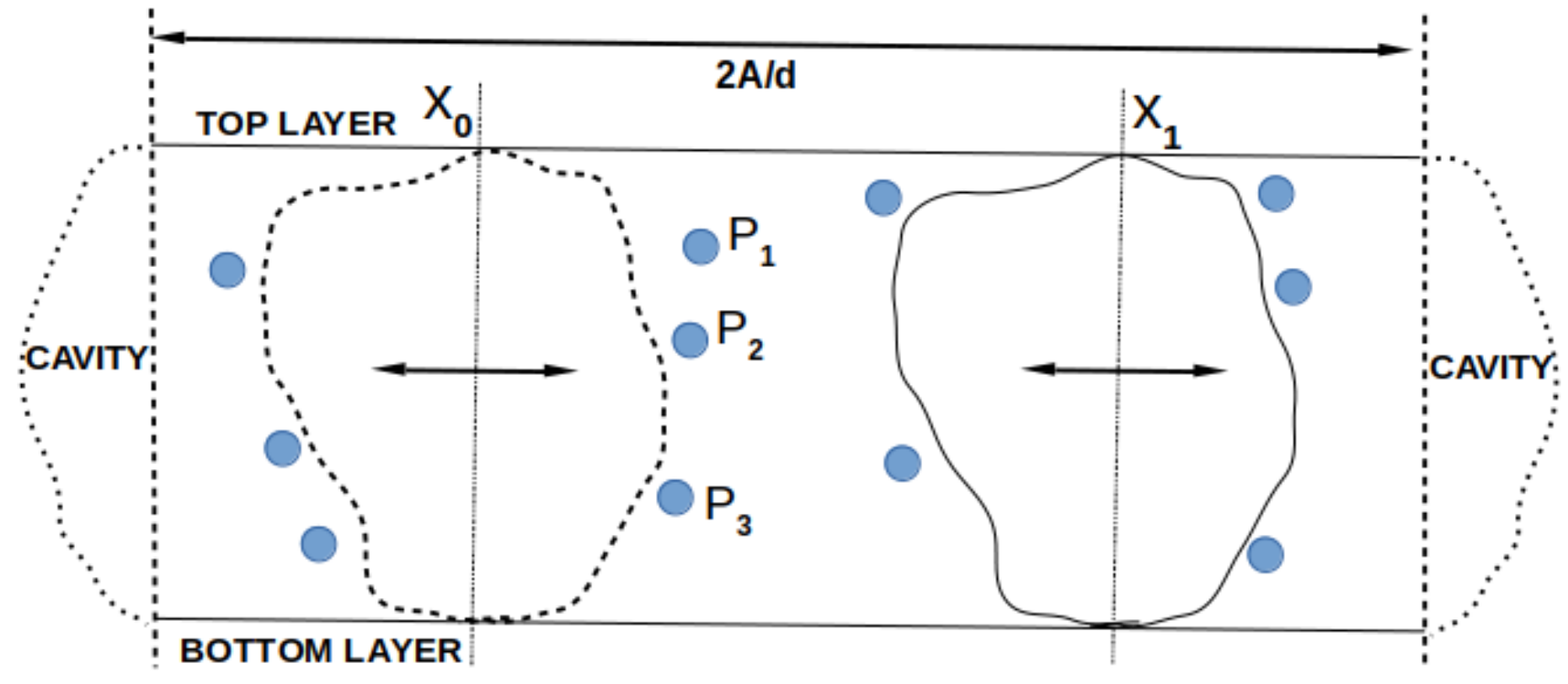}
	
	\caption{ A general shape ocillating horizontally with granular particles around it.\label{fig:genshape}}
\end{figure}

We modeled the rising behaviour of the intruder by considering an arbitrary shape of the intruder and $(x,y) \in $ (collection of points on the surface). In the granular medium, its position in the $x$ direction is defined as $x_{\text{center of mass}}$= $A$ $sin (\omega t)$ (figure \ref{fig:genshape}) owing to its horizontal oscillation. Now, if the orientation of an object is considered (with no rotational motion), then we can assign the axis such that $g= -|g|\hat{y}$ where $|g|$ is the magnitude of gravity. For the intruder to rise, an additional layer of particles has to form below the intruder as a result of free-fall in the cavity formed in the wake of the intruder. To form this additional bed, the particles surrounding the intruder have to drop below a certain point or threshold termed as ``point of no return" and defined as a certain y-coordinate ($y_{nr}$), in a sense, if the particle's $y$ position is below $y_{nr}$, it becomes the part of the additional bed of particles. Therefore, the point of no return can be defined as a point on the surface of the intruder where tangent to its surface is $\pi/2$. In cases with multiple points on such surfaces, we assume that it is the lowest point on the surface because of its proximity to bottom layer. Consider an intruder placed at a position $X_0$ interacting with the particles $P_1$, $P_2$, and $P_3$ (figure \ref{fig:genshape}). As the intruder moves from $X_0$ to $X_1$ position, it may be possible that particle $P_3$ will move down, whereas the particles $P_1$ and $P_2$ move upwards. So, we can say that the “point of no return” is somewhere in between the $P_2$ and $P_3$ position. This can be defined as the minimum $y$ for which $\frac{dy}{dx}=\infty$.

Therefore, any granular particle around the intruder whose $y<y_{nr}$ will eventually settle to the bottom of the cavity. Let us also consider, the $y-$position of the top surface of the intruder as $y_0$, then, a particle situated above the intruder has to move at least ($y_{0} - y_{nr} + \frac{d}{2}$) in one-half time period `$T$' of oscillation to be a part of the additional bed at the bottom. If one were to focus on the one-half oscillations such that at $t=0$, it exists at one end of the cavity, and at $t=T/2$, it exists on the other end. When the intruder travels for time``$t$", the net-free fall into the cavity begins at``$-t$". Therefore, the total time for net free-fall is ``$2t$" or a free-fall distance of $2gt^2$. The differential number of particles that cross the point of no return before the intruder approaches $x$ at time $t$,
\begin{figure}
	\centering 
	\includegraphics[width=0.75\linewidth]{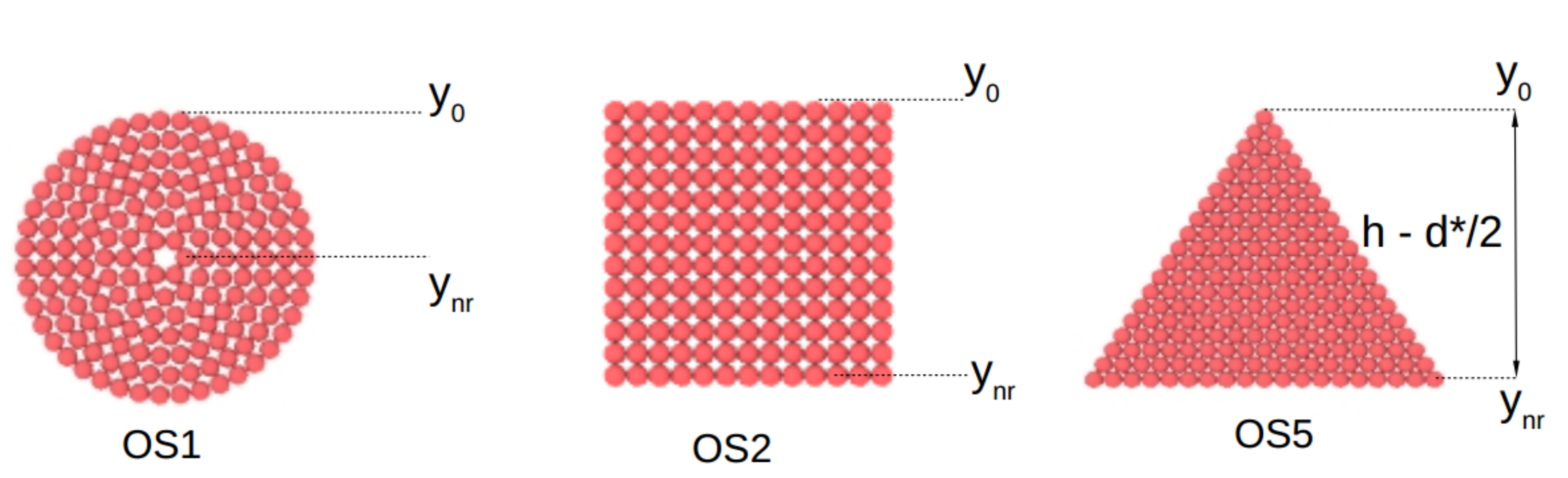}
	
	\caption{The shapes $OS1$, $OS2$ and $OS5$ with the points $y_{nr}$ and $y_0$ where the particles covering this distance of ($l=y_0-y_{nr}$) will fall beneath the base of the intruder.\label{fig:shapes}}
\end{figure}
\begin{equation}
dn = \frac{(2gt^{2}-l-\frac{d}{2})4\phi}{\pi d^2} dx; 
\end{equation}

\begin{equation}
dn = \frac{(2gt^{2}-l-\frac{d}{2})4\phi}{\pi d^2} \frac{2\pi A}{T} sin \big(\frac{2\pi t}{T}\big)  dt; 
\end{equation}

In equation 3.6, $l=y_0-y_{nr}$ is the distance the particles have to cover to fall beneath the base of the intruder, and ($2gt^2-l-d/2$) is the height of the zone of the particles that have crossed the point of no return and reached the bottom of the intruder. The first particle crosses the point of no return at $\sqrt{(l+\frac{d}{2})/2g}$, therefore, total particles reaching the bottom in one-half oscillation,

\begin{equation}
\int_{0}^{N}dn = \int_{\sqrt{\frac{l+\frac{d}{2}}{2g}}}^{T/2} \frac{(2gt^{2}-l-\frac{d}{2})4\phi}{\pi d^2} \frac{2\pi A}{T} sin \big(\frac{2\pi t}{T}\big)  dt; 
\end{equation}

\begin{equation}
N = \frac{4\phi 2\pi A}{\pi d^2 T} \int_{\sqrt{\frac{l+\frac{d}{2}}{2g}}}^{T/2} \big(2gt^{2}-l-\frac{d}{2}\big) sin \big(\frac{2\pi t}{T}\big)  dt; 
\end{equation}

Assuming $l^{'}= l+ \frac{d}{2}$ and $\Lambda$ to be the integral part,

\begin{equation}
\Lambda =\frac{T^2}{4\pi}\Bigg(\bigg(1-\frac{2}{\pi^2}\bigg)gT-\frac{2l'}{T}-\frac{2}{\pi^2}gT\cos\bigg(\frac{2\pi}{T}\sqrt\frac{l'}{2g}\bigg)-\frac{4}{\pi}g\sqrt\frac{l'}{2g}\sin\bigg(\frac{2\pi}{T}\sqrt\frac{l'}{2g}\bigg)\Bigg)
\end{equation}

If we are to assume, the particles in the cavity form a uniform layer at the bottom of cavity, then, the rate of rise $V_y$ can be given as elevation per half-oscillation and net oscillation as $N\pi (d^2/4) /(\phi 2A)$,

\begin{equation}
V_y = \frac{4\phi 2\pi A}{\pi d^2 T} \Lambda \frac{\pi d^2/4}{\frac{\phi 2 AT}{2}}  ; 
\end{equation}

\begin{equation}
V_y = \frac{2\pi}{T^{2}} \Lambda ; 
\end{equation}

\begin{equation}
V_y = \frac{1}{2}\Bigg(\bigg(1-\frac{2}{\pi^2}\bigg)gT-\frac{2l'}{T}-\frac{2}{\pi^2} gT\cos\bigg(\frac{2\pi}{T}\sqrt\frac{l'}{2g}\bigg)-\frac{4}{\pi}g\sqrt\frac{l'}{2g}\sin\bigg(\frac{2\pi}{T}\sqrt\frac{l'}{2g}\bigg)\Bigg)
\end{equation}

This equation describes $V_y$ with respect to $T$ for a given shape (fixed $l^{'}$).
For $V_y = 0$, we get 
\begin{equation}
T_{rise} = \sqrt{2l^{'}/g}
\end{equation}

\begin{equation}
	V_y = \frac{gT}{2}\Bigg(\bigg(1-\frac{2}{\pi^2}\bigg)-\bigg(\frac{T_{rise}}{T}\bigg)^2-\frac{2}{\pi^2} \cos\bigg(\pi\frac{ T_{rise}}{T}\bigg)-\frac{2}{\pi}\frac{T_{rise}}{T}\sin\bigg(\pi\frac{ T_{rise}}{T}\bigg)\Bigg)
	\label{eqn:fin}
\end{equation}

Additionally, $V_y$ cannot increase indefinitely with $T$ due to cavity filling up. At some $T_{max} \approx \sqrt{ \frac{8h}{g}}$
, it will start to decrease with
\begin{equation}
V_{y} \approx \frac{h-l}{T}
\end{equation}

showing an inverse relationship with timeperiod ($T$) at higher $T$ as observed in all the cases. Here, $``h"$ indicates the height of the oscillating object.

To summarise,
\begin{equation}
	V_y=
	\begin{cases}
		\frac{gT}{2}\Bigg(\bigg(1-\frac{2}{\pi^2}\bigg)-\bigg(\frac{T_{rise}}{T}\bigg)^2-\frac{2}{\pi^2} \cos\bigg(\pi\frac{ T_{rise}}{T}\bigg)-\frac{2}{\pi}\frac{T_{rise}}{T}\sin\bigg(\pi\frac{ T_{rise}}{T}\bigg)\Bigg)
,& \text{if } T_{rise}<T<T_{max}\\
		\frac{h-l}{T},              & \text{if } T>T_{max}
	\end{cases}
	\end{equation}

\begin{table}
	\centering
	\begin{tabular}{c c c c}
		\rule{0pt}{0.5ex}
		Shape	&	$l$ or $(l^{'}-d/2)$	&	$T_{rising}(cavity model)$ & $T_{rising}(observed)$\\
		\\
		OS1             & $3.50$ & 2.82 & 2.5-3.0\\
		OS2             & $6.50$ & 3.74 & 3.0-4.0\\
		OS3             & $9.50$ & 4.47 & 3.0-4.0\\
		OS4             & $4.50$ & 3.16 & 2.0-3.0\\
		OS5             & $8.22$ & 4.17 & 4.0-5.0\\
	\end{tabular}
	\caption{Comparison between value of $T_{rising}$ from theoretical estimate and the numerical simulations for an oscillating amplitude $A/d=5$.  \label{table:table2}}
\end{table}

\begin{figure}
	\centering
	\includegraphics[width=0.8\textwidth]{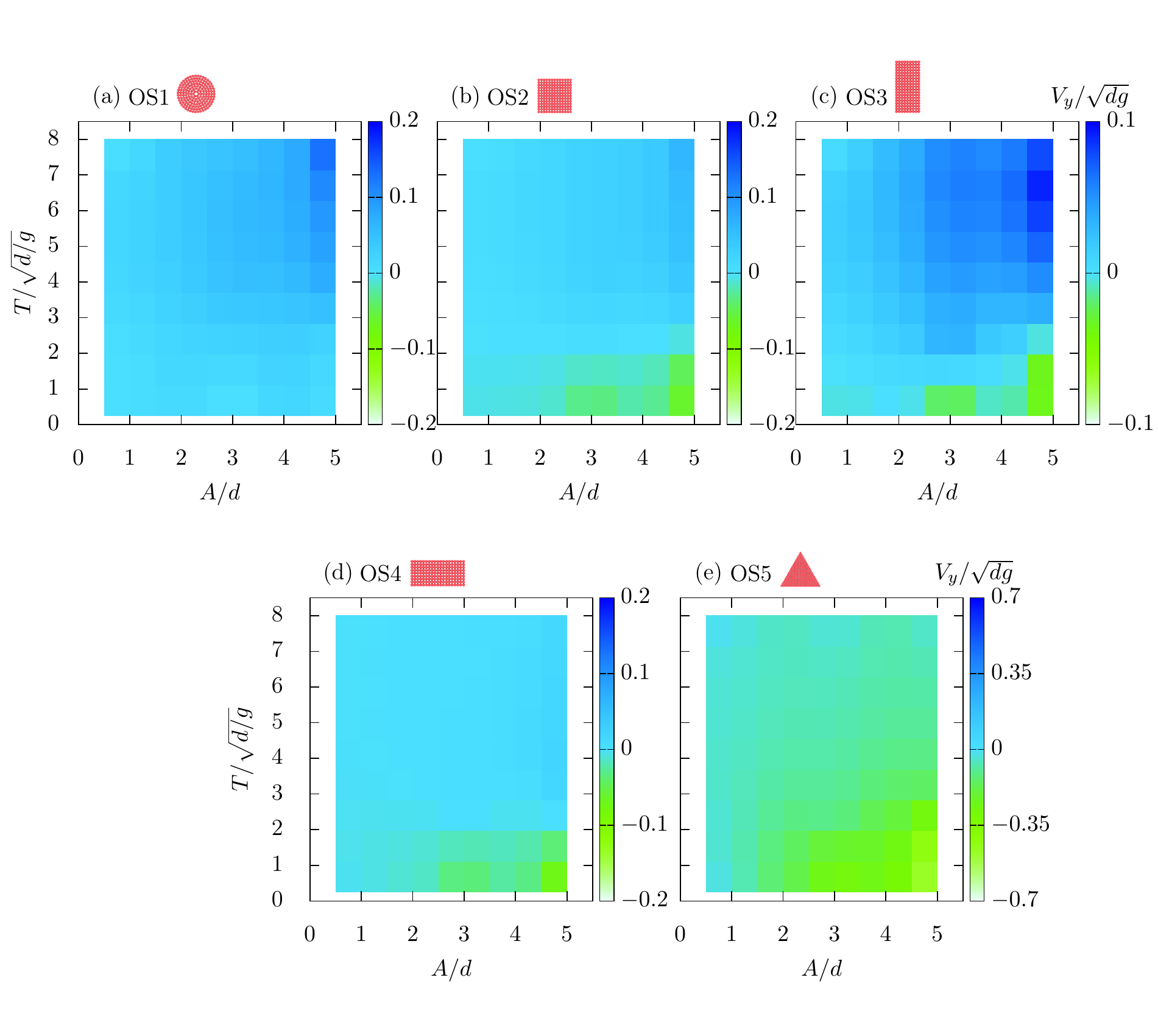}
	\caption{The colormap represents the rise rate $V_{y}$ of the intruder (a) disk, (b) square, (c) narrow-rectangle, (d) wide-rectangle, and (e) triangle for all the oscillated $T$ and $A$.\label{colrmap}} 
\end{figure}

From Eqn. \ref{eqn:fin}, we can say that an intruder can not sink when it is oscillated above its critical time period. The model will hold as long as the length of cavity is greater than one particle diameter, i.e., $2A>d$ or $A>d/2$, declaring $A_{min} \approx d/2$ for all the intruder which is consistent with our results. The sinking regime is induced by the fluidization of granular bed of the cavity and is more complicated to predict while taking into account the shape of the intruder. We have shown the vertical velocity by colormaping it as a function of $A$ and $T$ in figure \ref{colrmap} for all the intruder shapes.

\section{\label{sec:conclusions} Conclusions}

In the present work, we analyzed the vertical dynamics of a horizontally oscillating intruder (various shapes considered) via a numerical simulation method. Depending on the oscillation amplitude and time-period, the intruder can rise, sink, or almost remain in its initial position.
 
\subsection{Vertical displacement of oscillating intruder}

The vertical displacement of the intruder is sensitive to its oscillation parameters. We found that the intruder object rises in the granular medium at a high time period and sinks when oscillated at a low time period for $A/d > 1$. However, the triangular intruder is observed to have a negligible rise  even when it is oscillated  at a high time period. The final vertical settling position of the intruder changes non-monotonically with the time period of oscillation.

\subsection{Rate of rise or sink of intruder with time-period}

The shape of the intruder has a significant influence on its rate of climb or sink within the granular medium. We observed a maxima and minima for the rate of rise and sink for all the intruder shapes. The rate of sink was found to be the highest for $T/\sqrt{d/g}=1$ for all the intruder shapes when oscillated at $A/d =$ 3 and 5. Among all the shapes, we found that the disk ($OS1$) has the higher rate of rise, and triangle ($OS5$) has the higher rate of sink.

\subsection{Average rate of vertical displacement of an intruder with phase angle}

 The average rate of vertical displacement of the intruder varies with phase angle (Figure 7). The phase angle $\phi=0$ or $10\phi/\pi=0$ represents the centre point of oscillation when the intruder is moving from left to right, $\phi=\pi/2$ or $10\phi/\pi=5$ indicates the rightmost point of oscillation. The $V_y$ increases with $\phi$ where the intruder gradually climbs over the particles. While there is a gradual decrease in $V_y$ at certain range of phase angle which can be attributed to greater number of particles in the path of oscillating intruder providing a greater resistance and hampering the intruder's positive vertical displacement.

\subsection{Mean flow fields}

The mean flow fields are reported for different phase angles around the intruder. The velocity field around the intruder shows that flow velocity is more on one side of the oscillating direction for the different phase angles. The pressure field around each intruder is  asymmetric for different phase angles due to the wake formation. The area fraction field is observed to be less on one side and more on the other side due to the wake formation either to the left side of the intruder or to the right side, depending on its direction of motion. At the same time, the area fraction in the regions above and below the intruder is nearly equal to that of the bulk.

\subsection{Cavity Model}

We developed a theoretical model to predict motion of intruders oscillating in the granular medium at a specific amplitude and time period. The model was able to identify a critical time period, below which an oscillating intruder will not rise. For all the intruder shapes, a point of no return was proposed; if the granular particle passes this point, it will enter the cavity. These particles settling in the cavity will help the intruder to climb on them and rise.

\subsection{Scope for future work}

It will be interesting to study (i) the effect of the nature of the constituent granular particles on the rise/sink dynamics of oscillating intruders,  (ii) the impact of orientation for a triangular  intruder on its rise or sink (iii) effect of the  aspect ratio of the cylinder intruder on its vertical dynamics, and (iv) the rise and sink dynamics of horizontally oscillating intruder in a three-dimensional granular medium. \\\\

\textbf{Acknowledgement}

We thank PARAM-ISHAN, the high-performance computing  facility of IITG, where the simulations were performed.\\

\textbf{Declaration of Interests:} The authors report no conflict of interest.

\bibliographystyle{jfm}
\bibliography{jfm-wiggle.bib}

\begin{thebibliography}{33}
\expandafter\ifx\csname natexlab\endcsname\relax\def\natexlab#1{#1}\fi
\def\au#1{#1} \def\ed#1{#1} \def\yr#1{#1}\def\at#1{#1}\def\jt#1{\textit{#1}}
  \def\bt#1{#1}\def\bvol#1{\textbf{#1}} \def\vol#1{#1} \def\pg#1{#1}
  \def\publ#1{#1}\def\arxiv#1{#1}\def\org#1{#1}\def\st#1{\textit{#1}}

\bibitem[Brilliantov {\em et~al.\/}(1996)Brilliantov, Spahn, Hertzsch \&
  P{\"o}schel]{Brilliantov}
{\sc \au{Brilliantov, Nikolai~V}, \au{Spahn, Frank}, \au{Hertzsch, Jan-Martin}
  \& \au{P{\"o}schel, Thorsten}} \yr{1996}  \at{Model for collisions in
  granular gases}.  \jt{Physical review E}  \bvol{53}~(5),  \pg{5382}.

\bibitem[Clark {\em et~al.\/}(2012)Clark, Kondic \&
  Behringer]{PhysRevLett.109.238302}
{\sc \au{Clark, Abram~H.}, \au{Kondic, Lou} \& \au{Behringer, Robert~P.}}
  \yr{2012}  \at{Particle scale dynamics in granular impact}.  \jt{Phys. Rev.
  Lett.}  \bvol{109},  \pg{238302}.

\bibitem[Clark {\em et~al.\/}(2014)Clark, Petersen \&
  Behringer]{PhysRevE.89.012201}
{\sc \au{Clark, Abram~H.}, \au{Petersen, Alec~J.} \& \au{Behringer, Robert~P.}}
  \yr{2014}  \at{Collisional model for granular impact dynamics}.  \jt{Phys.
  Rev. E}  \bvol{89},  \pg{012201}.

\bibitem[Cundall \& Strack(1979)]{cundall1979discrete}
{\sc \au{Cundall, Peter~A} \& \au{Strack, Otto~DL}} \yr{1979}  \at{A discrete
  numerical model for granular assemblies}.  \jt{geotechnique}  \bvol{29}~(1),
  \pg{47--65}.

\bibitem[Dhiman {\em et~al.\/}(2020)Dhiman, Kumar, Reddy \&
  Gupta]{dhiman2020origin}
{\sc \au{Dhiman, Manish}, \au{Kumar, Sonu}, \au{Reddy, K~Anki} \& \au{Gupta,
  Raghvendra}} \yr{2020}  \at{Origin of the long-ranged attraction or repulsion
  between intruders in a confined granular medium}.  \jt{Journal of Fluid
  Mechanics}  \bvol{886}.

\bibitem[Ding {\em et~al.\/}(2011)Ding, Gravish \& Goldman]{ding2011drag}
{\sc \au{Ding, Yang}, \au{Gravish, Nick} \& \au{Goldman, Daniel~I}} \yr{2011}
  \at{Drag induced lift in granular media}.  \jt{Physical review letters}
  \bvol{106}~(2),  \pg{028001}.

\bibitem[Glasser \& Goldhirsch(2001)]{glasser2001scale}
{\sc \au{Glasser, BJ} \& \au{Goldhirsch, I}} \yr{2001}  \at{Scale dependence,
  correlations, and fluctuations of stresses in rapid granular flows}.
  \jt{Physics of Fluids}  \bvol{13}~(2),  \pg{407--420}.

\bibitem[Goldman(2014)]{RevModPhys.86.943}
{\sc \au{Goldman, Daniel~I.}} \yr{2014}  \at{Colloquium: Biophysical principles
  of undulatory self-propulsion in granular media}.  \jt{Rev. Mod. Phys.}
  \bvol{86},  \pg{943--958}.

\bibitem[Goldman \& Hu(2010)]{goldman2010wiggling}
{\sc \au{Goldman, Daniel~I.} \& \au{Hu, David~L.}} \yr{2010}  \at{Wiggling
  through the world: The mechanics of slithering locomotion depend on the
  surroundings}.  \jt{American Scientist}  \bvol{98}~(4),  \pg{314--323}.

\bibitem[Guillard {\em et~al.\/}(2014)Guillard, Forterre \&
  Pouliquen]{guillard2014lift}
{\sc \au{Guillard, Fran{\c{c}}ois}, \au{Forterre, Yo{\"e}l} \& \au{Pouliquen,
  Olivier}} \yr{2014}  \at{Lift forces in granular media}.  \jt{Physics of
  Fluids}  \bvol{26}~(4),  \pg{043301}.

\bibitem[Hu {\em et~al.\/}(2009)Hu, Nirody, Scott \& Shelley]{Hu10081}
{\sc \au{Hu, David~L.}, \au{Nirody, Jasmine}, \au{Scott, Terri} \& \au{Shelley,
  Michael~J.}} \yr{2009}  \at{The mechanics of slithering locomotion}.
  \jt{Proceedings of the National Academy of Sciences}  \bvol{106}~(25),
  \pg{10081--10085},  \arxiv{arXiv:
  https://www.pnas.org/content/106/25/10081.full.pdf}.

\bibitem[Huang {\em et~al.\/}(2016)Huang, Ran \&
  Blumenfeld]{PhysRevE.94.062906}
{\sc \au{Huang, Ling}, \au{Ran, Xianwen} \& \au{Blumenfeld, Raphael}} \yr{2016}
   \at{Vertical dynamics of a horizontally oscillating active object in a
  two-dimensional granular medium}.  \jt{Phys. Rev. E}  \bvol{94},
  \pg{062906}.

\bibitem[Kumar {\em et~al.\/}(2017)Kumar, Reddy, Takada \&
  Hayakawa]{kumar2017scaling}
{\sc \au{Kumar, Sonu}, \au{Reddy, K~Anki}, \au{Takada, Satoshi} \&
  \au{Hayakawa, Hisao}} \yr{2017}  \at{Scaling law of the drag force in dense
  granular media}.  \jt{arXiv preprint arXiv:1712.09057} .

\bibitem[Li {\em et~al.\/}(2013)Li, Zhang \& Goldman]{li2013terradynamics}
{\sc \au{Li, Chen}, \au{Zhang, Tingnan} \& \au{Goldman, Daniel~I}} \yr{2013}
  \at{A terradynamics of legged locomotion on granular media}.  \jt{science}
  \bvol{339}~(6126),  \pg{1408--1412}.

\bibitem[Liu {\em et~al.\/}(2021)Liu, Ran, Cheng, Tang, Zhou \&
  Blumenfeld]{liu2021locomotion}
{\sc \au{Liu, Ping}, \au{Ran, Xianwen}, \au{Cheng, Qi}, \au{Tang, Wenhui},
  \au{Zhou, Jingyuan} \& \au{Blumenfeld, Raphael}} \yr{2021}  \at{Locomotion of
  self-excited vibrating and rotating objects in granular environments}.
  \jt{Applied Sciences}  \bvol{11}~(5),  \pg{2054}.

\bibitem[Maladen {\em et~al.\/}(2009)Maladen, Ding, Li \&
  Goldman]{maladen2009undulatory}
{\sc \au{Maladen, Ryan~D}, \au{Ding, Yang}, \au{Li, Chen} \& \au{Goldman,
  Daniel~I}} \yr{2009}  \at{Undulatory swimming in sand: subsurface locomotion
  of the sandfish lizard}.  \jt{science}  \bvol{325}~(5938),  \pg{314--318}.

\bibitem[Maladen {\em et~al.\/}(2011{\natexlab{{\em a\/}}})Maladen, Ding,
  Umbanhowar \& Goldman]{maladen2011undulatory}
{\sc \au{Maladen, Ryan~D}, \au{Ding, Yang}, \au{Umbanhowar, Paul~B} \&
  \au{Goldman, Daniel~I}} \yr{2011{\natexlab{{\em a\/}}}}  \at{Undulatory
  swimming in sand: experimental and simulation studies of a robotic sandfish}.
   \jt{The International Journal of Robotics Research}  \bvol{30}~(7),
  \pg{793--805}.

\bibitem[Maladen {\em et~al.\/}(2011{\natexlab{{\em b\/}}})Maladen, Umbanhowar,
  Ding, Masse \& Goldman]{maladen2011granular}
{\sc \au{Maladen, Ryan~D}, \au{Umbanhowar, Paul~B}, \au{Ding, Yang}, \au{Masse,
  Andrew} \& \au{Goldman, Daniel~I}} \yr{2011{\natexlab{{\em b\/}}}} Granular
  lift forces predict vertical motion of a sand-swimming robot.  \bt{In {\em
  2011 IEEE International Conference on Robotics and Automation\/}},  \pg{pp.
  1398--1403}. IEEE.

\bibitem[Marvi {\em et~al.\/}(2014)Marvi, Gong, Gravish, Astley, Travers,
  Hatton, Mendelson~III, Choset, Hu \& Goldman]{marvi2014sidewinding}
{\sc \au{Marvi, Hamidreza}, \au{Gong, Chaohui}, \au{Gravish, Nick}, \au{Astley,
  Henry}, \au{Travers, Matthew}, \au{Hatton, Ross~L}, \au{Mendelson~III,
  Joseph~R}, \au{Choset, Howie}, \au{Hu, David~L} \& \au{Goldman, Daniel~I}}
  \yr{2014}  \at{Sidewinding with minimal slip: Snake and robot ascent of sandy
  slopes}.  \jt{Science}  \bvol{346}~(6206),  \pg{224--229}.

\bibitem[Mazouchova {\em et~al.\/}(2010)Mazouchova, Gravish, Savu \&
  Goldman]{mazouchova2010utilization}
{\sc \au{Mazouchova, Nicole}, \au{Gravish, Nick}, \au{Savu, Andrei} \&
  \au{Goldman, Daniel~I}} \yr{2010}  \at{Utilization of granular solidification
  during terrestrial locomotion of hatchling sea turtles}.  \jt{Biology
  letters}  \bvol{6}~(3),  \pg{398--401}.

\bibitem[Meirion-Griffith \& Spenko(2010)]{meirion2010empirical}
{\sc \au{Meirion-Griffith, Gareth} \& \au{Spenko, Matthew}} \yr{2010} An
  empirical study of the terramechanics of small unmanned ground vehicles.
  \bt{In {\em 2010 IEEE Aerospace Conference\/}},  \pg{pp. 1--6}. IEEE.

\bibitem[Ping {\em et~al.\/}(2020)Ping, Ran \& Blumenfeld]{ping2020sink}
{\sc \au{Ping, Liu}, \au{Ran, Xianwen} \& \au{Blumenfeld, Raphael}} \yr{2020}
  \at{Sink-rise dynamics of horizontally oscillating active matter in granular
  media: Theory}.  \jt{arXiv preprint arXiv:2006.04160} .

\bibitem[Plimpton(1995)]{PLIMPTON19951}
{\sc \au{Plimpton, Steve}} \yr{1995}  \at{Fast parallel algorithms for
  short-range molecular dynamics}.  \jt{Journal of Computational Physics}
  \bvol{117}~(1),  \pg{1 -- 19}.

\bibitem[Raibert {\em et~al.\/}(2008)Raibert, Blankespoor, Nelson \&
  Playter]{raibert2008bigdog}
{\sc \au{Raibert, Marc}, \au{Blankespoor, Kevin}, \au{Nelson, Gabriel} \&
  \au{Playter, Rob}} \yr{2008}  \at{Bigdog, the rough-terrain quadruped robot}.
   \jt{IFAC Proceedings Volumes}  \bvol{41}~(2),  \pg{10822--10825}.

\bibitem[Reddy {\em et~al.\/}(2021)Reddy, Kumar \& Reddy]{reddy2021granular}
{\sc \au{Reddy, A Vamsi~Krishna}, \au{Kumar, Sonu} \& \au{Reddy, K~Anki}}
  \yr{2021}  \at{Granular particle-shape heterogeneous mixtures discharging
  through a silo}.  \jt{Journal of Fluid Mechanics}  \bvol{912}.

\bibitem[Sharpe {\em et~al.\/}(2013)Sharpe, Ding \&
  Goldman]{sharpe2013environmental}
{\sc \au{Sharpe, Sarah~S}, \au{Ding, Yang} \& \au{Goldman, Daniel~I}} \yr{2013}
   \at{Environmental interaction influences muscle activation strategy during
  sand-swimming in the sandfish lizard scincus scincus}.  \jt{Journal of
  Experimental Biology}  \bvol{216}~(2),  \pg{260--274}.

\bibitem[Sharpe {\em et~al.\/}(2015)Sharpe, Kuckuk \&
  Goldman]{sharpe2015controlled}
{\sc \au{Sharpe, Sarah~S}, \au{Kuckuk, Robyn} \& \au{Goldman, Daniel~I}}
  \yr{2015}  \at{Controlled preparation of wet granular media reveals limits to
  lizard burial ability}.  \jt{Physical biology}  \bvol{12}~(4),  \pg{046009}.

\bibitem[Silbert {\em et~al.\/}(2001)Silbert, Erta{\c{s}}, Grest, Halsey,
  Levine \& Plimpton]{silbert2001granular}
{\sc \au{Silbert, Leonardo~E}, \au{Erta{\c{s}}, Deniz}, \au{Grest, Gary~S},
  \au{Halsey, Thomas~C}, \au{Levine, Dov} \& \au{Plimpton, Steven~J}} \yr{2001}
   \at{Granular flow down an inclined plane: Bagnold scaling and rheology}.
  \jt{Physical Review E}  \bvol{64}~(5),  \pg{051302}.

\bibitem[Stukowski({2010})]{ovito}
{\sc \au{Stukowski, Alexander}} \yr{{2010}}  \at{{Visualization and analysis of
  atomistic simulation data with OVITO-the Open Visualization Tool}}.
  \jt{{MODELLING AND SIMULATION IN MATERIALS SCIENCE AND ENGINEERING}}
  \bvol{{18}}~({1}).

\bibitem[Tripura {\em et~al.\/}(2021)Tripura, Kumar, Reddy \&
  Talbot]{Bitang2021drag}
{\sc \au{Tripura, Bitang~Kwrung}, \au{Kumar, Sonu}, \au{Reddy, K.~Anki} \&
  \au{Talbot, Julian}} \yr{2021}  \at{Role of shape on the forces on an
  intruder moving through a dense granular medium}.  \jt{Particulate Science
  and Technology}  \bvol{0}~(0),  \pg{1--11},  \arxiv{arXiv:
  https://doi.org/10.1080/02726351.2021.1983905}.

\bibitem[Urquhart(2010)]{URQUHART201023}
{\sc \au{Urquhart, James}} \yr{2010}  \at{Lizard-inspired robot can ‘swim’
  through sand}.  \jt{New Scientist}  \bvol{206}~(2766),  \pg{23}.

\bibitem[Zhang {\em et~al.\/}(2013)Zhang, Qian, Li, Masarati, Hoover,
  Birkmeyer, Pullin, Fearing \& Goldman]{zhang2013ground}
{\sc \au{Zhang, Tingnan}, \au{Qian, Feifei}, \au{Li, Chen}, \au{Masarati,
  Pierangelo}, \au{Hoover, Aaron~M}, \au{Birkmeyer, Paul}, \au{Pullin, Andrew},
  \au{Fearing, Ronald~S} \& \au{Goldman, Daniel~I}} \yr{2013}  \at{Ground
  fluidization promotes rapid running of a lightweight robot}.  \jt{The
  International Journal of Robotics Research}  \bvol{32}~(7),  \pg{859--869}.

\bibitem[Zhu {\em et~al.\/}(2021)Zhu, Yang, Li, Wang, Shui \&
  Chen]{ZHU2021104629}
{\sc \au{Zhu, Liangliang}, \au{Yang, Pengfei}, \au{Li, Feng}, \au{Wang, Kai},
  \au{Shui, Langquan} \& \au{Chen, Xi}} \yr{2021}  \at{On the snake-like
  lateral un-dulatory locomotion in terrestrial, aquatic and sand
  environments}.  \jt{Journal of the Mechanics and Physics of Solids}
  \bvol{157},  \pg{104629}.

\end{thebibliography}

\end{document}